\documentclass[psfig]{aa} 

\usepackage{times}
\usepackage{graphics}
\usepackage{amssymb}
\usepackage{epsfig}

\begin{document}


\title{
Infrared to millimetre photometry of ultra-luminous IR galaxies\thanks{Based on observations with the 
Infrared Space Observatory ISO, the James Clerk Maxwell Telescope JCMT, the Swedish ESO Submillimetre 
Telescope SEST and at the Calar Alto Observatory. ISO is an ESA project with instruments funded by ESA 
Member States (especially the PI countries France, Germany, the Netherlands and the United Kingdom) and 
with the participation of ISAS and NASA.}: new evidence favouring a 3--stage dust model.
}

\author{
Ulrich~Klaas\inst{1}
\and
Martin~Haas\inst{1}
\and
Sven~A.H.~M\"uller\inst{1,2}
\and
Rolf~Chini\inst{2}
\and
Bernhard~Schulz\inst{3}
\and
Iain~Coulson\inst{4}
\and \\
Hans~Hippelein\inst{1}
\and
Karsten~Wilke\inst{1}
\and
Marcus~Albrecht\inst{2}
\and
Dietrich~Lemke\inst{1}
}

\offprints{U.\ Klaas (klaas@mpia.de)}

\institute{
Max--Planck--Institut f\"ur Astronomie (MPIA), K\"onigstuhl \/17, \/D--69117 Heidelberg, Germany, 
\and
Astronomisches Institut, Ruhr-Universit\"at Bochum, \/D--44780 Bochum, Germany
\and
ISO Data Centre, Astrophysics Division, Space Science Dep. of ESA,
Villafranca, P.O. Box 50727, E--28080 Madrid, Spain
\and
Joint Astronomy Centre, 660 N. Aohuku Place, University Park, Hilo 96720, Hawaii, USA
}

\date{Received 21 December 2000; accepted 3 October 2001}

\authorrunning{U. Klaas et al.} 
\titlerunning{Ultra-luminous IR galaxies}

\abstract{Infrared to millimetre spectral energy distributions (SEDs) have been obtained for 
41 bright ultra-luminous infrared galaxies (ULIRGs). The observations were carried out with ISOPHOT 
between 10 and 200\,$\mu$m and supplemented for 16 sources with JCMT/SCUBA at 450 and 850\,$\mu$m and 
with SEST at 1.3\, mm. In addition, seven sources were observed at 1.2 and 2.2\,$\mu$m with the 2.2 m 
telescope on Calar Alto. These new SEDs represent the most complete set of infrared photometric templates 
obtained so far on ULIRGs in the local universe. \\
The SEDs peak at 60-100\,$\mu$m and show often a quite shallow Rayleigh-Jeans tail. Fits with one single 
modified blackbody yield a high FIR opacity and small dust emissivity exponent $\beta$\,$<$\,2. However, 
this concept leads to conflicts with several other observational constraints, like the low PAH extinction 
or the extended filamentary optical morphology. A more consistent picture is obtained using several dust 
components with $\beta$\,=\,2, low to moderate FIR opacity and cool (50\,K\,$>$\,T\,$>$\,30\,K) to cold 
(30\,K\,$>$\,T\,$>$\,10\,K) temperatures. This provides evidence for two dust stages, the cool starburst 
dominated one and the cold cirrus-like one. The third stage with several hundred Kelvin warm dust is 
identified in the AGN dominated ULIRGs, showing up as a NIR-MIR power-law flux increase. While AGNs and SBs 
appear indistinguishable at FIR and submm wavelengths, they differ in the NIR-MIR. This suggests that the 
cool FIR emitting dust is not related to the AGN, and that the AGN only powers the warm and hot dust. 
In comparison with optical and MIR spectroscopy, a criterion based on the SED shapes and the NIR colours is 
established to reveal AGNs among ULIRGs. Also the possibility to recognise evolutionary trends among the 
ULIRGs via the relative amounts of cold, cool and warm dust components is investigated.
}
\maketitle

\begin{table*}
\caption{\label{tab_fluxes}
        Measured flux densities in Jy as a function of wavelength in $\mu$m. The objects are 
        listed by IRAS name, except when another names is more common, then the IRAS name is given in 
        the footnote. The statistical errors from signal processing are typically about 5\,\% at 
        100--150\,$\mu$m, and reach in the other passbands 10--20\% of the quoted fluxes depending 
        mainly on the source/sky brightness contrast as seen by the apertures used. In case of 
        uncertainties larger than 30\,\% 3\,$\sigma$ upper limits are listed. The apertures used are 
        30$\arcsec\,\times$\,30$\arcsec$ at 1.2 and 2.2\,$\mu$m, circular with 52$\arcsec$ diameter 
        at 10--25\,$\mu$m, 46$\arcsec\,\times$\,46$\arcsec$ at 60 and 90\,$\mu$m, 
        184$\arcsec\,\times$\,184$\arcsec$ at 120--200\,$\mu$m, and circular with 8$\arcsec$, 15$\arcsec$,
        24$\arcsec$ diameters at 450, 850, 1300\,$\mu$m, respectively.}
\scriptsize
\begin{tabular}{lrrrrrrrrrrrrrrr}
\hline\noalign{\smallskip}
{\bf Name}&{\bf 1.2}&{\bf 2.2}&{\bf 10}&{\bf 12}&{\bf 15}&{\bf 25}&{\bf 60}&{\bf 90}&{\bf 120}&{\bf 150}&{\bf 180}&{\bf 200}&{\bf 450}&{\bf 850}&{\bf 1300}\\
\hline\noalign{\smallskip} 
00199-7426     &  &  &   0.057&$<$0.120&   0.09& 0.25&      &     &  7.50& 5.34&   4.21&   3.00&        &        &          \\ 
00262+4251$^*$ & 0.0054& 0.0055&$<$0.087&   0.043&   0.10& 0.36&  3.36& 2.60&  2.90& 2.40&   2.40&$<$2.10&        &        &          \\ 
00406-3127     &  &  &        &        &       &     &      &     &  0.77& 0.63&   0.44&   0.32&        &        &          \\ 
03068-5346     &  &  &$<$0.063&   0.040&   0.05& 0.15&  3.60& 3.34&  3.33& 2.78&   1.72&   1.35&        &        &          \\ 
03158+4227     &  &  &$<$0.165&   0.085&$<$0.21& 0.25&  5.25& 3.88&  3.19& 1.75&$<$1.79&   1.05&        &        &          \\ 
03538-6432     &  &  &        &        &       &     &      &     &  1.37& 1.07&   0.85&   0.62&        &        &          \\ 
04232+1436     &0.0055&0.0106&$<$0.108&   0.065&   0.07& 0.39&  3.88& 4.14&  3.50& 2.30&   1.60&   1.10&        &        &          \\ 
05189-2524     &  &  &   0.600&   0.800&   1.00& 3.08& 13.02&10.70& 10.00& 6.80&   4.36&   3.84&        &        & $<$0.036 \\ 
06035-7102     &  &  &   0.090&   0.130&   0.15& 0.53&  5.86& 5.05&  5.15& 2.90&   2.01&   1.35&        &        &          \\ 
06206-6315     &  &  &$<$0.072&   0.040&   0.11& 0.24&  4.84& 4.20&  4.57& 2.90&   1.87&   1.50&        &        &          \\ 
12112+0305     &  &  &   0.065&   0.100&   0.20& 0.44&  9.94& 8.95&  8.31& 5.87&   3.93&   2.93&        &        &          \\ 
MK231$^a$      &  &  &   1.425&   2.400&   2.90& 8.66& 31.68&27.34& 24.32&14.74&   9.75&   6.88&        &        &          \\ 
MK273$^b$      &  &  &   0.100&   0.250&   0.50& 2.07& 27.45&23.78& 19.99&13.10&   8.69&   7.40&        &        &          \\ 
MK463$^c$      &  &  &   0.425&   0.580&   0.80& 1.79&  2.33& 1.94&  1.60& 0.99&$<$0.84&   0.45&        &        &          \\ 
14348-1447     &  &  &$<$0.096&   0.108&$<$0.27& 0.42&  8.21& 7.23&  6.66& 5.61&   3.76&   2.69&    0.21&   0.024& $<$0.009 \\ 
14378-3651     &  &  &$<$0.276&$<$0.399&$<$0.36& 0.41&  8.53& 7.49&  5.90& 3.60&   2.80&   2.45&        &        & $<$0.010 \\ 
15245+1019     &  &  &$<$0.084&   0.051&$<$0.18& 0.22&  5.25& 5.43&  5.58& 3.60&   1.94&   1.31&        &        &          \\ 
15250+3609     &  &  &$<$0.267&   0.238&   0.56& 1.31&  8.63& 6.66&  4.16& 2.37&   1.60&   1.16&        &        &          \\ 
Arp220$^d$     &  &  &   0.147&   0.600&   1.14& 8.28&113.35&111.50&109.00&87.89&  63.95&  54.81&        &        &          \\ 
15462-0450$^*$ &  &  &   0.064&   0.100&   0.15& 0.38&  2.81&     &      & 2.80&   2.44&   $<$2.5&        &        &          \\ 
16090-0139     &  &  &$<$0.252&   0.100&$<$0.21& 0.20&  6.70& 5.80&  4.00& 2.75&   2.00&   1.21& $<$0.13&$<$0.017&          \\ 
NGC6240$^e$    &  &  &   0.259&   0.750&   1.00& 3.31& 23.60&26.70& 25.90&18.91&  12.73&   9.00&    1.00&   0.150&          \\ 
17208-0014     &  &  &   0.080&   0.200&   0.25& 1.32& 32.22&31.90& 30.00&23.00&  17.50&  12.50&    1.07&   0.155&          \\ 
17463+5806     &  &  &        &        &       &     &      &     &  0.63& 0.58&   0.37&   0.23&        &        &          \\ 
18090+0130$^*$ & 0.0200& 0.0195&   0.187&   0.300&$<$0.38& 0.59&      &     & 16.63&14.50&  10.48&   9.74&        &        &          \\ 
18470+3234     & 0.0027& 0.0036&$<$0.081&   0.135&   0.15& 0.44&  4.15& 3.55&  3.20& 2.45&   1.70&   1.25&        &        &          \\ 
19254-7245$^f$ &  &  &   0.123&   0.200&   0.40& 1.32&  5.57& 5.26&  4.27& 3.06&   2.30&   1.61&        &        &    0.012 \\ 
19458+0944$^*$ &  &  &$<$0.396&$<$0.540&$<$0.77& 0.26&      & 5.82&      & 7.61&       &   7.05&        &         &         \\ 
20046-0623     & 0.0039& 0.0040&$<$0.084&   0.125&$<$0.39& 0.47&  4.50& 4.71&  3.69& 3.18&   2.79&   1.40& $<$0.16&$<$0.026&          \\ 
20087-0308     &  &  &$<$0.117&   0.070&   0.10& 0.25&  5.54& 4.85&  5.96& 4.05&   2.95&   2.30& $<$0.10&$<$0.010&          \\ 
20100-4156     &  &  &$<$0.090&   0.090&   0.14& 0.34&  6.53& 5.11&  4.51& 3.36&   2.09&   1.86&        &        & $<$0.014 \\ 
20414-1651     &  &  &$<$0.168&$<$0.117&$<$0.18& 0.25&  6.16& 4.22&  3.60& 2.56&   1.54&$<$1.35&        &        &          \\ 
ESO286-19$^g$  &  &  &$<$0.204&   0.275&   0.50& 2.11& 13.90& 9.26&  6.73& 4.14&   2.56&   1.80&        &        &          \\ 
21130-4446     &  &  &$<$0.090&   0.055&$<$0.45& 0.15&  4.03& 3.60&  4.50& 3.94&   3.14&   2.90&        &        & $<$0.023 \\ 
21504-0628     & 0.0055& 0.0057&   0.040&   0.050&   0.15& 0.28&  3.33& 2.82&  2.80& 1.52&   1.26&   0.70& $<$0.11&$<$0.016&          \\ 
22491-1808     &  &  &$<$0.063&   0.080&   0.11& 0.50&  6.80& 5.45&  3.55& 2.70&   1.95&   1.70& $<$0.20&   0.019&          \\ 
ESO148-2$^h$   &  &  &   0.126&   0.250&   0.36& 1.55& 12.50&11.36& 10.00& 6.40&   4.70&   3.50&        &        &          \\ 
23230-6926     &  &  &   0.035&   0.060&   0.10& 0.29&  4.80& 3.98&  4.10& 3.00&   2.00&   1.70&        &        &    0.010 \\ 
23365+3604     & 0.0075& 0.0065&$<$0.120&   0.080&   0.15& 0.65&  7.78& 6.76&  6.75& 4.81&   3.35&   3.01&    0.17&   0.020&          \\ 
23389-6139     &  &  &$<$0.075&   0.031&$<$0.26& 0.20&  3.88& 3.30&  4.03& 2.83&   1.99&   1.52&        &        & $<$0.005 \\ 
23515-2917     &  &  &        &        &       &     &      &     &  0.80& 0.64&   0.48&   0.33&        &        &          \\

\hline\noalign{\smallskip}
\end{tabular}
\\
$^{*}$ ISO data partly distorted and uncertain;    
$^{a}$ 12540+5708,    
$^{b}$ 13428+5608,    
$^{c}$ 13536+1836,    
$^{d}$ 15327+2340,    
$^{e}$ 16504+0228,    
$^{f}$ 'Super-Antennae',    
$^{g}$ 20551-4250,    
$^{h}$ 23128-5919    
\end{table*}

\section{Introduction}

Ultra-luminous IR galaxies (ULIRGs) emit the bulk of their energy 
in the mid- and far-infrared with L$_{\rm IR} >$ 10$^{\rm 12}$\,L$_{\odot}$. 
Optical and near-infrared images show a disturbed morphology and signatures 
of interaction and merging. Spectra show emission lines characteristic for strong 
starbursts, shocks (LINERs) and in some cases also Seyfert types (AGNs). 
Sanders et al.\ (1988a) suggested that the ULIRGs form an evolutionary link 
between starburst galaxies and quasars. The current literature and the research 
state are excellently reviewed by Sanders \& Mirabel (1996), and most recently by 
Genzel \& Cesarsky (2000).

Here we focus on the following topics: 
\begin{itemize}
\item[1)] While the luminosity of an ULIRG originates from warm T\,$>$\,100\,K and 
          cool T\,$>$\,30\,K dust, the bulk of the dust in normal spiral galaxies is 
          much colder at T\,$\approx$\,15\,K. Do ULIRGs also have such cold cirrus-like 
          dust, or is all the dust already heated up to T\,$>$\,30\,K? 
          This is observationally related to: How does the IR SED continue beyond 
          100\,$\mu$m into the submm range? For bright but less luminous IRAS galaxies a 
          three component model -- with cirrus, starburst and AGN dust -- has been 
          proposed and successfully applied by Rowan-Robinson \& Crawford (1989).
\item[2)] What can we learn about the evolution of ULIRGs from such a cold dust stage? 
          How far can the merging state be inferred from the relative amounts of cool and 
          cold dust? For example, with respect to the Sanders et al. (1988a) conjecture 
          this means: Do AGN-ULIRGs have less cold dust than starburst dominated SB-ULIRGs? 
\item[3)] Can the dust mass and opacity provide clues in favour of or against hidden AGNs? 
          How does this compare to the near- and mid-IR emission of ULIRGs whose optical 
          and MIR spectra favor their power to be dominated by either AGNs or SBs? 
          Observationally this relates to measuring the NIR-MIR SEDs, which are currently  
          known only for a few ULIRGs. 
\end{itemize}

These topics introduce new aspects for exploring the ULIRGs. In this paper, based on IR 
to mm photometry, we present a sample of observational templates, and confine the discussion 
to a few basic implications, which may further stimulate more detailed analyses.

\section{The sample}

From the Queen Mary \& Westfield College IRAS galaxy catalogue (QIGC, 17711 entries, 
Rowan-Robinson et al., 1991) we chose the 48 brightest sources with (1) F$_{\rm 60\mu m} >$ 
3\,Jy and (2) L$_{\rm FIR}$ = L$_{\rm 40-120\,\mu m} >$ 10$^{\rm 12}$\,L$_{\odot}$ 
(based on H$_{\rm 0}$ = 50\,km/s/Mpc). Throughout this paper, however, we use 
H$_{\rm 0}$ = 75\,km/s/Mpc, thus, some of the sources reach only about 
10$^{\rm 11.5}$\,L$_{\odot}$. In the course of the ISO mission 37 out of these 48 sources 
could be observed (77\,\%), all objects having z $<$ 0.2. Thus, the sample is representative 
for the nearby bright ULIRGs. In coordination with the other ISO Guaranteed Time programmes the 
two slightly less luminous objects NGC 6240 and Mrk 463 were included.  
Later we observed four more objects with higher redshifts (0.4 $>$ z $>$ 0.2, 
F$_{\rm 60\,\mu m} \approx$ 1\,Jy) which were visible to the satellite.

\section{Observations and data reduction}

\subsection{ISOPHOT 10\,--\,200\,$\mu$m data}
\label{sect_ISOPHOTobs}

The observations were carried out with ISOPHOT (Lemke et al. 1996), 
the photometer on board ISO (Kessler et al. 1996). A description
of the ISOPHOT observation templates (AOTs) is given in Laureijs et al. (2000).
The MIR 10 to 25\,$\mu$m observations were done in triangular chopped mode 
(AOT P03) with a 52$\arcsec$ aperture and a chopper throw of 60$\arcsec$ 
which matched the compact size of the targets well. 
The FIR 60 to 200\,$\mu$m observations were performed with the array
cameras C100 and C200 in the ``sparse map'' off--on mode (AOT P37-39)
with background positions 3-4 $\arcmin$ north of the source. Beam sizes
used for photometry are given in Table~\ref{tab_fluxes}. The four faint 
sources at higher redshift were observed in mini-map mode (AOT P22) 
in a 4\,$\times$\,2 raster and restricted to the filters at 120, 150, 180, 
and 200 $\mu$m.

The data were reduced using the PHT Interactive Analysis tool (PIA\footnote
{PIA is a joint development by the ESA Astrophysics Division and the ISOPHOT
consortium led by the Max--Planck--Institut f\"ur Astronomie, Heidelberg.}) 
(V7.3.3e) in standard processing mode, together with the calibration data set 
V4.0 (ISOPHOT Data User Manual V4.0, Laureijs et al. 1998). This includes 
correction for non-linearity of the electronics, deglitching (removal of data 
disturbed by cosmic particle events), and correction for signal dependence on 
the reset interval time. To handle the signal transients in the time series of 
bright FCS illuminations, only the last half was taken, when the values approached 
the final signal level. The calibration of detector responsivity and its changes 
was performed using associated measurements of the thermal fine calibration source 
(FCS) on board.  

For the chopped measurements with the P1 and P2 detectors at wavelengths 
10-25 $\mu$m we inspected the sequence of chopper plateaux and removed  
outliers, with the criterion that the uncertainty was larger than the average 
three sigma of the measurements. By this procedure typically the first 
chopper plateau of a series and those with residual glitches were removed.

Finally the fluxes were corrected for aperture/beam size effects. 
For the 3\,$\times$\,3 pixel C100 array (60 and 90\,$\mu$m) 
the fluxes were derived with two methods using (1) only the central pixel 
(46$\arcsec\,\times$\,46$\arcsec$) and (2) the whole array 
(138$\arcsec\,\times$138$\arcsec$). Both methods yield essentially the 
same flux (with some higher noise for the whole array), providing evidence that 
the objects are point-like and not resolved in the FIR. 

The error propagation in ISOPHOT data reduction is described in Laureijs \& Klaas
(1999). The statistical errors derived from signal processing are about 5-20\,\%, 
depending on the wavelength range and object brightness (see caption to 
Tab.~\ref{tab_fluxes}), but systematic errors due to absolute calibration accuracy 
are estimated to be 30\,\% (Klaas et al. 1998b). To account for the overall 
uncertainty in the signal derivation as well as relative and absolute photometric 
calibration we have adopted a general photometric uncertainty of 30\,\%.

\subsection{SCUBA 450 and 850\,$\mu$m data}
\label{sect_SCUBAobs}

The 450 and 850 $\mu$m observations were obtained on July 1st and 3rd, 1999, 
using the Submillimetre Common User Bolometer Array (SCUBA, Holland et al. 1999) 
at the James Clerk Maxwell Telescope (JCMT) on Mauna Kea, Hawaii. 
In general, photometry with 1$\arcmin$ chopper throw, mini-jiggle and 
beam switching was performed. We used the SCUBA narrow band filters in order 
to minimize possible CO line contributions (see also sect.\,\ref{sect_seds}). The 
atmospheric transmission was determined every hour from measurement series at various 
zenith distances (sky dips). The conditions were excellent and stable 
($\tau$$_{850\,\mu m}$\,=\,0.134$\pm$0.002 on July 1st and 0.209$\pm$0.003 on 
July 3rd, respectively). $\tau$$_{450\,\mu m}$ was extrapolated from the 850\,$\mu$m 
skydips ($\tau$$_{450\,\mu m}$\,=\,0.65$\pm$0.01 on July 1st and 1.09$\pm$0.02 on 
July 3rd, respectively). Mars and Uranus served as standard calibrators, and repeated 
photometry measurements gave a reproducibility of better than 5\,\% and 14\,\% at 
850\,$\mu$m and 450\,$\mu$m, respectively. The observing time per source was 
10 -- 20\,min (actually the dome carousel driver was out of order, so that the targets 
could only be observed during their oblique rising passage across the dome slit pointing 
towards south-east). The data were reduced using the SCUBA User Reduction Facility 
(SURF) with special emphasis on identifying noisy bolometer pixels and removing the sky 
noise determined from off-source pixels. 

In order to check for possible extended submm flux, jiggle maps covering 2$\arcmin$ 
were obtained for three sources. These sources were NGC6240 and 17208-0014, both 
revealing a submm flux excess (as discussed below), and 20046-0623 providing the same 
good upper limits for different beams. None of the sources seems to be extended at the 
spatial resolution of about 8$\arcsec$ and 15$\arcsec$ FWHM, respectively. 
These test cases suggest that for the other sources the bulk of submm flux is also 
contained within our photometry beams and that some possible extended submm flux is 
small and lies within the total adopted uncertainties of about 20-30\,\%, in 
particular since NGC\,6240 and 17208-0014 lie in the low redshift range of our sample.

\subsection{SEST 1300 $\mu$m data}
\label{sect_SESTobs}

The $1300\,\mu$m observations were obtained between July 4th and 9th, 1999, using the 
MPIfR single channel facility bolometer (Kreysa 1990) at the Swedish ESO 
Submillimetre Telescope (SEST) on La Silla, Chile. The system provides a beamsize of 
24$\arcsec$ and was used in the ON-OFF mode with dual beam switching and a beam separation 
of 68$\arcsec$ in azimuth. Pointing and focus were checked at least every two hours and 
the pointing accuracy was always better than 4$\arcsec$. The atmospheric transmission as 
determined by sky dips was stable at about  $\tau_{1300} \approx$\, 0.1. Uranus 
served as a standard calibrator. The observing time per source was about 1 hour (on-source). 
From repeated calibrator measurements we estimate the absolute accuracy to be about 
20-30\,\%.

\begin{figure*}[ht!]
\vspace{1.5cm}
\hspace{1.5cm}\resizebox{16cm}{!}{\includegraphics{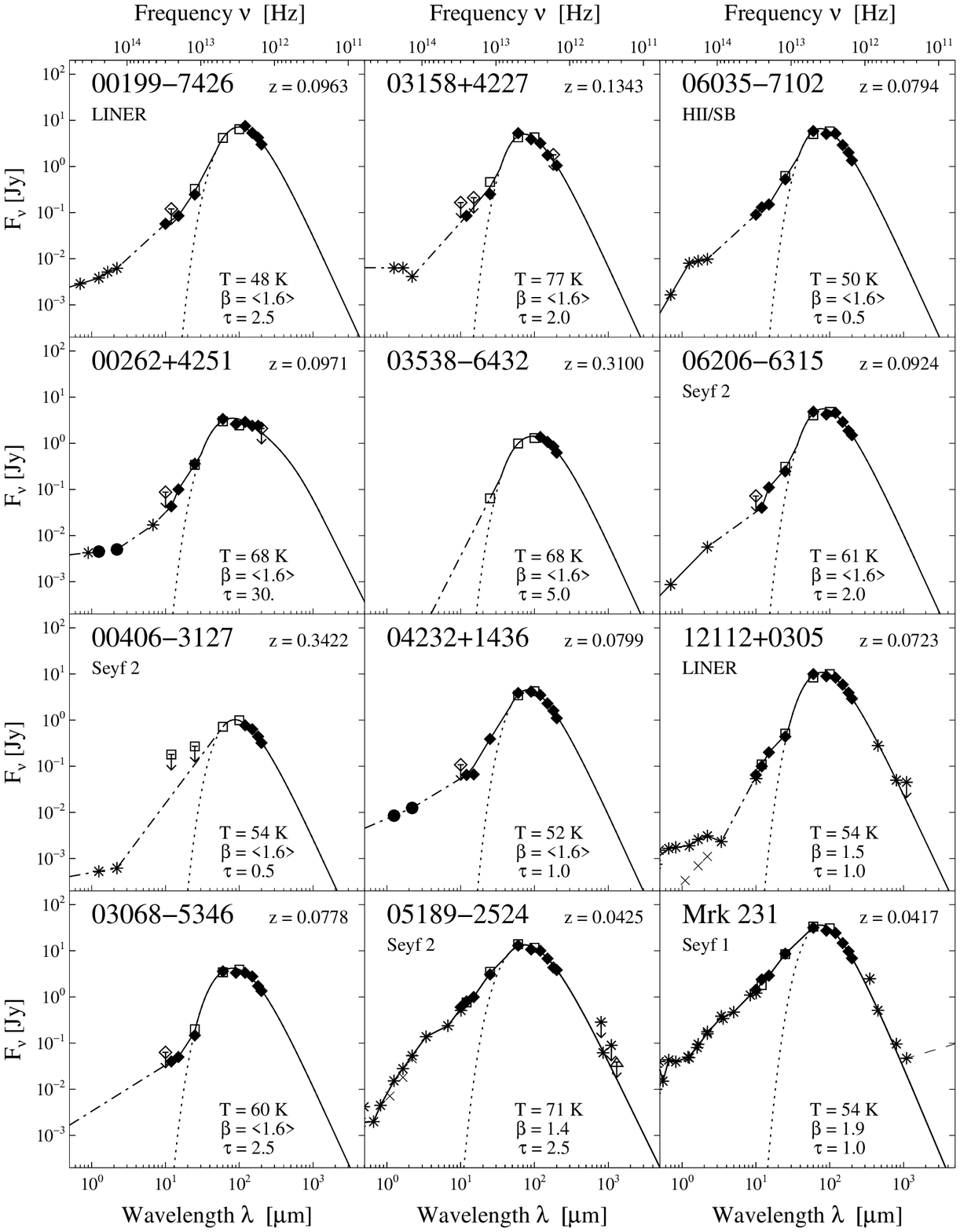} }
\vspace{1.5cm}
\caption{\label{fig_seds}}
\end{figure*}

\addtocounter{figure}{-1}
\begin{figure*}[ht!]
\vspace{1.5cm}
\hspace{1.5cm}\resizebox{16cm}{!}{\includegraphics{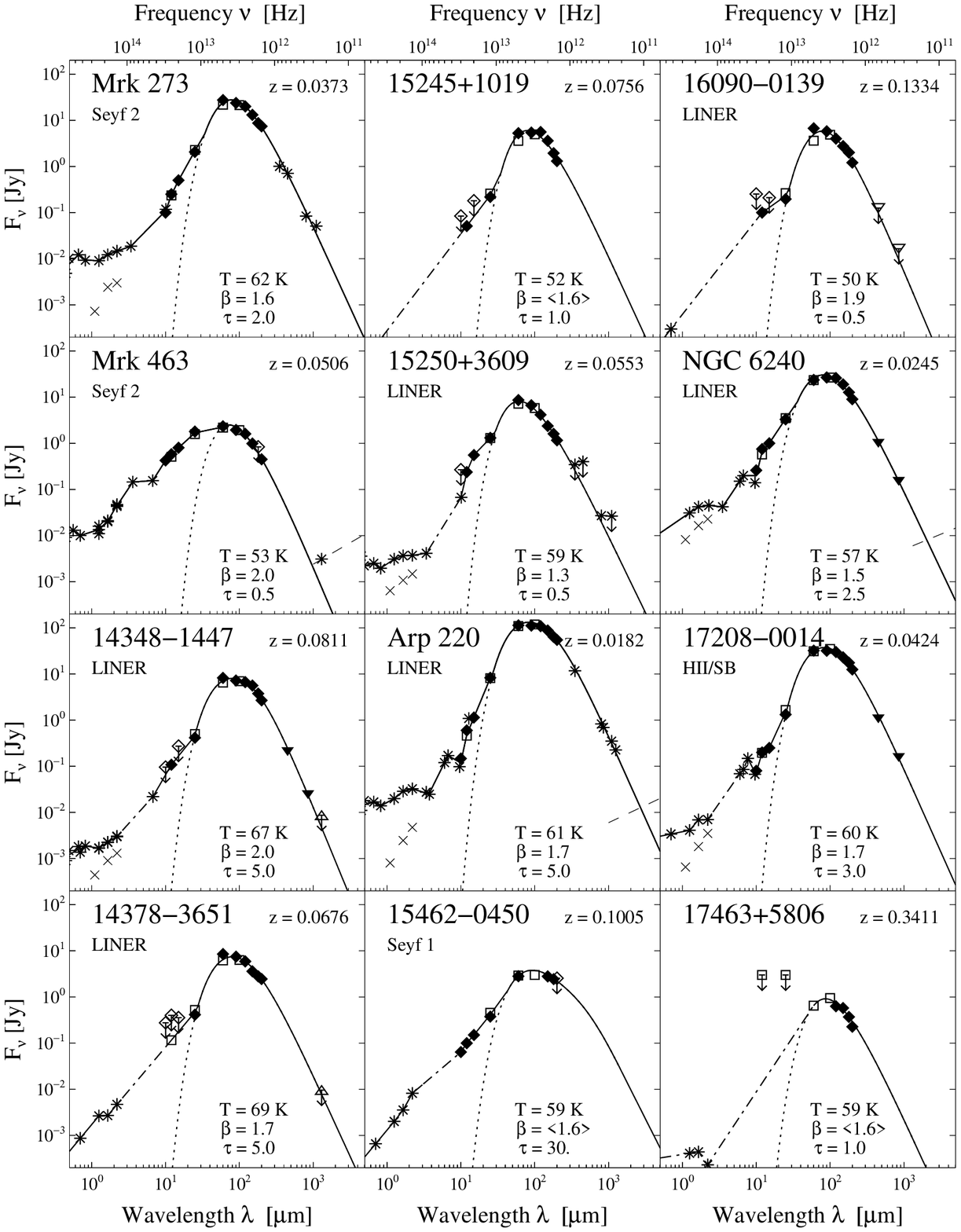} }
\vspace{1.5cm}
\caption{ continued.}
\end{figure*}

\addtocounter{figure}{-1}
\begin{figure*}[ht!]
\vspace{1.5cm}
\hspace{1.5cm}\resizebox{16cm}{!}{\includegraphics{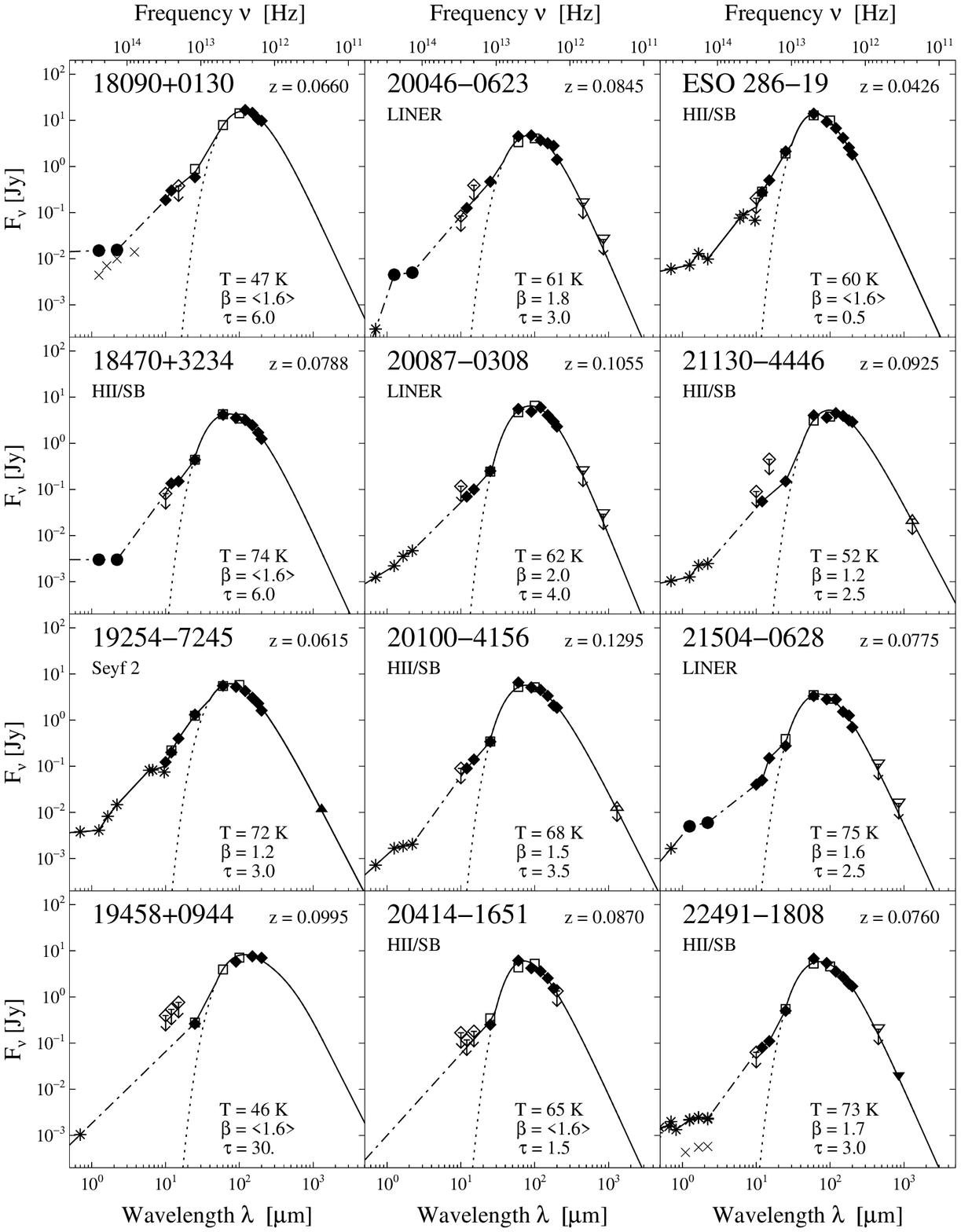} }
\vspace{1.5cm}
\caption{ continued.}
\end{figure*}

\addtocounter{figure}{-1}
\begin{figure*}[ht!]
\vspace{1cm}
\hspace{1.5cm}\resizebox{16cm}{!}{\includegraphics{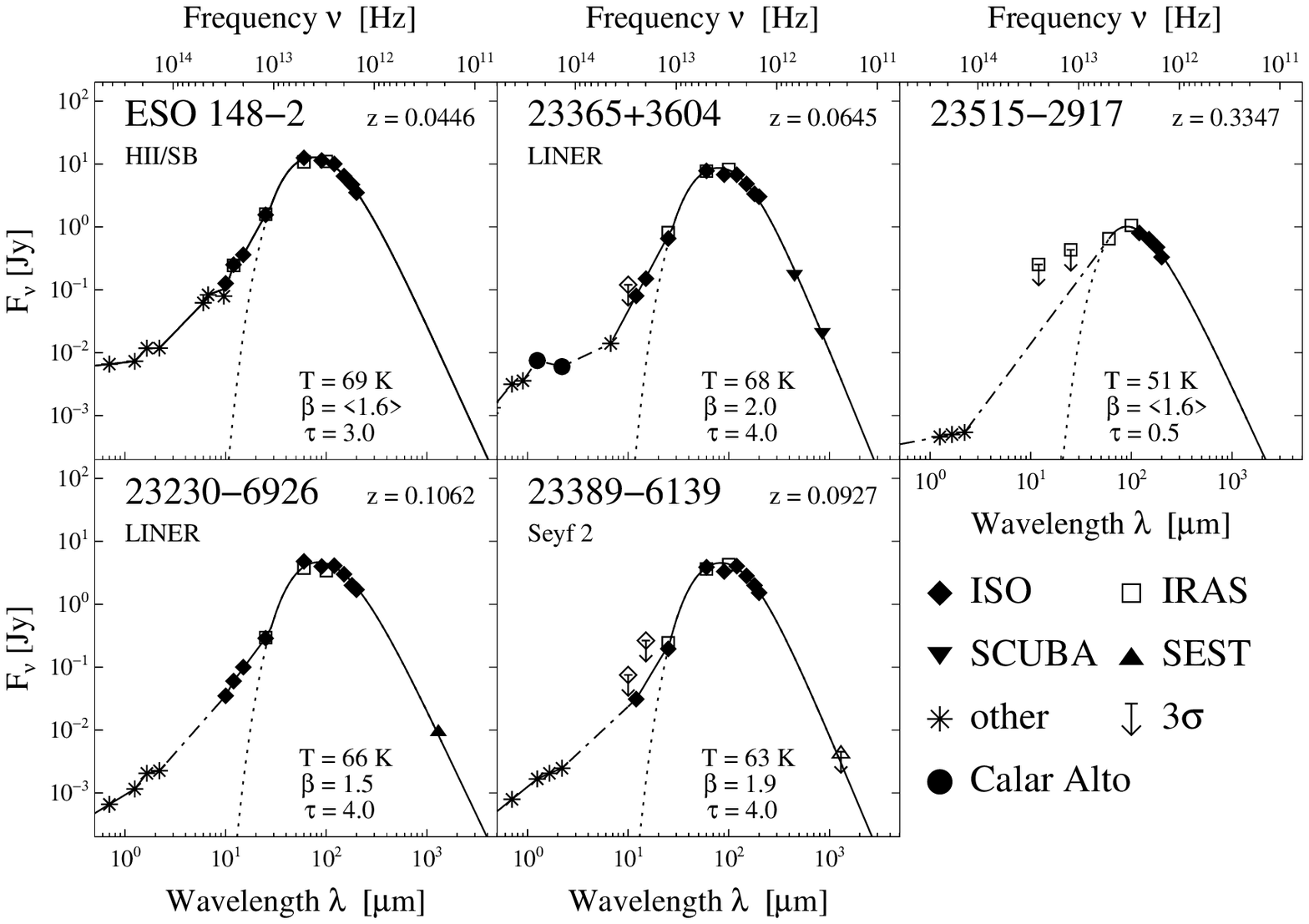} }
\vspace{-8cm}
\caption{ continued. Spectral energy distributions of ultra-luminous IR galaxies, 
ordered along the columns by RA. Upper limits are indicated by a downward 
arrow and in the case of ISO, SCUBA and SEST measurements in addition by open symbols. 
The redshifts and, if known, the optical spectral types are listed. The wavelength 
and frequency ranges are as observed and not corrected with regard to the rest frame 
of the objects. The dotted lines represent modified blackbody fits. Emissivity 
exponent $\beta$ and opacity $\tau_{\rm 100 \mu m}$ are listed in the plots. The
resulting temperatures T are corrected for redshift. For those sources with no 
submm/mm data available we used an average $\beta$\,=\,$\langle$\,1.6\,$\rangle$. 
The solid lines indicate the shapes of the SEDs, dash-dotted lines are used 
where they are uncertain. At NIR-MIR wavelengths the SED shape is obtained by 
connecting the data points, and at long FIR and submm wavelengths it is determined by
the modified blackbody fit. Luminosities were derived by integrating the SEDs between
10 and 1000\,$\mu$m. Dashed lines in the mm range indicate synchrotron extrapolations 
from cm observations. The literature data are from Sanders et al. (1988a,b), 
Crawford et al. (1996), Marx et al. (1994), Murphy et al. (1996), Rigopoulou et al. (1996), 
Duc et al. (1997), Rigopoulou et al. (1999) and Lisenfeld et al. (2000).
For some sources the NIR fluxes of the central regions are plotted with a ``$\times$'' 
(Diameter = 1$\arcsec$, from Scoville et al. (2000), except 5$\farcs$5 for 18090+0130 
from Hill et al. (1988)).}
\end{figure*}


\subsection{MAGIC 1.2 and 2.2 $\mu$m data}
\label{sect_NIRobs}

J and K-band images were obtained at the Calar Alto 2.2\,m telescope on August 21 and 22, 
2000, using the 256\,$\times$\,256 pixel NIR camera MAGIC (Herbst et al. 1993) 
with a pixel scale of 0$\farcs$6416. The observations were performed in standard dithering 
mode with a total exposure time per source of about 10 and 30 minutes in J and K, respectively. 
The seeing was about 2$\arcsec$ preventing the identification of spatial details in the 
sources (see Appendix A). But the conditions were photometric, in particular during the 
second night. The UKIRT standard stars FS2, FS27 and FS35 were observed for flux calibration.  
The data reduction followed the standard procedures. Based on the cross calibration of 
the standard stars, we estimate the photometric accuracy to be about 15\%. In the case of
IRAS\,18090+0130 the uncertainty is somewhat higher due to superposition of two stars.

\section{Results}

\subsection{Spectral energy distributions}
\label{sect_seds}

The fluxes from our observations are listed in Table\,\ref{tab_fluxes}. The 60 and 90\,$\mu$m 
values agree within 20\,\% with the IRAS 60 and 100 $\mu$m ones. For many sources, where IRAS 
provided only upper limits at 12 or 25\,$\mu$m, now the 10 -- 25\,$\mu$m fluxes could be 
measured. Also, for 16 sources out of 41, submm/mm fluxes and good upper limits are provided, 
as well as NIR 1.2 and 2.\,2\,$\mu$m fluxes for seven sources.

The spectral energy distributions (SEDs, as measured, not corrected for redshift) are shown 
in Fig.\,\ref{fig_seds}, supplemented by literature data. The remarkable features of the SEDs 
are: 

\begin{itemize}
\item[1)] For each galaxy the maximum of the SED can now be clearly determined. It lies between 
          60 and 100\,$\mu$m. The ISOPHOT long wavelength filters beyond 100\,$\mu$m clearly 
          outline the beginning of the Rayleigh-Jeans branch. In some cases  (e.g. NGC6240, 
          17208-0014) a somewhat plateau-like broad maximum is revealed, 
          suggesting a high opacity even in the FIR and/or the presence of various cool to cold 
          dust components (see also Fig.\,\ref{fig_multi_bb_examples}).  
\item[2)] For 20 sources the shape of the Rayleigh-Jeans branch can now be determined by the mm 
          and submm data points. As discussed in Sections~\ref{sect_ISOPHOTobs} 
          and~\ref{sect_SCUBAobs}, within the error budget the effects of different instrumental 
          beams and a possible contribution of extended flux appear to be negligible. In particular 
          for those cases (e.g. NGC6240, 17208-0014), where ISOPHOT already indicated a broad 
          maximum, the submm measurements confirm the high flux prediction from the 100-200\,$\mu$m 
          measurements. Also, where ISOPHOT indicated a steep Rayleigh-Jeans branch (e.g. 16090-0139, 
          23365+3604, 23389-6139) the mm and submm fluxes and upper limits are low, demonstrating 
          the consistency between the instruments.
\item[3)] Shortward of the maximum at around 60-100\,$\mu$m the SEDs exhibit two basic shapes: 
          \begin{itemize}
          \item[i)] a flat NIR plateau followed by a jump-like flux increase at about 10\,$\mu$m 
	            (e.g. Arp220 and 12112+0305). In some cases also indications of PAH emission 
                    around 7.7\,$\mu$m and silicate 9.7\,$\mu$m absorption features (e.g. Arp220 
                    and NGC6240) are recognized.
          \item[ii)] a power-law-like flux increase from the NIR to the MIR (e.g. Mrk463) or 
	             FIR (e.g. Mrk231). Spectral PAH emission or silicate absorption features may 
                     be present, e.g. for Mrk463, compare also with Fig. 3 in Rigopoulou et al. 
                     (1999), but due to the high continuum level they are diluted in the broad
                     band photometry.
          \end{itemize}
\end{itemize}

\subsection{CO line and synchrotron emission contamination of submm/mm fluxes}
\label{sect_contamin}

The determination of the Rayleigh Jeans branch allows a detailed analysis with respect
to the dust emissivity $\lambda^{-\beta}$ and the opacity $\tau_{\rm 100\mu m}$, as carried
out in the next section, under the condition that the emission is of thermal nature. Therefore, 
beforehand one has to check possible contamination of the submm and mm fluxes by CO lines and/or 
synchrotron emission:

\begin{itemize}
\item[1)] For all our sources observed at 1300\,$\mu$m with SEST the CO\,(2-1) line at 230\,GHz 
          (1304\,$\mu$m) moves out of the filter band pass due to their redshifts of z $\ga$ 0.05.
\item[2)] The  CO\,(3-2) line at 345\,GHz (869\,$\mu$m) may contribute to the SCUBA 850\,$\mu$m 
          fluxes for those sources with z $\la$ 0.04. Direct CO\,(3-2) observations are only 
          available for Arp220 (Mauersberger et al. 2000). Therefore, we estimated CO\,(3-2) line
          strengths from the CO\,(1-0) fluxes (Solomon 1997, Downes and Solomon 1998, Gao \& 
          Solomon 1999), adopting an intensity ratio R = I$_{\rm CO\,(3-2)}$\,/\,I$_{\rm CO\,(1-0)}$ 
          = 0.9 as found for Arp220. The strength of the redshifted CO\,(3-2) line, however, is 
          reduced according to the transmission of the SCUBA 850\,$\mu$m filter which lies 
          between 55\,\% and 30\,\% for our sources with z $\la$ 0.04. It turns out that only 
          four sources have a CO\,(3-2) line contribution to the 850\,$\mu$m flux which exceeds 5\,\%:
          Arp220 and Mrk\,273 (both 11\,\%), Mrk\,231 (22\,\%), and NGC\,6240 (35\,\%).
          These contributions are smaller or of the order of our adopted photometric uncertainties.
          Therefore we decided not to correct for the CO\,(3-2) line contribution.
\item[3)] In 20 cases with available radio fluxes (Condon et al. 1990 and Crawford et al. 1996) 
          extrapolations with spectral indices between 0.5 and 1.0 from the cm range towards shorter 
          wavelengths are far below ($<$\,10\%) our measured submm and mm fluxes (many of the 
          extrapolations lie even below the range plotted in Fig.\,\ref{fig_seds}). The only 
          exception is Mrk463 (with two Seyfert nuclei) where the 1.3 mm flux is variable, thus 
          dominated by synchrotron emission (Chini et al. 1989a and Marx et al. 1994.)
\end{itemize}

Our conclusion is therefore, that, like the FIR emission, the bulk of the submm/mm flux
is of thermal nature being emitted by dust.

\subsection{Dust parameters}
\label{sect_dustparam}

In order to characterise the dust emission, the SEDs are fitted with modified blackbodies.
Such fits, however, are not unique. They rely largely on the mass absorption coefficient 
$\kappa$ and its wavelength dependence $\beta$, both still being a matter of debate. 
Values of $\beta$ between 1 and 2 are commonly used (e.g. Hildebrand 1983). In case of a flat 
Rayleigh-Jeans tail the SEDs can also be modelled by several dust components. Since the 
interpretation of the dust emission as well as the derivation of the dust mass depend on the 
blackbody models used, we investigate the two main cases. They represent simplified formalisms, 
each relying on implicit assumptions, and a realistic description probably lies between these two 
extremes. In the following two subsections the FIR-submm range is investigated, 
and the MIR part is addressed in the third subsection.

\subsubsection{Single modified blackbody}
\label{sect_singlebb}

\begin{figure}
\hspace{-1cm}
\resizebox{10cm}{!} {\includegraphics {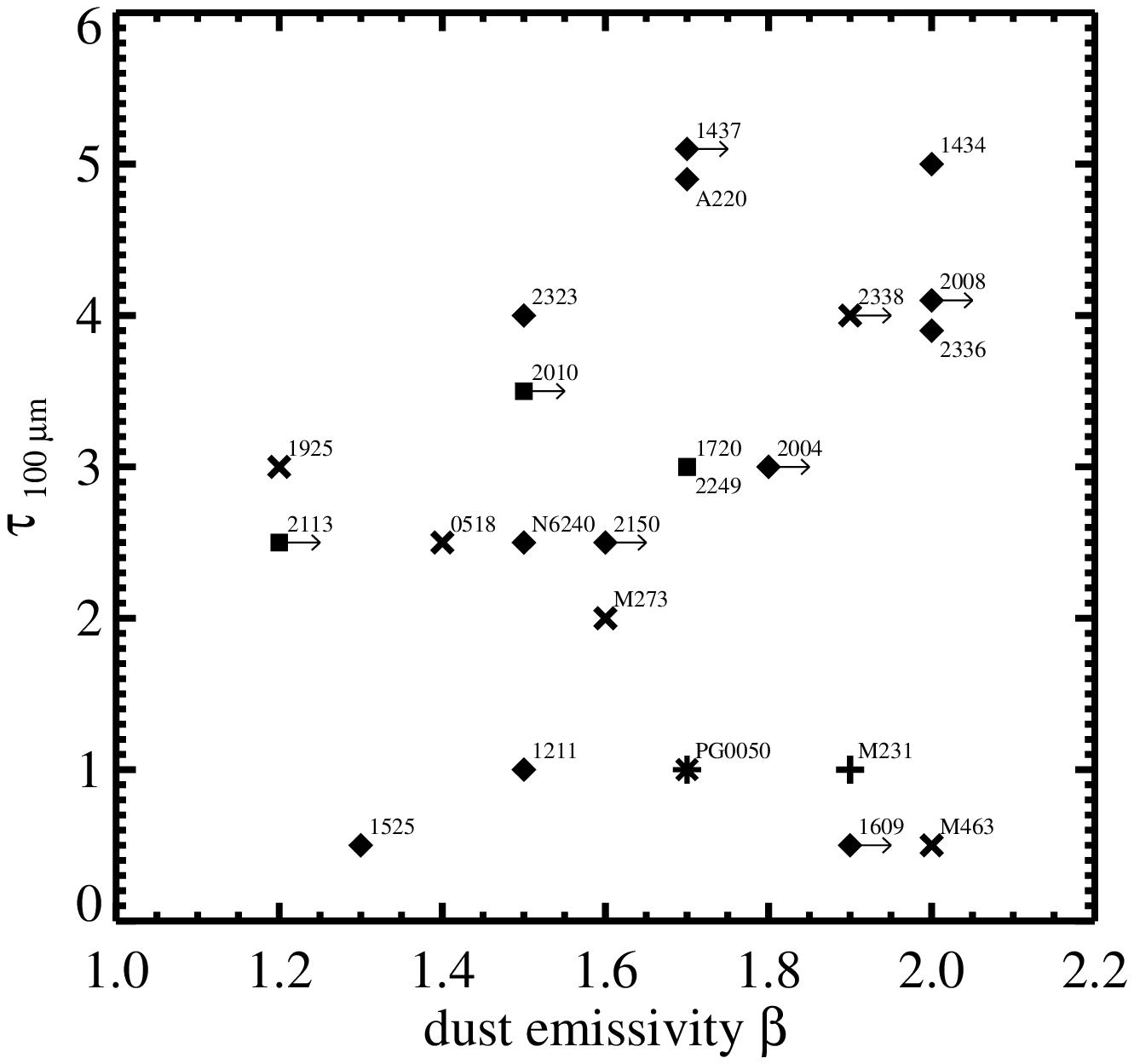} } 
\caption{\label{fig_tau_beta_comparison} 
         Distribution of $\tau_{ 100\,\mu m}$ versus $\beta$ for those sources with measured 
         submm/mm fluxes. Different symbols correspond to optical spectral types as in Table 4: 
         + Seyfert\,1, $\times$ Seyfert\,2, $\blacklozenge$ LINER, $\blacksquare$ HII/SB, and 
         $\bullet$ for not classified; the quasar PG\,0050+124 denoted by $\ast$ is also included 
         for comparison (from Haas et al. 2000a). Arrows indicate lower limit cases.}
\end{figure}

We used the following model:

\begin{equation}
\label{eqn_1}
S_{\lambda} = {B_{\lambda}(T)} \cdot (1 - e^{- \tau_{\lambda} }) $, with $
\end{equation}
\begin{equation}
\label{eqn_2}
\tau_{\lambda} = \tau_{ 100\,\mu m} \cdot (100\,\mu m / \lambda )^{\beta} 
\end{equation}
  
The SEDs are fitted between 60 and 1300\,$\mu$m with a single modified blackbody 
leaving T, $\beta$ and $\tau_{\rm 100 \mu m}$ free (minimising $\chi$$^{\rm 2}$ in a grid 
search). Although the emissivity exponent $\beta$ is a free parameter in eqn.\,2, 
it is kept constant over the whole wavelength range. 
In order to keep the parameter space under control, we have decided not to introduce a $\lambda$ 
dependence of $\beta$. 

The relation of the parameters $\beta$, $\tau_{ 100 \mu m}$ and T with the SED shapes is: 
\begin{itemize}
\item[$\bullet$] $\beta$ corresponds to the slope of the Rayleigh-Jeans tail, 
                 whether it is flat (small $\beta$) or steep (large $\beta$);  
\item[$\bullet$] $\tau_{ 100\,\mu m}$ determines, whether the peak plateau of the SED is narrow 
                 (small $\tau_{ 100 \mu m}$) or broad (large $\tau_{ 100\,\mu m}$); 
\item[$\bullet$] Like for normal blackbodies (Wien's displacement law), T is related to the 
                 wavelength, at which the SED maximum is located.
\end{itemize}

Fits were performed only for the ``mm-subsample'', i.e. for those 22 sources with 
submm/mm fluxes available. We also used the 8 cases of upper limits\footnote{these are: 
14378-3651, 16090-0139, 20046-0623, 20087-0308, 20100-4156, 21130-4446, 21504-0628, 23389-6139}, 
treating them formally as detections, and after the fitting procedure taking into account that 
they provide lower limits for $\beta$. In some cases, where the ISO 60 or 90\,$\mu$m fluxes 
show a large error or deviate strongly from the IRAS 60 and 100\,$\mu$m fluxes, we also included 
the IRAS data (e.g. for 16090-0139).

The fitted parameters $\beta$, $\tau_{ 100\,\mu m}$ and T are listed in 
Table\,\ref{tab_luminosities}, together with $\chi^{2}$. A visual impression of the quality of
the fits is given in Fig.~\ref{fig_seds}. In summary, the results for the mm-subsample are: 
\begin{itemize}
\item[$\bullet$] \quad 1.2 $<$ $\beta$ $\la$ 2.0~~~(Fig.~\ref{fig_tau_beta_comparison})
\item[$\bullet$] \quad 0.5 $<$ $\tau_{ 100\,\mu m}$ $<$ 5~~~(Fig.~\ref{fig_tau_beta_comparison})
\item[$\bullet$] \quad 50\,K $<$ T $<$ 70\,K. 
\end{itemize}
with an uncertainty of about 0.2 for $\beta$, 25\% for $\tau$ and 5-10\,K for T. 
The quoted uncertainties have been estimated considering ``bent'' SEDs with the FIR fluxes
reduced and the submm fluxes increased by their measurement uncertainty and vice versa. It
should also be noted that for sources with significant MIR flux (an extreme case is Mrk\,463) 
additional uncertainties in the determination of T and/or $\tau_{\rm 100\,\mu m}$ are introduced
when using the full 60\,$\mu$m flux, which is partially due to a dust component not considered
here.

The quoted parameter values should be considered with some tolerance and their interdependence 
borne in mind:

\begin{itemize}
\item[$\bullet$] For two SEDs with maxima located at similar wavelengths, T decreases with $\beta$. 
                 Due to the steeper fall-off at long wavelengths the width of the peak plateau 
                 can only be maintained, if the maximum is shifted to longer wavelengths. A
                 complementary pair are 19254-7245 ($\beta$\,=\,1.2) and 20046-0623 ($\beta$\,=\,1.8).
\item[$\bullet$] For two SEDs with similar $\beta$, but largely different $\tau_{\rm 100\,\mu m}$,
                 T increases with $\tau_{\rm 100\,\mu m}$. Due to the broader width, but similar
                 shape on the long wavelength side, the maximum is shifted towards shorter
                 wavelengths. A complementary pair is Mrk\,231 and 14348-1447.
\end{itemize}

On the other hand, there seems to be no correlation between $\tau_{\rm 100\,\mu m}$ and $\beta$ 
as can be seen from Fig.~\ref{fig_tau_beta_comparison}.

The dust parameters were determined via equation (1) only for the mm-subsample. 
For the remaining sources with wavelength coverage limited to 200\,$\mu$m (IR-subsample)
$\beta$ could not be fitted reliably (as we found from tests with the mm-subsample using only 
the 60-200\,$\mu$m fluxes). For the IR-subsample we kept $\beta$ fixed using the average value 
$\beta$\,=\,1.6 derived from the mm-subsample. Then $\tau_{ 100\,\mu m}$ and T could be determined 
reasonably well from the 60-200\,$\mu$m fluxes alone. (Exceptions are 00262+4251, 15462-0450,
18090+0130 and 19458+0944 which have less complete spectral coverage due to bad quality measurements
as flagged in Table~\ref{tab_fluxes}. In these cases $\tau_{ 100\,\mu m}$ was fixed to 6.0). The 
resulting values lie in the same range as for the mm-subsample (Table~\ref{tab_luminosities} and
Fig.~\ref{fig_seds}). As a check, we fitted also $\tau_{ 100\,\mu m}$ and T of the mm-subsample 
with a fixed $\beta$\,=\,1.6 using only the 60-200\,$\mu$m fluxes. The results are basically 
consistent with those obtained from the longer wavelength coverage, except for the sources 
with extremely low or high true $\beta$. Hence, in the discussion below we can mostly use the full 
sample, and only where $\beta$ plays a role, we confine it to the mm-subsample.

\subsubsection{Multiple modified blackbodies}
\label{sect_multiplebb}

As derived in the previous section, for the majority of the ``mm-subsample'' sources (11 out 
of 14, not having lower limits for $\beta$, one of them having $\tau_{100\,\mu m}$ = 5, 
Fig.~\ref{fig_tau_beta_comparison}) it is not possible to fit the FIR-submm SEDs properly with one single 
modified blackbody with an emissivity law of $\lambda$$^{\rm -2}$, rather the superposition of two 
or more modified blackbodies is required.

In the low opacity case Equation (1) can be approximated by 
\begin{equation}
\label{eqn_3}
S_{\lambda} = {B_{\lambda}(T)} \cdot \tau_{\lambda} \propto {B_{\lambda}(T)} \cdot \lambda ^{- \beta}
\end{equation}
 
\begin{figure*}[ht!]
\vspace*{0.0cm}
\begin{center}
\rotatebox{90} {\resizebox{6.5cm}{!} {\includegraphics {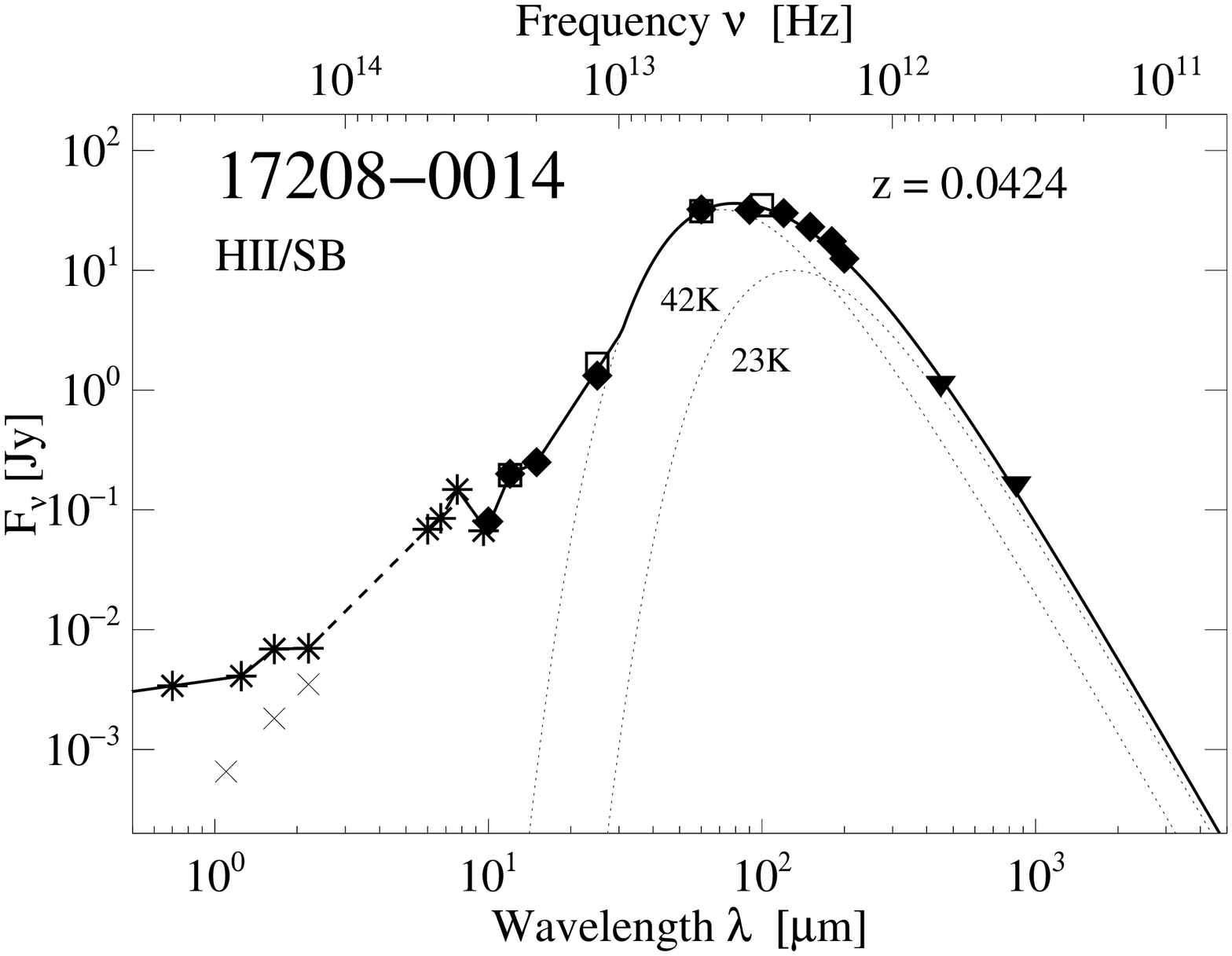} } }
\hspace*{0.01cm}
\rotatebox{90} {\resizebox{6.5cm}{!} {\includegraphics {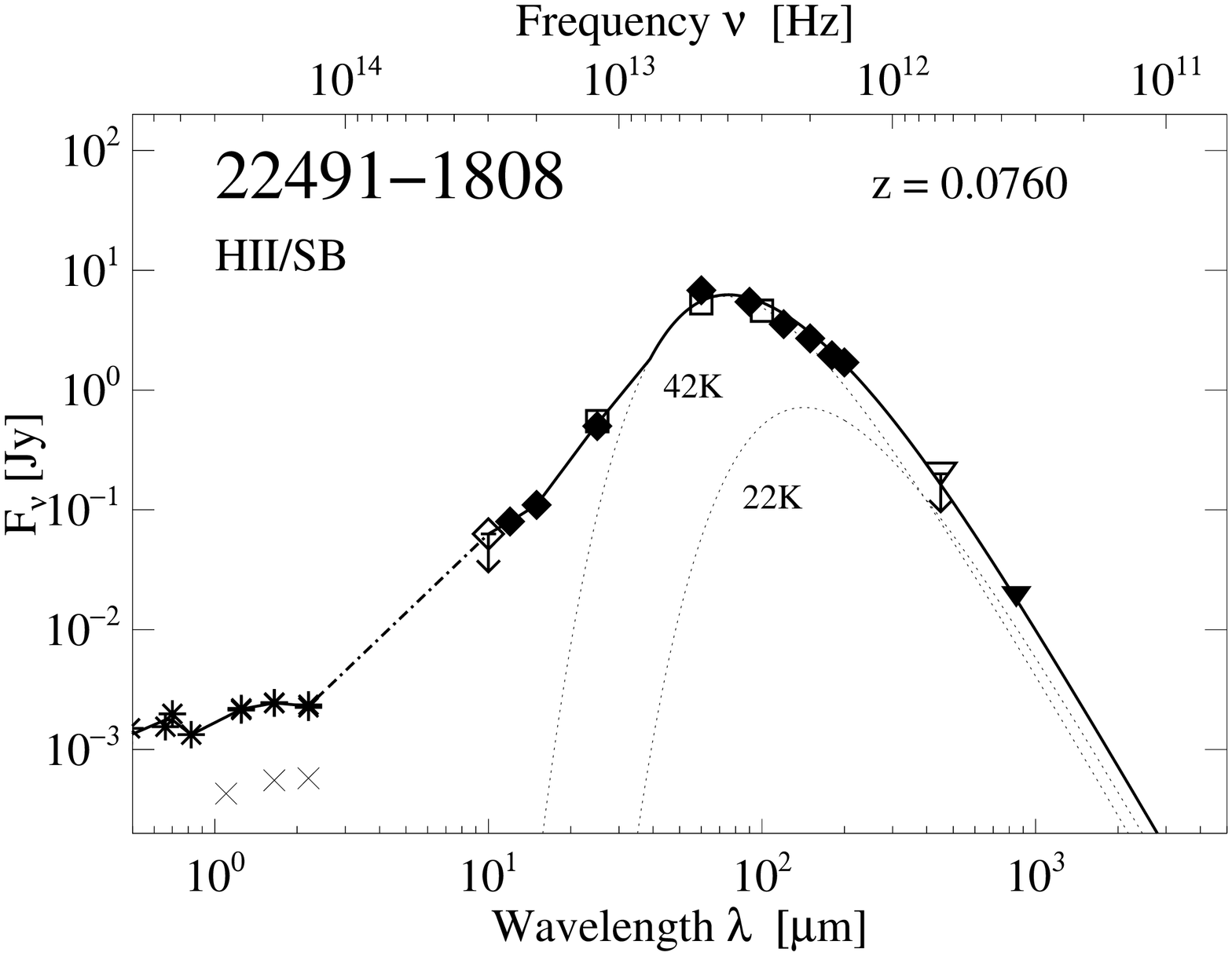} } }
\end{center}
\vspace*{-0.5cm}
\begin{center}
\rotatebox{90} {\resizebox{6.5cm}{!} {\includegraphics {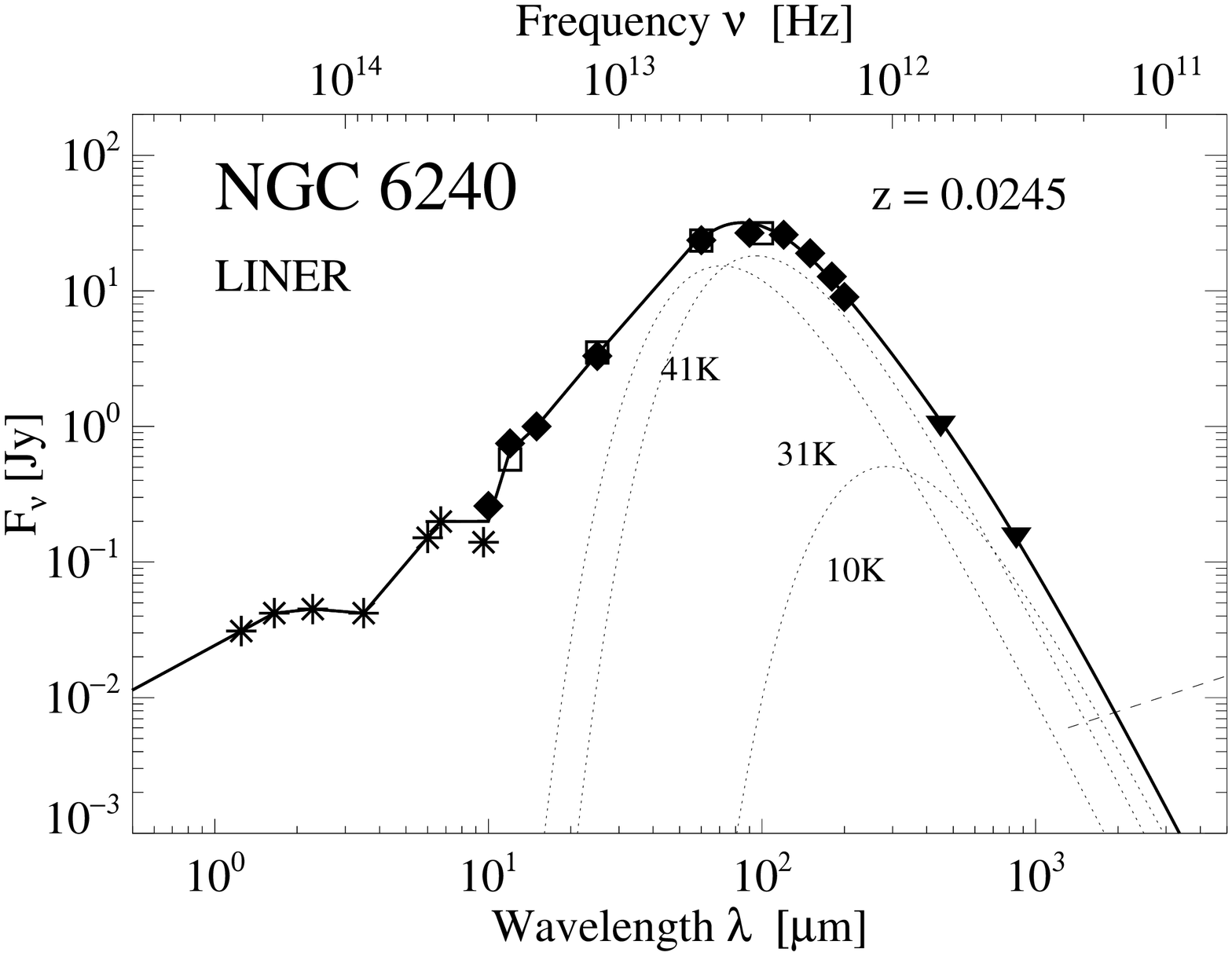} } }
\hspace*{0.01cm}
\rotatebox{90} {\resizebox{6.5cm}{!} {\includegraphics {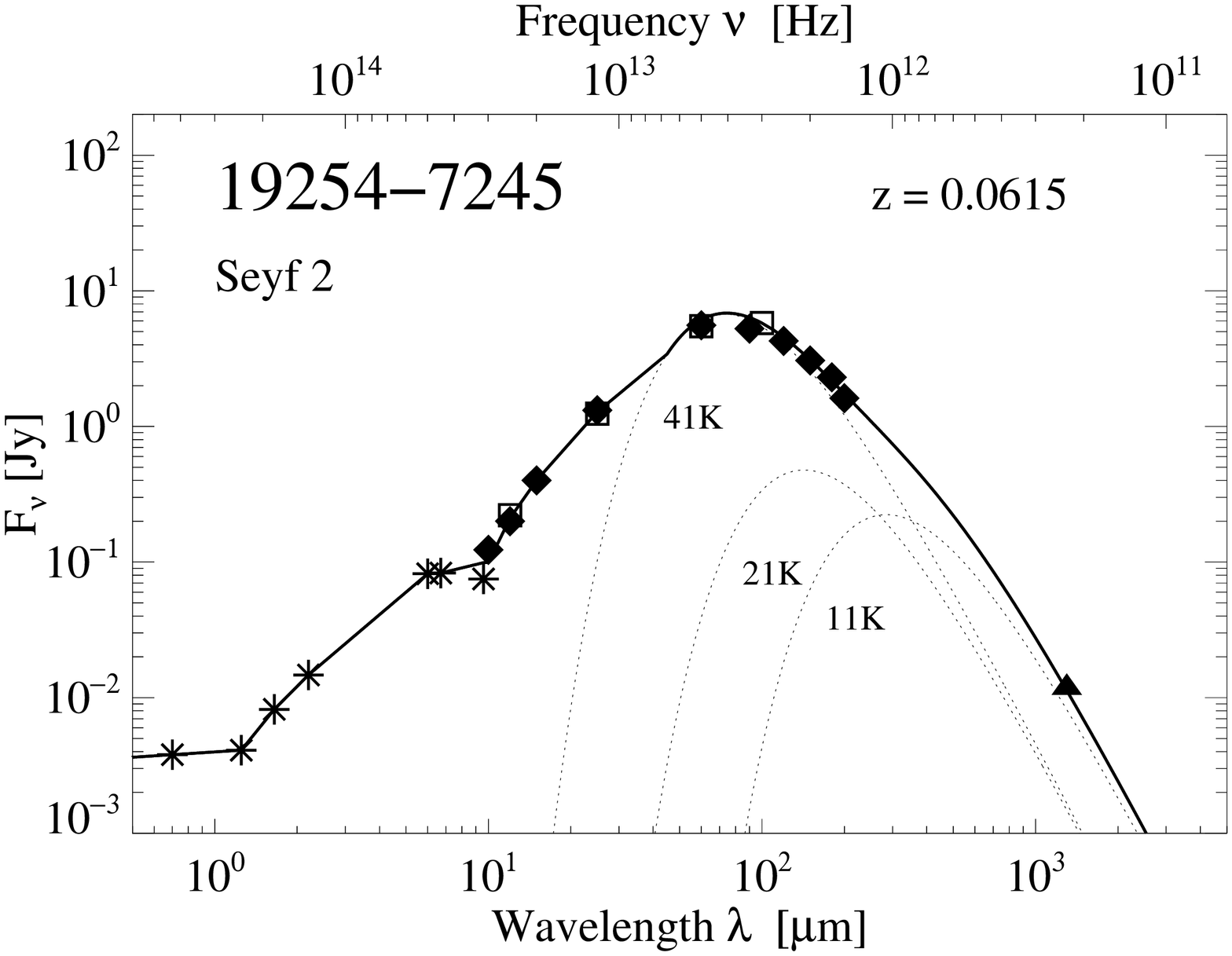} } }
\end{center}
\vspace*{-0.5cm}
\begin{center}
\rotatebox{90} {\resizebox{6.5cm}{!} {\includegraphics {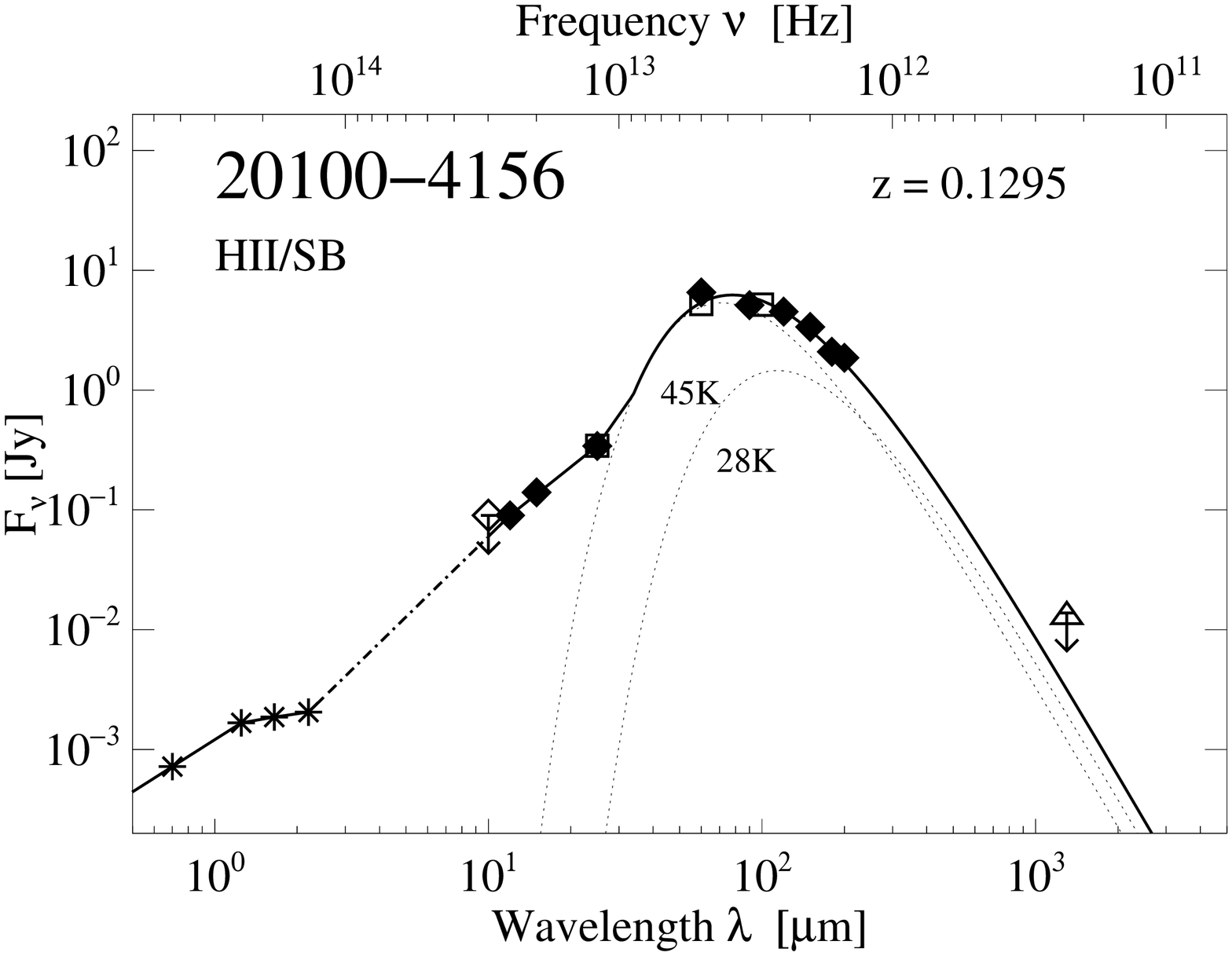} } }
\hspace*{0.01cm}
\rotatebox{90} {\resizebox{6.5cm}{!} {\includegraphics {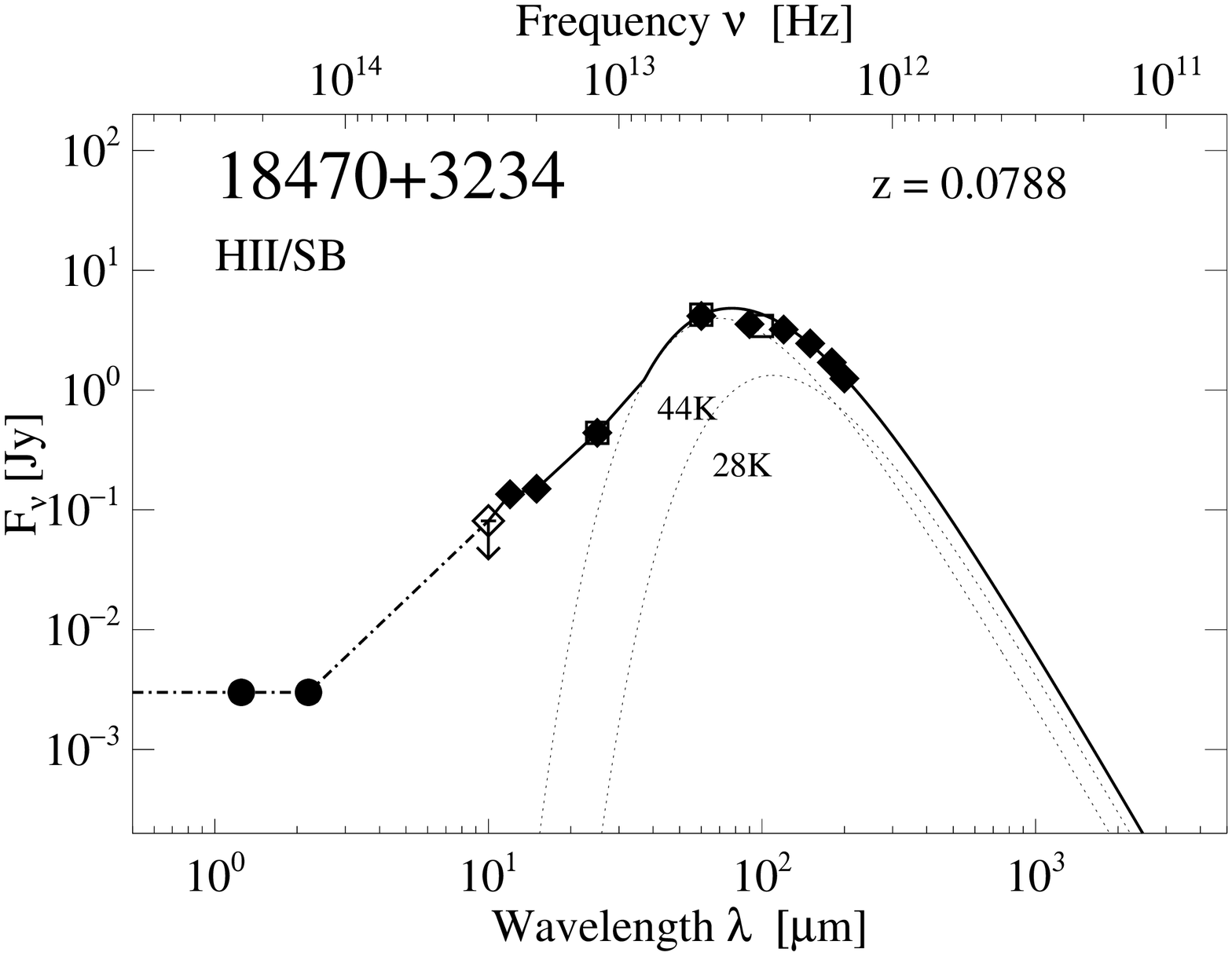} } }
\end{center}
\vspace*{0.0cm}
\caption{\label{fig_multi_bb_examples} 
         Examples for fits with several optically thin ($\tau_{100\,\mu m} \ll$ 1) modified blackbodies 
         (BBs) with an emissivity law of $\lambda$$^{\rm -2}$. Top row: superposition of two BBs; middle 
         row: possible superposition of three BBs; bottom row: fits can also be done for those SEDs with
         no submm data: approximation by two BBs.}
\end{figure*}

Based on Equation (3), implicitly assuming $\tau_{100\,\mu m} \ll$ 1, we fitted several modified 
blackbodies with free T and fixed $\beta$ = 2 to all observed SEDs, as illustrated for some examples in 
Fig.~\ref{fig_multi_bb_examples}. They have temperatures in the range between 30 and 50\,K (cool), 
and between 10 and 30\,K (cold). However, the decomposition into the various components is not 
unique, because of the sparse data coverage in the sub-mm range.

In the case of multiple blackbodies no direct conclusion about the opacity $\tau_{100\,\mu m}$ can 
be drawn (nevertheless, in Sect. 5.1.2 below, $\tau_{100\,\mu m}$ will be constrained using CO data). 
The most realistic case might be that of several blackbodies with $\beta$\,$\approx$\,2, and a range 
of opacities from low to partly high.

\subsubsection{Description of the NIR-MIR emitting dust}

The two basic SED shapes in the NIR-MIR outlined in Sect.~\ref{sect_seds}
can be formally fitted by a superposition of several warm dust components. 
For the cases with flat NIR plateau the maximum temperatures are about 100-150\,K 
(e.g. Klaas et al. 1997, 1998a). The power-law-like SEDs can be approximated by a suite 
of blackbodies up to the dust grain evaporation temperatures of about 1000-1500\,K (the hotter
the blackbody the less dust mass is involved). Modelling of the continuum is hampered by the 
presence of strong spectral features like PAH emission and silicate absorption. Using higher 
spectral resolution, Laurent et al. (2000) investigated this spectral part quantitatively.

\subsection{Luminosities}
\label{sect_luminosities}

Table~\ref{tab_luminosities} lists the luminosities derived within various bandpasses 
in the rest frame of the objects by integrating the spectral energy distribution as outlined
by the thick solid and dash-dotted lines shown in Fig.\,\ref{fig_seds} for the indicated 
wavelength ranges. On the Rayleigh-Jeans tail and around the SED maximum this comprises the single 
blackbody curve obtained with the eqn. (1) fit, and shortward thereof the lines connecting the data 
values by linear interpolation. 

The total IR-submm luminosity L$_{\rm 10-1000\,\mu m}$ is dominated by the FIR in the wavelength 
range 40-150\,$\mu$m, while the 150-1000\,$\mu$m submm range plays a minor role 
(L$_{\rm 150-1000\,\mu m}$\,$<$\,0.1\,$\cdot$\,L$_{\rm 40-150\,\mu m}$) as well as the 10-40\,$\mu$m 
MIR range (except for Mrk\,463 and the z\,$>$\,0.3 sources which are MIR dominant). The luminosities 
L$_{\rm 8-1000\,\mu m}$ extrapolated from the four IRAS bands (formula cf. Table 1 in 
Sanders \& Mirabel 1996) typically slightly overestimate our IR-submm luminosity values by about 15\,\%; 
nevertheless this is still a good agreement. 

The MIR/FIR luminosity ratio has a median value of about 0.3. Thus, the sample of bright nearby ULIRGs 
preferentially comprises objects with cool MIR/FIR colours (compared with quasars having 
L$_{\rm  MIR}$\,/\,L$_{\rm FIR}$\,$>$\,1, cf. Haas et al. 2000a). Though the luminosity range of the 
ULIRG sample spans about one decade, there is no trend of luminosity with optical spectral type or 
MIR/FIR colours. 

For sources without submm/mm observations available, the submm luminosity 
L$_{\rm submm}$\,=\,L$_{\rm 150-1000 \mu m}$ is extrapolated using the average value $\beta$\,=\,1.6 
(Sect.~\ref{sect_singlebb}), while the actual L$_{\rm submm}$ depends on the actual value of $\beta$. 
A check on the mm-subsample shows that, for the case of minimum $\beta$\,$\approx$\,1.2 or maximum 
$\beta$\,$\approx$\,2, L$_{\rm submm}$ obtained using the average $\beta$\,=\,1.6 can deviate from the 
true value by factors of 2 and 0.5, respectively.

\subsection{Dust masses}
\label{sect_dust_masses}

In order to derive the dust masses, we used the standard approach based on Hildebrand (1983) and 
further developed  by various authors (e.g. Chini et al. 1986, Kr\"ugel et al. 1990):

\begin{equation}
\label{eqn_4}
M_{\rm dust} = \frac{ 4 \pi \cdot a^3 \cdot \rho } { 3 \cdot a^2 \cdot Q_{em}(850 \mu m, \beta) } \cdot \frac{ D ^{2} \cdot S_{850 \mu m} } { B_{850 \mu m}(T) } 
\end{equation}
with \newline
$\bullet$ average grain size a = 0.1 $\mu$m \newline
$\bullet$ grain density $\rho$ = 3 g cm$^{\rm -3}$ \newline
$\bullet$ distance D \newline
$\bullet$ flux S$_{850 \mu m}$ (in the restframe of the object)\newline
$\bullet$ Planck function B$_{850 \mu m}$(T) at restframe 850\,$\mu$m, and  \newline
$\bullet$ dust grain emission efficiency Q$_{em}$(850$\mu$m, $\beta$) where \newline

\begin{equation}
\label{eqn_5}
Q_{em} (\lambda, \beta ) = Q_0 \cdot a \cdot \{ 250 / \lambda \} ^ \beta
\end{equation}
with Q$_{\rm 0}$ = 40 cm$^{\rm -1}$ for $\lambda$\,=\,250\,$\mu$m.  \newline 
The first fraction in eqn. 4 is conveniently summarized yielding 

\begin{equation}
\label{eqn_6}
M_{\rm dust} = \frac { 1 } { \kappa_{850 \mu m} ( \beta ) } \cdot \frac{ D ^{2} \cdot S_{850 \mu m} } { {B_{850 \mu m}(T)}  } 
\end{equation}

with \newline

\begin{equation}
\label{eqn_7}
\kappa_{850 \mu m}(\beta) = \kappa_{850 \mu m}^{\beta = 2} \cdot \{ 250 \mu m / 850 \mu m \}^{\beta - 2} 
\end{equation}

with \newline \smallskip
$\bullet$ \quad $\kappa_{850 \mu m}^{\beta = 2} = 0.865\,cm^{2}\,g^{-1}$ \newline \smallskip 
which corresponds to \newline \smallskip
$\bullet$ \quad $\kappa_{1300 \mu m}^{\beta = 2} = 0.4\,cm^{2}\,g^{-1}$ \newline 

Our ``choice'' of $\kappa$ is consistent with that favoured by Kr\"ugel et al. (1990, see their eqn. 10), 
and Lisenfeld et al. (2000). Note that we account for $\beta$ in the wavelength dependence of Q$_{\rm em}$ 
and $\kappa$, respectively. In the case of $\beta$\,$<$\,2 this leads to a larger value of $\kappa$, 
hence to dust masses which are smaller than for $\beta$\,=\,2.

\begin{table*}[ht!]

\caption[] {Infrared luminosities, dust temperatures, dust masses and IR source sizes. \\
The luminosity distance is determined as D$_{\rm L}$ = c\,/\,H$_{\rm 0}$\,*\,(z\,+\,z$^{\rm 2}$\,/\,2), 
i.e. q$_{\rm 0}$\,=\,0, with H$_{\rm 0}$\,=\,75\,km/s/Mpc. 
The infrared luminosities L$_{\rm MIR}$ (3-40$\,\mu$m), L$_{\rm FIR}$ (40-150$\,\mu$m) and L$_{\rm submm}$ 
(150-1000$\,\mu$m) are determined from the SED curves shown in Fig.\,\ref{fig_seds}, by integrating over
the indicated wavelength range shifted to the rest frame of the objects. The dust mass is derived for 
three models of the IR emission:
\begin{itemize}
\item[1)] M$_{\rm d}^{\rm \tau free}$ for the single modified blackbody with free T, $\tau$ and $\beta$ 
          as described by eqn.\,1 (minimising $\chi$$^{\rm 2}$) and shown in Fig.\,\ref{fig_seds} 
          ($\beta$ values in brackets $<\,>$ means that a fixed average of $\beta$\,=\,1.6 is used, lower 
          limits are indicated by a $>$ sign).
\item[2)] M$_{\rm d(FIR)}^{\rm \beta\,=\,2}$ for a single modified blackbody with fixed $\beta\,=\,2$ 
          fitted by eqn.\,3 to the FIR 60-200\,$\mu$m range (for details see sect.~\ref{sect_dust_masses}).
\item[3)] M$_{\rm d(total)}^{\rm \beta\,=\,2}$ for multiple modified blackbodies with fixed $\beta\,=\,2$.
\end{itemize}
\vspace{-2mm}
The sizes of the dust emitting regions are given for the two cases of a single blackbody as (1) brightness 
radius r$_{\rm b}$, and (2) as ``smallest transparent radius'' r$_{\rm \tau}$, at which 
$\tau$$_{\rm 100\,\mu m}$\,=\,1 (for details see sect.\,\ref{sect_dust_size}). For comparison M(H$_{\rm 2}$) 
is listed (using the galactic conversion factor M(H$_{\rm 2}$) = 4.6\,$\cdot$\,L$_{\rm CO}$).
   \label{tab_luminosities}
}
\scriptsize
\begin{tabular}{@{}l@{\hspace{1mm}}|@{\hspace{1mm}}c@{\hspace{2mm}}c@{\hspace{1mm}}|@{\hspace{1mm}}r@{\hspace{2mm}}r@{\hspace{2mm}}r@{\hspace{1mm}}|@{\hspace{1mm}}r@{\hspace{2mm}}r@{\hspace{2mm}}c@{\hspace{2mm}}r@{\hspace{2mm}}r@{\hspace{2mm}}r@{\hspace{2mm}}r@{\hspace{1mm}}|@{\hspace{1mm}}r@{\hspace{2mm}}r@{\hspace{2mm}}r@{\hspace{2mm}}r@{\hspace{1mm}}|@{\hspace{1mm}}r@{\hspace{3mm}}|@{\hspace{1mm}}r@{\hspace{1mm}}}
\hline\noalign{\smallskip}
{\bf Name} & {\bf z}& {\bf D}& {\bf L$_{\rm MIR}$} & {\bf L$_{\rm FIR}$} & {\bf L$_{\rm submm}$}& {\bf T} & {\bf $\tau$$_{\rm 100}$} & {\bf $\beta$} & $\chi$$^{\rm 2}$ & {\bf M$_{\rm d}^{\rm \tau free}$} & {\bf r$_{\rm b}$} & {\bf r$_{\rm b}$} & {\bf T} & {\bf M$_{\rm d(FIR)}^{\rm \beta\,=\,2}$} & {\bf r$_{\rm \tau}$}  & {\bf r$_{\rm \tau}$}& {\bf M$_{\rm d(total)}^{\rm \beta\,=\,2}$}  & M(H$_{\rm 2}$)\\
\noalign{\smallskip}
 &  & & 10-40$\mu$m & 40-150 & 150-1000 & &   & &   &  & & & & single BB & & & several BBs  &   \\
\noalign{\smallskip}
       &    & [Mpc] & [10$^{\rm 9}$ L$_{\odot}$]& [10$^{\rm 9}$ L$_{\odot}$] & [10$^{\rm 9}$ L$_{\odot}$]& [K] & & & & [10$^{\rm 6}$ M$_{\odot}$] & [pc] & [$\arcsec$] & [K] & [10$^{\rm 6}$ M$_{\odot}$] & [pc] & [$\arcsec$] & [10$^{\rm 6}$ M$_{\odot}$]  & [10$^{\rm 9}$ M$_{\odot}$]   \\
\noalign{\smallskip}\hline\noalign{\smallskip}
00199-7426 &0.0963& 403& 307&1139&130& 48&2.50&$<$1.6$>$&1.07&109&390&0.199 &32&160&1097&0.6& 549              &      \\
00262+4251*&0.0971& 407& 342& 675& 77& 68&6.00&$<$1.6$>$&1.24& 28&131&0.066 &36& 48& 602&0.3& 762              &29$^a$\\
00406-3127 &0.3422&1602&    &2480&101& 53&0.50&$<$1.6$>$&1.04& 51&688&0.088 &43& 59& 666&0.1& 157              &      \\
03068-5346 &0.0778& 323& 126& 481& 37& 60&2.50&$<$1.6$>$&1.07& 22&172&0.109 &36& 32& 488&0.3& 152              &      \\
03158+4227 &0.1343& 573& 685&1724& 71& 77&2.00&$<$1.6$>$&1.09& 26&220&0.079 &43& 42& 563&0.2& 115              &      \\
03538-6432 &0.3100&1431&    &2877&188& 68&5.00&$<$1.6$>$&1.06& 72&263&0.038 &39&110& 910&0.1& 519              &      \\
04232+1436 &0.0799& 332& 244& 540& 34& 51&1.00&$<$1.6$>$&1.04& 23&275&0.170 &37& 33& 495&0.3&  73              &41$^a$\\
05189-2524 &0.0425& 173& 460& 500& 33& 70&2.50&     1.4 &1.07& 15&128&0.152 &38& 26& 443&0.5& 448\parbox{0cm}{$^{\rm +}$}&23$^b$\\
06035-7102 &0.0794& 330& 376& 785& 40& 49&0.50&$<$1.6$>$&1.09& 29&434&0.271 &38& 35& 513&0.3&  82              &38$^b$\\
06206-6315 &0.0924& 386& 271& 878& 54& 61&2.00&$<$1.6$>$&1.09& 30&225&0.120 &37& 50& 611&0.3& 130              &52$^b$\\
12112+0305 &0.0723& 299& 264&1052& 68& 53&1.00&     1.5 &1.06& 42&355&0.244 &36& 71& 731&0.5& 856              &      \\
Mrk231     &0.0417& 170&1215&1219& 58& 54&1.00&     1.9 &1.02& 22&371&0.449 &50& 28& 456&0.5& 131\parbox{0cm}{$^{\rm +}$}&35$^a$\\ 
Mrk273     &0.0373& 152& 248& 714& 47& 62&2.00&     1.6 &1.14& 27&208&0.282 &36& 47& 596&0.8& 104\parbox{0cm}{$^{\rm +}$}              &23$^a$\\
Mrk463     &0.0506& 207& 347& 136&  5& 52&0.50&     2.0 &1.06&  4&188&0.187 &40&  5& 198&0.2&  12              &      \\
14348-1447 &0.0811& 337& 294&1037& 79& 67&5.00&     2.0 &1.12& 37&194&0.118 &35& 81& 781&0.5& 219\parbox{0cm}{$^{\rm +}$}              &64$^c$\\
14378-3651 &0.0676& 279& 199& 651& 49& 69&5.00&  $>$1.7 &1.10& 26&139&0.103 &36& 44& 579&0.4& 423              &15$^b$\\
15245+1019 &0.0756& 314& 138& 621& 39& 51&1.00&$<$1.6$>$&1.15& 27&300&0.197 &37& 38& 534&0.4&  97              &      \\
15250+3609 &0.0553& 227& 304& 461& 17& 59&0.50&     1.3 &1.10& 10&215&0.195 &44& 11& 289&0.3&  28              &      \\
Arp220     &0.0182&  73& 190& 820& 86& 61&5.00&     1.7 &1.09& 66&214&0.600 &40& 32& 488&1.4&1479\parbox{0cm}{$^{\rm +}$}              &32$^a$\\
15462-0450*&0.1005& 422& 397& 763&109& 59&6.00&$<$1.6$>$&1.12& 52&179&0.087 &34& 73& 741&0.4&1498              &      \\
16090-0139 &0.1334& 569& 635&1944& 85& 49&0.50&  $>$1.9 &1.15& 43&645&0.233 &40& 66& 708&0.3& 182              &56$^a$\\
NGC6240    &0.0245&  99& 157& 347& 28& 57&2.50&     1.5 &1.07& 21&154&0.320 &33& 36& 520&1.1& 580\parbox{0cm}{$^{\rm +}$}              &37$^a$\\
17208-0014 &0.0424& 173& 235&1226&107& 60&3.00&     1.7 &1.09& 64&274&0.326 &34&116& 936&1.1& 602\parbox{0cm}{$^{\rm +}$}              &32$^a$\\
17463+5806 &0.3411&1596&    &2192& 87& 59&1.00&$<$1.6$>$&1.10& 40&434&0.056 &43& 57& 658&0.1& 142              &      \\
18090+0130*&0.0660& 273& 322&1100&193& 46&6.00&$<$1.6$>$&1.06&173&319&0.241 &28&322&1558&1.2&1746              &      \\
18470+3234 &0.0788& 327& 270& 537& 35& 74&6.00&$<$1.6$>$&1.07& 14& 92&0.058 &38& 26& 442&0.3& 113              &      \\
19254-7245 &0.0615& 253& 375& 473& 32& 72&3.00&     1.2 &1.04& 15&108&0.088 &38& 24& 429&0.3&1265\parbox{0cm}{$^{\rm +}$}              &35$^b$\\
19458+0944*&0.0995& 418& 602&1343&307& 46&6.00&$<$1.6$>$&1.21&226&373&0.184 &28&388&1711&0.8&6405              &55$^a$\\
20046-0623 &0.0845& 352& 344& 658& 47& 60&3.00&  $>$1.8 &1.18& 26&188&0.110 &34& 57& 654&0.4& 123              &      \\
20087-0308 &0.1055& 444& 337&1370&103& 61&4.00&  $>$2.0 &1.10& 50&254&0.118 &35&110& 911&0.4& 239\parbox{0cm}{$^{\rm +}$}              &74$^a$\\
20100-4156 &0.1295& 551& 622&1919&135& 67&3.50&  $>$1.5 &1.07& 65&252&0.094 &40& 86& 803&0.3& 337              &      \\
20414-1651 &0.0870& 363& 265& 836& 38& 65&1.50&$<$1.6$>$&1.12& 19&206&0.117 &40& 29& 471&0.3&  88              &      \\
ESO286-19  &0.0426& 174& 272& 460& 16& 60&0.50&$<$1.6$>$&1.05&  8&225&0.266 &44& 11& 289&0.3&  28              &22$^b$\\
21130-4446 &0.0925& 387& 130& 772& 98& 52&2.50&  $>$1.2 &1.19& 71&262&0.140 &32&111& 916&0.5& 932              &      \\
21504-0628 &0.0775& 322& 214& 425& 21& 75&2.50&  $>$1.6 &1.10&  8&106&0.068 &40& 16& 343&0.2&  39              &      \\
22491-1808 &0.0760& 315& 268& 661& 40& 73&3.00&     1.7 &1.10& 16&143&0.093 &39& 32& 488&0.3& 213\parbox{0cm}{$^{\rm +}$}              &31$^b$\\
ESO148-2   &0.0446& 182& 248& 502& 33& 68&3.00&$<$1.6$>$&1.05& 17&136&0.154 &36& 34& 504&0.6&  83              &18$^b$\\
23230-6926 &0.1062& 447& 327&1038& 88& 66&4.00&     1.5 &1.12& 42&187&0.086 &36& 66& 705&0.3&1565              &      \\
23365+3604 &0.0645& 266& 250& 704& 47& 68&4.00&     2.0 &1.12& 21&159&0.123 &36& 47& 595&0.5& 108\parbox{0cm}{$^{\rm +}$}              &39$^a$\\
23389-6139 &0.0927& 388& 182& 753& 56& 63&4.00&  $>$1.9 &1.12& 26&173&0.092 &35& 59& 668&0.4& 135              &      \\
23515-3127 &0.3347&1562&    &2323&105& 50&0.50&$<$1.6$>$&1.04& 62&759&0.100 &42& 58& 662&0.1& 184              &      \\
\hline\noalign{\smallskip}									    
\end{tabular}\\	      
$^{*}$ measurements partly distorted and uncertain, therefore $\tau$$_{\rm 100 \mu m}$ has been fixed to 6.0 \\   
$^+$ good determination of total dust mass, data used for estimate of gas-to-dust ratio (see Sect.\,\ref{sect_opacity_for_multiple_bb}) \\
$^a$ from  Solomon et al. 1997, Table2 \\
$^b$ from Mirabel et al. 1990, recalculated by Gao \& Solomon 1999 \\
$^c$ from Sanders et al. 1991, recalculated by Gao \& Solomon 1999 \\
\end{table*}

Table\,\ref{tab_luminosities} lists the dust masses derived from flux values at 850\,$\mu$m, where the 
emission is certainly transparent (and no mass is ``hidden''). For the further discussion, 
three different kinds of dust mass are determined for each galaxy:

\begin{itemize}
\item[(1)] For the assumption that the FIR emission can be modelled by one single blackbody with free 
           $\tau$, $\beta$ and T, as described by eqn.\,1 and shown in Fig.\,\ref{fig_seds}. 
\end{itemize}
For the case of several components with fixed $\beta$\,=\,2 (eqn.\,3)  the temperature decomposition 
is not unique (Sect.~\ref{sect_multiplebb}), and the uncertainty in the derived total dust mass becomes 
large, in particular for those sources without any submm/mm data. Therefore Table\,\ref{tab_luminosities} 
lists two estimates: 
\begin{itemize}
\item[(2)] The total dust mass M$_{\rm d(total)}^{\rm \beta\,=\,2}$ associated with the multiple 
           ``optically thin'' blackbodies.
\item[(3)] The dust mass associated with the ``bulk FIR emission component''  
           M$_{\rm d(FIR)}^{\rm \beta\,=\,2}$.
\end{itemize}

While M$_{\rm d(total)}^{\rm \beta\,=\,2}$ may be considered as a maximum dust mass, 
M$_{\rm d(FIR)}^{\rm \beta\,=\,2}$ is derived under the assumption of one single optically thin 
$\lambda$$^{\rm -2}$ modified blackbody of T\,$\approx$\,35--40\,K fitted to the FIR 60-200\,$\mu$m range. 
It contributes the bulk of the luminosity and may be considered as a firm lower limit for the dust mass, 
in particular if other colder dust components exist. M$_{\rm d(total)}^{\rm \beta\,=\,2}$ contains 
contributions from cold dust components, it is typically much larger than 
M$_{\rm d(FIR)}^{\rm \beta\,=\,2}$.

For all cases, the dust mass does not show any correlation with the total, mid- or far-infrared 
luminosity. But the dust mass is quite well correlated with the submm luminosity L$_{\rm 150-1000 \mu m}$. 

\subsection{Size of dust emitting regions}
\label{sect_dust_size}

The smallest possible extent of the FIR emitting region is listed in Tab.\,\ref{tab_luminosities}. For the 
case of an opaque blackbody (eqn.\,1), the brightness radius r$_{\rm b}$ is determined via
\begin{equation}
\label{eqn_8}
r_{b} =  \left\{ \frac {D ^{2} \cdot S_{100 \mu m} } { \pi \cdot B_{100 \mu m}(T) \cdot (1 - e^{- \tau_{100 \mu m} })  } \right\}^{\rm 1/2}
\end{equation}

For the case of a transparent blackbody with fixed $\beta$\,=\,2 (eqn.\,3), the low opacity condition 
$\tau_{ 100\,\mu m}$\,$<$\,1 requires that the extent of the dust emission cannot be smaller than a minimum 
size. Such a ``smallest transparent radius'' can be roughly estimated as follows: the dust mass (eqn.\,6) -- 
as determined for the bulk FIR emitting component at T\,$\approx$\,30-40\,K -- is distributed homogeneously 
in a ``minimum face-on disk'' of radius r$_{\rm \tau}$. Thereby the following standard conversions are used: 
0.1\,M$_{\odot}$\,$\cdot$\,pc$^{\rm -2}$ $\Leftrightarrow$ $\tau$$_{\rm V}$\,=\,0.4 (Whittet 1992), and 
$\tau_{\rm 100\,\mu m}$\,=\,0.006\,$\cdot$\,$\tau$$_{\rm V}$ (Mathis et al.1983). Thus 
\begin{equation}
\label{eqn_9}
r_{\tau} = \left\{ \frac {M_{dust} [M_{\odot}]} { \tau_{\rm 100 \mu m} \cdot 41.7 \cdot \pi } \right\}^{\rm 1/2} [pc]
\end{equation}

Typically r$_{\tau}$, for $\tau_{ 100 \mu m}$\,$=$\,1, is a factor 2-5 larger than r$_{b}$ 
(see Tab.\,\ref{tab_luminosities}), and we will discuss further below whether a dust emitting region as 
large as that given by r$_{\tau}$ is still consistent with other data.

\section{Discussion}

\subsection{Evidence for cold cirrus-like dust}
\label{evidence_colddust}

The observations show that, though the SEDs usually peak at 60-100$\mu$m, they can stay on a plateau before 
the onset of the Rayleigh-Jeans tail.
The analysis of Sect.~\ref{sect_singlebb} and~\ref{sect_multiplebb} showed that the observed SEDs
can be modelled by either one single tepid (50-75\,K) blackbody with $\beta$\,$\le$\,2, 
partly inferring a high opacity even in the FIR, or by several cool to cold transparent dust bodies 
with $\beta$\,=\,2. Which of these scenarios (or which combination) might more realistically apply for 
the ULIRGs?

In the following discussion, we will check for observational evidence for the existence of cold 
cirrus-like dust in ULIRGs being in line with the multiple blackbody model. One prerequisiste to allow a 
fit of the SEDs with several blackbodies is that $\beta$\,$\approx$\,2. Even in the case of several 
blackbodies the opacity cannot be low at all, since the derived dust mass and the size of the emission area 
exceed other constraints. In order to reduce the amount of dust mass and size, the opacity of some of the 
blackbodies must be increased (whereby the dust temperature also rises, resulting in a lower dust mass and 
a smaller region). Thus, we end up with the picture of several blackbodies with $\beta$\,=\,2, which are 
partly optically thick. 

\subsubsection{Evidence against the single blackbody model}
\label{sect_against_single_bb}

The following arguments can be brought forward against the concept of one single blackbody to describe 
the FIR-submm SEDs of most of our ULIRGs (i.e all of our mm-subsample except Arp220):

\begin{itemize}
\item[1)] In general, the optical morphology is disturbed and suggests a variety of patchy dust complexes 
          with less dense regions in between. Since in such a scenario various dust temperatures are 
          also expected, multiple blackbodies appear to provide a better physical description than one 
          single blackbody.
\item[2)] For a variety of objects several authors found observational evidence that $\beta \approx$ 2
          in the FIR-submm range ($\lambda$\,$>$\,100\,$\mu$m): 1.8\,$\la$\,$\beta$\,$\la$\,2.0 was found 
          for interstellar dust in our Galaxy (e.g. Mathis et al. 1983, Lagache et al. 1998), 
          $\beta$\,$\approx$\,2.0 for normal spiral galaxies (e.g. Bianchi et al. 1999) and the archetypal 
          starburst galaxy M\,82 (e.g. Thuma et al. 2000), and $\beta$\,=\,2.0$\pm$0.2 for active (Markarian) 
          galaxies (Chini et al. 1989b). In contrary to that, Dunne et al. (2000) found an average value
          $\beta$\,$\approx$\,1.3 from their SCUBA Local Universe Galaxy Survey. However, one weakness of
          this study appears to be the restriction to three wavelength points at 60, 100, and 850\,$\mu$m 
          without probing the maximum and the start of the Rayleigh Jeans branch in sufficient detail.
          One common object between their sample and ours is Arp\,220. While their single temperature fit
          provides $\beta$\,=\,1.2 our single modified blackbody fit to a much better sampled SED gives 
          $\beta$\,=\,1.7. Also, the spread of $\beta$ between 1.2 and 2.0 in our ULIRG sample would suggest 
          a large variety of dust properties, like grain size distribution, among the same type of object with
          probably similar evolution histories. It appears much more plausible to explain the SEDs by a
          composition of dust components all with consistent emissivity laws of $\beta \approx$ 2.

\item[3)] We developed the following quantitative method to check the consistency of the derived 
          $\tau_{ 100\,\mu m}$ values from the single blackbody fit with other extinction sensitive 
          quantities. One is strength of the PAH 7.7\,$\mu$m feature which is ubiquitous in ULIRGs 
          (Rigopoulou et al. 1999). There must be a sufficient number of UV photons to excite the PAHs,
          however, an intense starburst is not a prerequisite, as can be seen from the wide distribution of
          PAH emission in our Galaxy (Mattila et al. 1996) and in NGC\,891 (Mattila, Lehtinen \& Lemke 1999). 
          Mattila, Lehtinen \& Lemke (1999) found that the PAH distribution is similar to the one of large 
          dust grains and neutral molecular clouds. It is reasonable to assume that the PAH carriers are 
          mixed with other constituents of the ISM, at least on the spatial scales of kpcs we are looking at
          with the resolution of our observations. If the PAHs are not cospatial with the dust component 
          emitting the bulk of the FIR emission, then there is a second noticable dust component per se.
          
          In the case of all the FIR emission being irradiated by one single blackbody, we would expect a 
          mixture of the PAH carriers with the FIR emitting dust and a strong decrease of the 
          7.7\,$\mu$m PAH strength (published by Rigopoulou et al. 1999, see examples in 
          Fig.\,\ref{fig_pah_spectra} and Tab.\,\ref{tab_pah_fluxes}) relative to the submm and FIR fluxes 
          with increasing $\tau_{ 100\,\mu m}$. This decrease should follow the direction of the 
          A$_{\rm V}$-vector in Fig.\,\ref{fig_pah_to_fir_vs_tau}, (using the galactic extinction curve by 
          Mathis et al. (1983)). However, Fig.\,\ref{fig_pah_to_fir_vs_tau} shows that the normalised 
          7.7\,$\mu$m PAH fluxes are largely independent of $\tau_{ 100 \mu m}$, except for Arp220 which is 
          further discussed in Haas et al. (2001).

\begin{figure}[ht!]
\hspace{-1cm}
\resizebox{10cm}{!}{\includegraphics{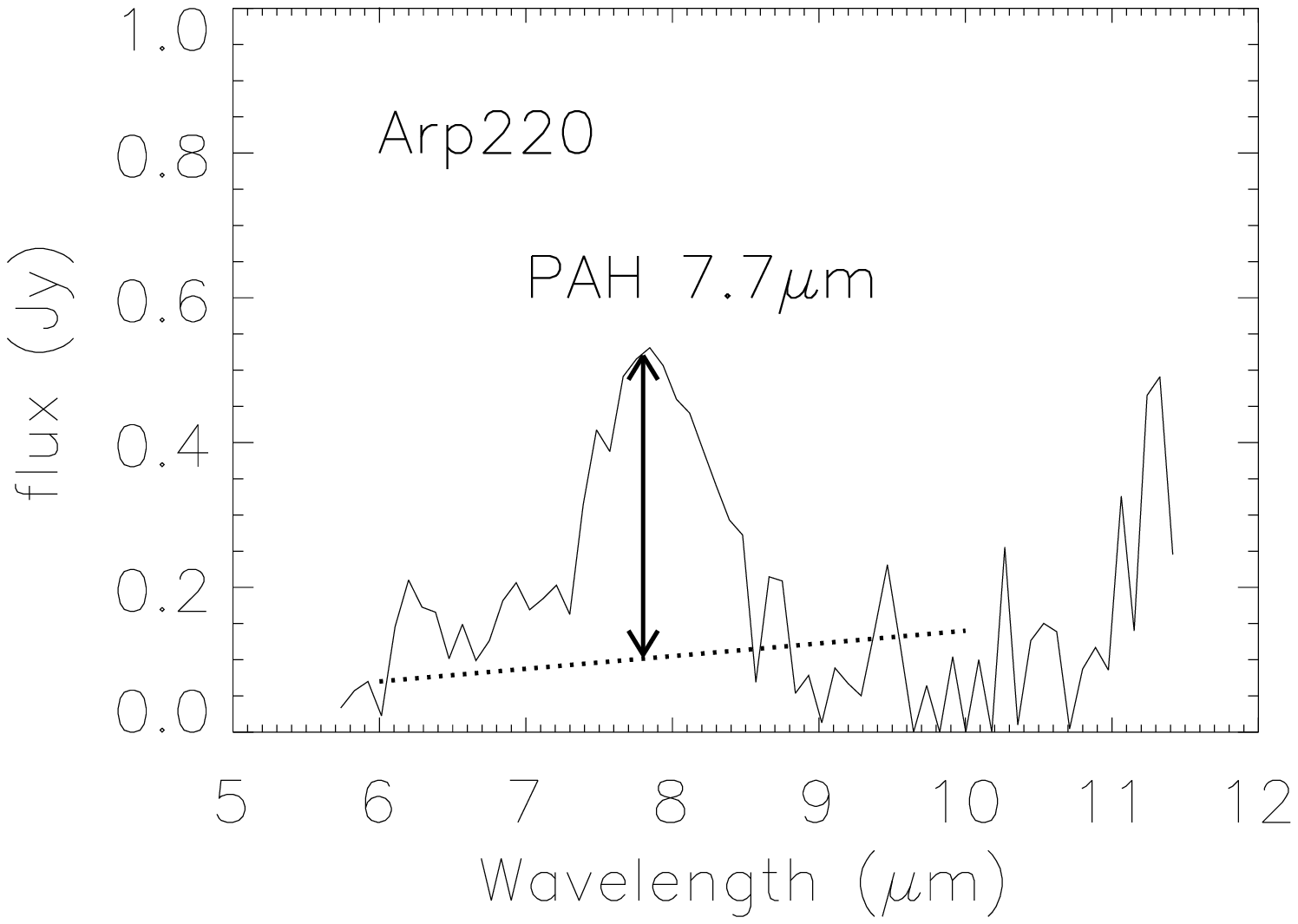}}
\vspace{-0.0cm}
\hspace{-1cm}
\resizebox{10cm}{!}{\includegraphics{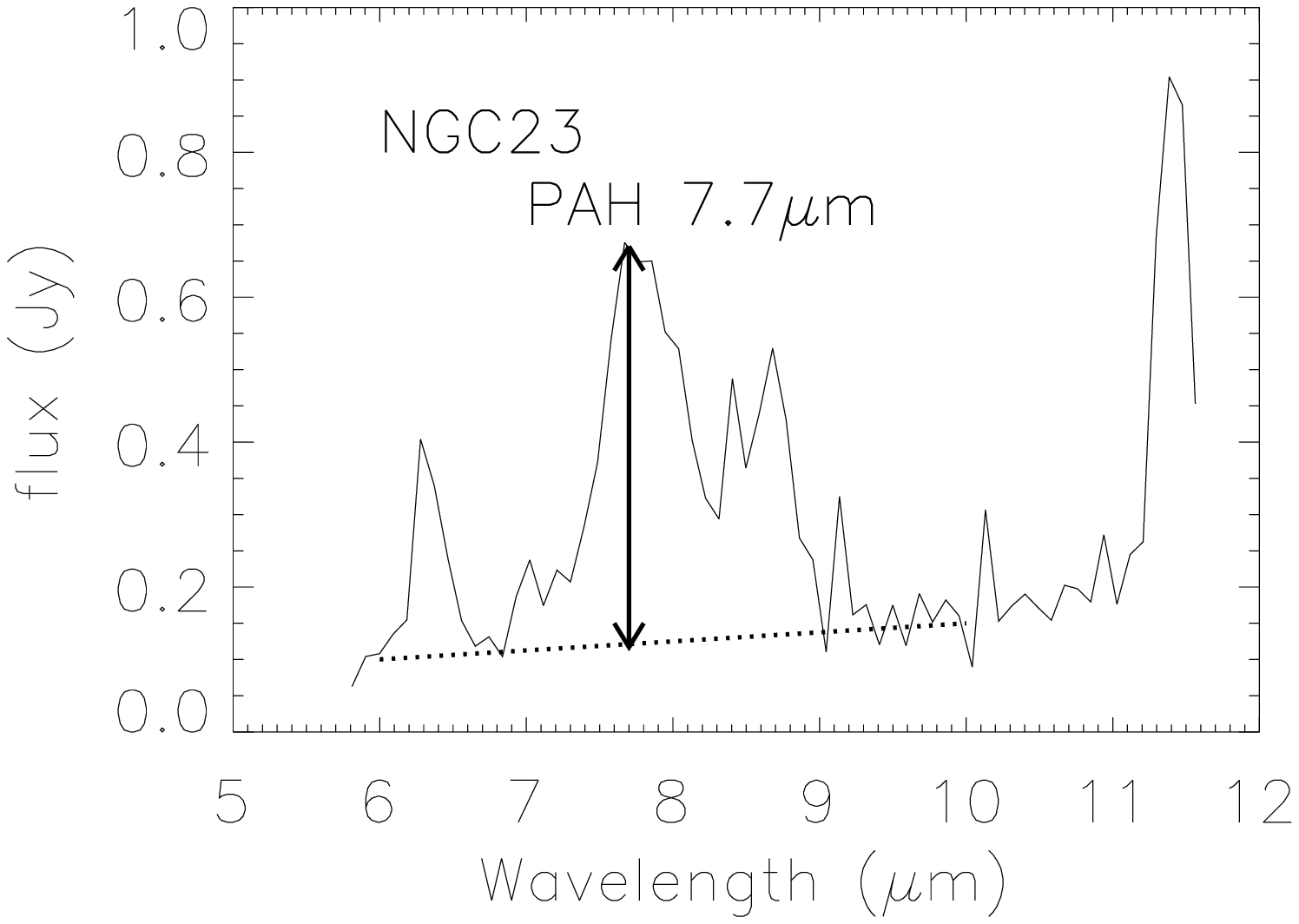}}
\caption{\label{fig_pah_spectra} 
Examples of ISOPHOT-S 5.8-11.6\,$\mu$m spectra for the ULIRG Arp220 and the comparison galaxy NGC\,23. 
The dotted line indicates the continuum subtracted for the estimate of the PAH 7.7\,$\mu$m peak flux 
strength. Arp220 has a strong silicate 9.7\,$\mu$m absorption feature (e.g. Dudley 1999), which is 
accounted for in the continuum estimate.}
\end{figure}

\begin{table}[ht!]
\footnotesize{
\begin{center}
 \caption[]{ PAH 7.7\,$\mu$m peak flux strength, 100\,$\mu$m and 850\,$\mu$m continuum fluxes used for 
             the assessment of flux ratios in Fig.\,\ref{fig_pah_to_fir_vs_tau}. Measured fluxes are in Jy. 
             The MIR continuum around the PAH line was subtracted, also the extrapolated synchrotron continuum
             from the 850\,$\mu$m flux. Uncertainties are less than 30\,\%. For the ULIRG sample PAH fluxes 
             are from Rigopoulou et al. (1999), 850\,$\mu$m fluxes are from Rigopoulou et al. (1996), 
             Lisenfeld et al. (2000) and from this work. For the sample of normal comparison galaxies PAH 
             fluxes are from this work and 850\,$\mu$m fluxes are from Dunne et al. (2000). The 100\,$\mu$m 
             fluxes are all from IRAS. 
   \label{tab_pah_fluxes}
}
\begin{tabular}{l|rrr}
\noalign{\smallskip}
 ULIRGs & PAH         &  100    & 850      \\
        & 7.7\,$\mu$m & $\mu$m  & $\mu$m   \\
\hline					        					        
 05189-2524   & 0.220 &   11.73 &    0.048 \\
 UGC\,5101    & 0.186 &   21.24 &    0.143 \\
 12112+0305   & 0.080 &    9.73 &    0.030 \\
 Mrk231       & 0.289 &   30.33 &    0.045 \\
 Mrk273       & 0.193 &   24.58 &    0.077 \\
 14348-1447   & 0.068 &    7.60 &    0.016 \\
 15250+3609   & 0.128 &    5.80 &    0.027 \\
 Arp220       & 0.414 &   126.7 &    0.744 \\
 NGC\,6240    & 0.420 &   28.13 &    0.137 \\
 17208-0014   & 0.274 &   35.66 &    0.119 \\
 19254-7245   & 0.084 &    5.38 &    0.029 \\
 20100-4156   & 0.041 &    5.20 & $<$0.019 \\
 22491-1808   & 0.043 &    5.06 &    0.012 \\
 23365+3604   & 0.086 &    8.00 &    0.014 \\
 23389-6139   & 0.028 &    4.33 & $<$0.009 \\
\hline					        
\noalign{\smallskip}
  Comparison Galaxies &  &  &  \\
\hline					                                                     
   Arp148     & 0.250 & 10.99   & 0.092   \\       
MGC+02-04-025 & 0.380 &  9.60   & 0.039   \\       
   Mrk331     & 0.650 & 20.86   & 0.132   \\       
   NGC\,~~~23 & 0.560 & 14.96   & 0.144   \\          
   NGC\,~~695 & 0.700 & 13.80   & 0.136   \\       
   NGC\,1122  & 0.600 & 15.15   & 0.084   \\       
   NGC\,1667  & 0.400 & 16.54   & 0.163   \\       
   NGC\,5256  & 0.350 & 10.35   & 0.082   \\       
   NGC\,5653  & 0.700 & 21.86   & 0.205   \\       
   NGC\,5713  & 1.000 & 36.27   & 0.359   \\       
   NGC\,5962  & 0.300 & 20.79   & 0.317   \\       
   NGC\,6052  & 0.270 & 10.18   & 0.095   \\       
   NGC\,7591  & 0.400 & 13.52   & 0.135   \\       
   NGC\,7592  & 0.450 & 10.50   & 0.108   \\       
   NGC\,7674  & 0.250 &  7.91   & 0.108   \\       
   NGC\,7679  & 0.350 & 10.65   & 0.093   \\       
   UGC\,2238  & 0.600 & 15.22   & 0.104   \\       
   UGC\,2369  & 0.300 & 11.10   & 0.072   \\       
   UGC\,2982  & 0.900 & 17.32   & 0.176   \\       
   UGC\,8387  & 0.500 & 24.90   & 0.113   \\

  \end{tabular}\\			 
\end{center}
}
\end{table}

\begin{figure}[ht!]
\resizebox{9cm}{!}{\includegraphics{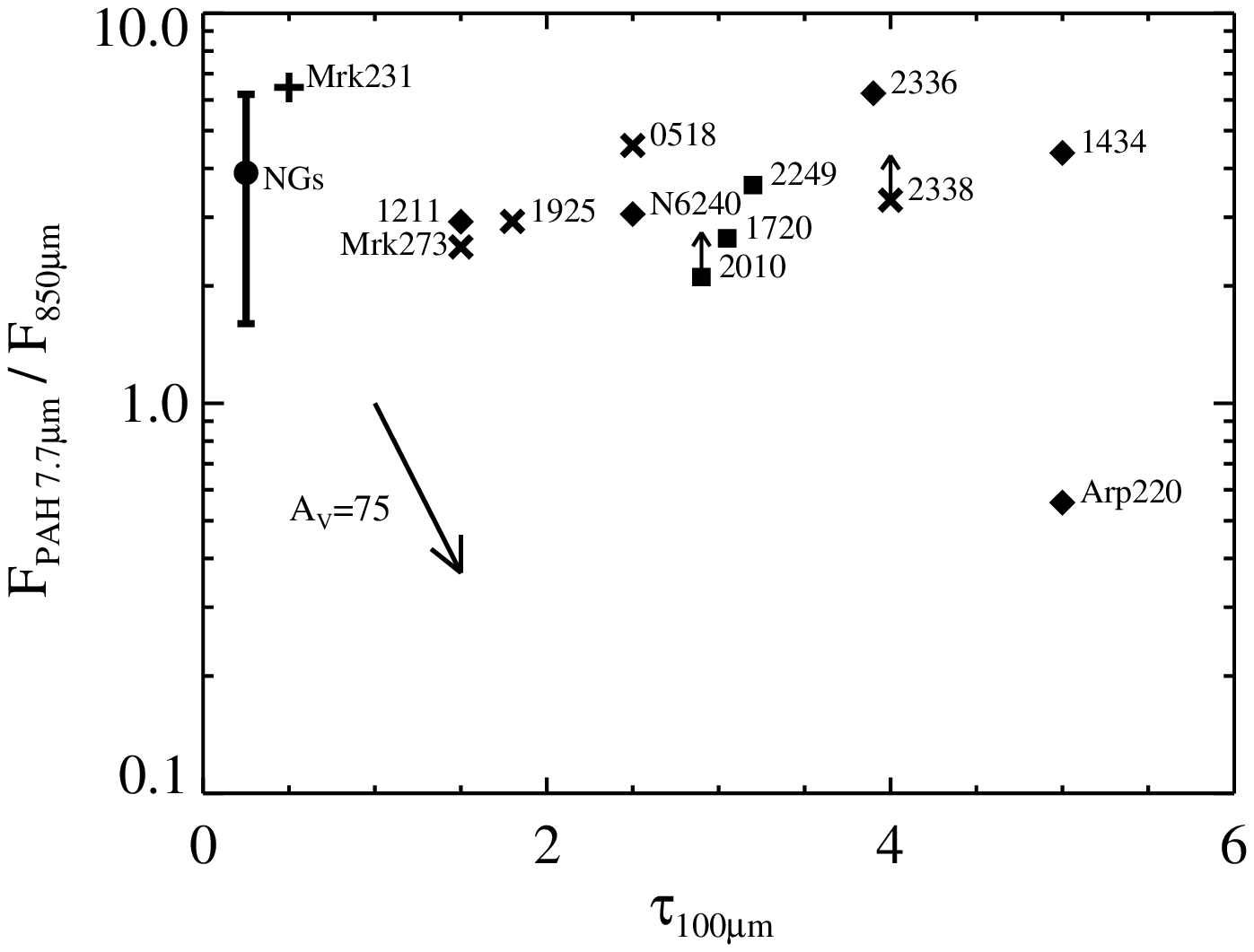}}
\resizebox{9cm}{!}{\includegraphics{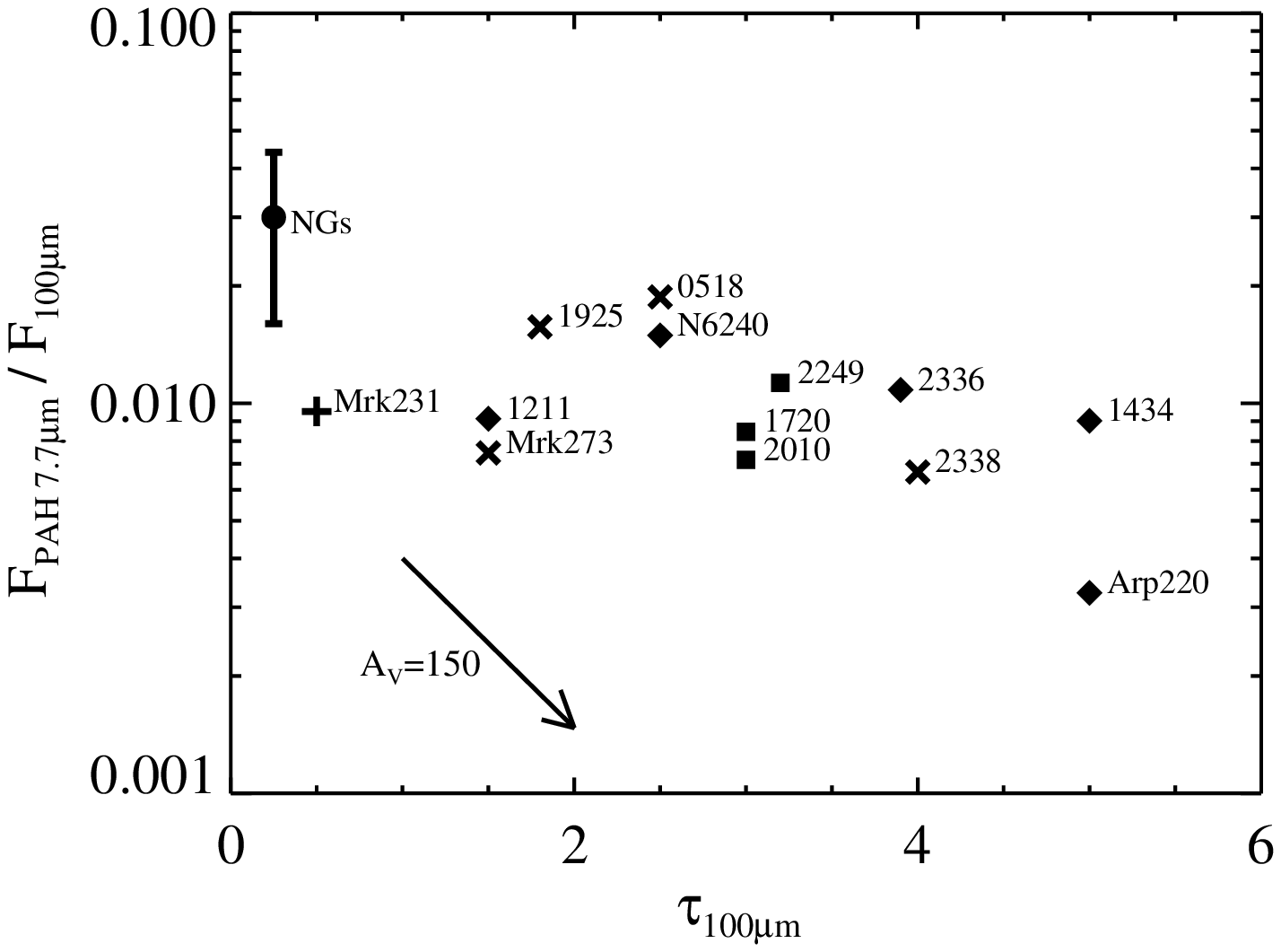}}
\caption{\label{fig_pah_to_fir_vs_tau} 
Normalised 7.7\,$\mu$m PAH flux versus opacity $\tau_{ 100 \mu m}$ 
as formally derived from the single blackbody fits.
Different symbols for each spectral type  as in Fig.\,\ref{fig_tau_beta_comparison}. 
The arrows show the effect of extinction, 
if the PAH carriers are mixed with the FIR opaque dust.
The filled circle and the thick vertical bar on the left-hand side 
show the range for normal galaxies ``NGs'' (see text).
}
\end{figure}

          Further clues on the relation between PAH and FIR-submm emitting dust comes from the comparison of 
          the ULIRGs with normal galaxies: We examined the PAH\,/\,100\,$\mu$m and PAH\,/\,850\,$\mu$m flux 
          ratios of the ULIRGs with that for normal galaxies, for which both 850\,$\mu$m and PAH data are 
          available. Among the SCUBA sample of 104 galaxies obtained by Dunne et al. (2000), 20 sources were 
          also measured with ISOPHOT-S (by several observers) covering a similar area (24$\arcsec$) as the 
          SCUBA observations. We derived the PAH\,7.7\,$\mu$m fluxes from the ISO data archive products 
          processed with OLP Version 9.0 (Laureijs et al. 2000). The values are listed in 
          Tab.\,\ref{tab_pah_fluxes} and an example for the MIR spectra is shown in 
          Fig.\,\ref{fig_pah_spectra}. Note, that now we consider only the vertical distribution in 
          Fig.\,\ref{fig_pah_to_fir_vs_tau} and ignore the horizontal one: The range for the normal galaxies 
          is indicated by the thick vertical bars on the left-hand side. Strikingly the normal galaxies 
          populate a similar PAH\,/\,850\,$\mu$m range (3.9\,$\pm$\,2.3) as the ``typical'' ULIRGs 
          (3.7\,$\pm$\,1.5, excluding Arp220), but the PAH\,/\,100\,$\mu$m range (0.03\,$\pm$\,0.014) lies 
          higher than for the ULIRGs (0.01\,$\pm$\,0.004). This difference cannot be explained by PAH 
          destruction in the ULIRGs, since it should be reflected in the PAH\,/\,850\,$\mu$m ratio as well. 
          Neither can this difference be due to extinction, because the shift of the normal galaxies 
          with respect to the ULIRGs must then be larger at 850\,$\mu$m than at 100\,$\mu$m (see the length 
          and orientation of the extinction arrows in Fig.~\ref{fig_pah_to_fir_vs_tau}. The difference of the 
          PAH\,/\,100\,$\mu$m ratio between normal galaxies and ULIRGs can be best explained by the higher 
          dust temperatures of the ULIRGs of about T\,=\,30-60\,K compared with T\,=\,15-30\,K in normal 
          galaxies, which increases the 100\,$\mu$m flux with respect to that at 850\,$\mu$m. This behaviour 
          also suggests that the PAH carriers not only trace the dust in active regions, but are even more 
          related to the entire dust content.
\item[4)] We compared the dust mass (M$_{\rm d}^{\rm \tau free}$ in Tab.\,\ref{tab_luminosities}) with 
          the molecular gas mass derived from integrated CO luminosities (M(H$_{\rm 2}$) in 
          Tab.\,\ref{tab_luminosities}). The resulting gas-to-dust mass ratio lies in the range between 500 
          and 2500, with an average of about 1500\,$\pm$\,500. This is much higher than the canonical value 
          of about 150 for normal galaxies (e.g. Stickel et al. 2000). This discrepancy becomes even higher, 
          when considering the additional contribution of the atomic gas. Since it is unlikely that
          the CO-to H$_{\rm 2}$ conversion factor for ULIRGs is so much different from that for normal 
          galaxies, a larger dust mass than M$_{\rm d}^{\rm \tau free}$ is required for ULIRGs in order to 
          match the canonical value. Such a suitably high dust mass can be achieved with the combination of 
          several dust components.
\end{itemize}

In summary, we listed several items where the single blackbody model appears to be insufficient or
inconsistent. This can be solved by the application of multiple blackbodies. In order to place several 
blackbodies underneath the Rayleigh-Jeans tail, $\beta$\,=\,2 seems to provide a reasonable choice.

\subsubsection{Constraining the FIR opacity for the multiple blackbody model}
\label{sect_opacity_for_multiple_bb}

For a low FIR opacity and several cold to cool blackbodies with $\beta$\,=\,2, the estimated total
dust mass M$_{\rm d(total)}^{\rm \beta\,=\,2}$ is considerably higher (on average by a factor of ten)
than for the single optically thick blackbody (cf. Table~\ref{tab_luminosities}). This is because of the 
contribution from the additional cold component (M $\propto$ T$^{-6}$ for $\beta$ = 2).
For the mm-subsample M$_{\rm d(total)}^{\rm \beta\,=\,2}$ lies in the range $\approx$ 10$^{\rm 8}$ -- 
10$^{\rm 9}$ M$_{\odot}$. As mentioned in Sect.~\ref{sect_dust_masses}, the uncertainty for 
M$_{\rm d(total)}^{\rm \beta\,=\,2}$ is quite large, because the decomposition into several blackbodies is 
not unique. Nevertheless, when comparing the M$_{\rm d(total)}^{\rm \beta\,=\,2}$ for the most reliable fits 
(marked by ``+'' in Tab.\,\ref{tab_luminosities}) with the molecular gas mass derived from integrated CO 
luminosities (Tab.\,\ref{tab_luminosities}), the gas-to-dust mass ratio lies in the range between 30 and 300 
with an average of 165 $\pm$ 120, close to the canonical value of about 150. The multiple blackbody model 
seems to yield a consistent total dust mass.
 
Stricter constraints on the dust mass can be obtained, when comparing it with the extension of CO gas 
inferred from interferometric data (Downes \& Solomon 1998). In order to keep the opacity, i.e. column 
density, low, the extent of the dust region must exceed a minimum size. As a check, we distribute the dust
mass associated with the bulk FIR emission, M$_{\rm d(FIR)}^{\rm \beta\,=\,2}$ (Sect.~\ref{sect_dust_masses}) 
evenly within a disk of minimal radius r$_{\tau\,=\,1}$, so that $\tau_{ 100\,\mu m}$\,=\,1 (r$_{\rm \tau}$ 
in Tab.\,\ref{tab_luminosities}). In each object the radius r($\tau_{\rm 100\,\mu m}$\,=\,1) is smaller 
than the radius of the optical-NIR image sizes. With the possible exception of Arp220, the extent of low 
opacity dust is consistent with the size of the CO disk, where available (from Tables 3\,\&\,4 in 
Downes \& Solomon 1998): Mrk231\,:\,0$\farcs$9, Mrk 273\,:\,0$\farcs$9-3$\farcs$1, 17208-0014\,:\,1$\farcs$8, 
23365+3604\,:\,1$\farcs$0. For $\tau_{ 100 \mu m}$\,=\,0.3, corresponding to A$_{\rm V}$\,$\approx$\,50, 
the minimal radius r$_{\rm \tau}$ is about a factor two larger than listed in Tab.\,\ref{tab_luminosities}, 
and exceeds the CO disk sizes. If the dust and the molecular gas are well mixed, then the opacity of the 
bulk FIR dust component cannot be low ($\tau_{ 100 \mu m}$\,$\le$\,0.3), rather it must be ``moderately'' 
high ($\tau^{\rm bulk}_{ 100 \mu m}$ \,$\approx$\,1, corresponding to A$_{\rm V}$\,$\approx$\,150) in many 
ULIRGs. Furthermore, with regard to our previous findings on the PAH/850\,$\mu$m and PAH/100\,$\mu$m ratios, 
which required relatively low extinction in the MIR (except for Arp220), we conclude that the bulk 
FIR dust component cannot contain the majority of the PAH carriers. The relative constancy of the 
PAH\,/\,850\,$\mu$m flux ratio with regard to normal galaxies is preserved, since the cold component
dominates at 850\,$\mu$m, as can be seen from the examples in Fig.~\ref{fig_multi_bb_examples}.

In summary, an emissivity exponent $\beta$\,=\,2 works well for the ULIRGs, as in normal spirals and 
Markarian galaxies. In some ULIRGs the single component fits already give $\beta$\,=\,2, and for those 
with $\beta$$_{\rm fitted}$\,$<$\,2 several blackbodies  with $\beta$\,=\,2 can be used. However, in order 
to match the dust mass and the size of the emitting regions with the constraints from the CO observations,
the opacity of the bulk FIR dust component might be moderately high 
(0.3\,$\la$\,$\tau^{\rm bulk}_{\rm 100\mu m}$\,$\approx$\,1).

\subsubsection{The proposed scenario: 
Cold cirrus dust and cool, partly opaque starburst dust}
\label{sect_scenario}

The previous discussion suggests that the FIR-submm SEDs are composed of several (at least two) 
$\lambda$$^{\rm -2}$ modified blackbodies with cool to cold temperatures. Actually, the dust might exhibit 
a continuous temperature range, but, when applying a formal model, within the uncertainties, a restriction 
to a few components works best. These dust components represent two stages: (1) the ``cirrus'' and 
(2) ``starburst'' stages.

The nuclear regions are probably surrounded by clumpy ``Super-Orion'' complexes providing the heating power 
for the starburst dust. The gaps in between the cloud complexes and the outer regions with lower density 
and weaker interstellar radiation field host the cirrus-like cold dust. In this picture the PAH 
emission comes preferentially from the submm emitting dust. Some fraction of the FIR peak dust 
emission in the nucleus or in dense clouds can be opaque in the FIR.

\subsection{Starburst and AGN heated dust}

In this section we assess for each optical spectral class (see Tab.\,\ref{tab_spectral_types}) the typical 
shape of the  IR-to-mm SEDs. While the SEDs of AGNs and SBs appear to be indistinguishable at FIR and submm 
wavelengths, they differ in the NIR-MIR. This suggests that the cool FIR emitting dust is not connected to 
the AGN, and that the AGN only powers the warm and hot dust. The difference at short IR wavelengths provides 
a NIR diagnostic tool based on the J-K colours, in order to reveal the presence of an AGN (with red colours, 
while SBs have blue colours). This tool is successful in the sense that if an AGN is seen by other 
diagnostics (optical or MIR spectroscopy or X-rays), then in most cases it is also seen via the NIR colours.

\begin{table}[!h]
\caption[] { \label{tab_spectral_types}
           Morphologies and spectroscopic classifications. The morphologies are from optical and NIR 
           images as referenced. The extinction E$_{\rm B-V}$ has been compiled from 
           H$_{\rm \alpha}$/H$_{\rm \beta}$ ratios. The optical spectral types distinguish between Seyfert 1, 
           Seyfert 2, LINER (galaxies with Low Ionisation Nuclear Emission Regions), and HII/starburst types, 
           as classified in the respective references. The MIR spectral types distinguish between 
           AGN\,=\,Seyfert\,1\,+\,Seyfert\,2 and SB\,=\,LINER\,+\,HII types (Lutz, Veilleux \& Genzel 1999). 
           The NIR classifications are derived from the J-K colours (AGN, SB, SB/n\,=\,SB for total galaxy 
           and red 1$\arcsec$ nucleus, SB/f\,=\,SB for total galaxy and 1$\arcsec$ nucleus). The X-ray 
           evidence for AGN is taken from Risaliti et al. (2000).}
\scriptsize
\begin{tabular}{@{}l@{\hspace{2mm}}l@{\hspace{2mm}}l@{\hspace{2mm}}l@{\hspace{2mm}}l@{\hspace{2mm}}l@{\hspace{2mm}}l}
\hline\noalign{\smallskip}
{\bf Name}       & {\bf Morph}     &{\bf E$_{\rm B-V}$}& {\bf Optical}   & {\bf MIR}       & {\bf NIR}  & {\bf X-ray} \\
                 &        .           & [mag]          & {\bf Sp-type}   & {\bf type}      & {\bf class}& {\bf AGN} \\
\hline\noalign{\smallskip}                                                             		    
00199-7426       & double $^{m1}$     &                & LINER? $^{s1}$  &                 & SB         & \\
00262+4251       & merger $^{m2,*}$   &                &                 &                 & SB$^{*}$   & \\
00406-3127$^{r1}$& merger $^{m6}$     &                & Seyf 2 $^{s6}$  &                 &            & \\
03068-5346       & double?$^{DSS}$    &                &                 &                 &            & \\
03158+4227       & compact$^{m2}$     &                &                 &                 &            & No  \\
03538-6432$^{r1}$& compact$^{DSS}$    &                &                 &                 &            & \\
04232+1436       & merger $^{*}$      &                &                 &                 & AGN$^{*}$  & \\
05189-2524       & merger $^{m3}$     &  2.03          & Seyf 2 $^{s3}$  &      AGN$^{a}$  & AGN        & Yes \\
06035-7102       & double $^{m1}$     &  1.34          & HII/SB $^{s1}$  &      SB         & SB         & \\
06206-6315       & double $^{m1}$     &  2.06          & Seyf 2 $^{s1}$  &      SB         & AGN        & \\
12112+0305       & double $^{m3}$     &  0.92          & LINER $^{s3}$   &      SB         & SB/n       & \\
Mrk 231          & merger $^{m3}$     &  0.70          & Seyf 1 $^{s3}$  &      AGN        & AGN        & Yes \\
Mrk 273          & merger $^{m3}$     &  1.22          & Sy2 $^{s3}$/LIN$^{s8}$&AGN        & SB/n       & No  \\
Mrk 463          & double $^{m4}$     &  0.60          & Seyf 2 $^{s4}$  &      AGN        & AGN        & Yes \\
14348-1447       & merger $^{m3}$     &  1.15          & LINER $^{s3}$   &      SB         & SB/n       & \\
14378-3651       & merger $^{m1}$     &  1.26          & LINER $^{s1}$   &                 & SB         & \\
15245+1019       & merger $^{m2}$     &                &                 &                 &            & \\
15250+3609       & ring gal. $^{m3}$  &  0.76          & LINER $^{s3}$   &      SB         & SB         & \\
Arp 220          & merger $^{m3}$     &  1.05          & LINER $^{s3}$   &      SB         & SB/n       & No  \\
15462-0450       & merger $^{m2}$     &  0.60          & Seyf 1 $^{s5}$  &                 & AGN        & \\
16090-0139       & merger $^{m1,2}$   &  1.55          & LINER $^{s1,5}$ &                 & SB         & \\
NGC 6240         & double $^{m5}$     &  1.65          & LINER $^{s5}$   &      SB         & SB/n       & Yes \\
17208-0014       & merger $^{m1,2}$   &  1.75          & HII/SB $^{s1}$  &      SB         & SB/n       & No  \\
17463+5806$^{r1}$& compact$^{DSS}$    &                &                 &                 &            & \\
18090+0130       & merger $^{*}$      &                &                 &                 & SB$^{*}$   & \\
18470+3234       & double $^{m2,*}$   &  1.13          & HII/SB $^{s5}$  &                 & SB$^{*}$   & \\
19254-7245       & double $^{m1,2}$   &  2.92          & Seyf 2 $^{s1}$  &      AGN        & AGN        & Yes \\
19458+0944       & double $^{m2}$     &                &                 &                 &            & \\
20046-0623       & double $^{m1,2,*}$ &                & LINER $^{*}$    &                 & SB         & \\
20087-0308       & merger $^{m1,2}$   &  1.43          & LINER $^{s1}$   &                 & SB         & \\
20100-4156       & double $^{m1}$     &  1.12          & HII/SB $^{s1}$  &      SB         & SB         & \\
20414-1651       & merger? $^{m1,2}$  &  1.31          & LINER $^{s1}$   &                 &            & \\
ESO 286-19       & merger $^{m1}$     &  1.07          & LINER $^{s1}$   &      SB         & SB         & No  \\
21130-4446       & merger $^{m1}$     &  0.50          & HII/SB $^{s1}$  &                 & SB         & \\
21504-0628       & merger $^{m1,*}$   &                & LINER $^{*}$    &                 & SB$^{*}$   & \\
22491-1808       & merger $^{m1}$     &  0.81          & HII/SB $^{s1}$  &      SB         & SB/f       & \\
ESO 148-2        & merger $^{m1}$     &  1.01          & HII/SB $^{s1}$  &      SB         & SB         & No  \\
23230-6926       & merger $^{m1}$     &  1.49          & LINER $^{s1}$   &                 & SB         & \\
23365+3604       & merger $^{m7,*}$   &  0.88          & LINER $^{s7}$   &      SB         & SB$^{*}$   & \\
23389-6139       & double $^{m1}$     &  3.06          & Seyf 2 $^{s1}$  &                 & SB         & \\
23515-3127$^{r1}$& compact$^{DSS}$    &                & Seyf 2 $^{s6}$  &                 &            & \\
\hline\noalign{\smallskip}
\end{tabular}
\\
$^{r1      }$ z\,$>$\,0.3 \\
$^{*       }$ this work, see Appendix     \\
$^{e0      }$ Boksenberg et al. 1977, Krabbe et al. 1997     \\
$^{m1,s1   }$ Duc et al. (1997)         \\
$^{m2      }$ Murphy et al. (1996)      \\
$^{m3,s3   }$ Sanders et al. (1988a)    \\
$^{m4      }$ Sanders et al. (1988b)    \\
$^{s4      }$ V\'eron-Cetty \& V\'eron (1985) \\
$^{m5      }$ Fried \& Schulz (1983) \\
$^{s5      }$ Kim et al. (1995) and Veilleux et al. (1995) \\
$^{m6      }$ Clements et al. (1996)    \\
$^{s6      }$ Allen et al. (1991)    \\
$^{m7, s7  }$ Klaas \& Els\"asser (1991)    \\
$^{s8      }$ Colina et al. (1999)    \\
$^{a       }$ Clavel et al. (2000)    
\end{table}

\subsubsection{Seyferts}

Our sample contains eight galaxies (with z\,$<$\,0.3) which have optical Seyfert spectra, i.e.\ are sources 
known to house an AGN. Two are Seyfert\,1s and six are Seyfert\,2s (see Tab.\,\ref{tab_spectral_types}).

The SEDs of the Seyferts show a variety of shapes in the NIR and MIR range:
\begin{itemize} 
\item[$\bullet$] On the one hand, a power-law-like flux increase is seen in the Seyfert\,1s (Mrk231 and 
                 15462-0450) and the Seyfert\,2s 05189-2524, 19254-7245 and Mrk463 (and probably also 
                 06206-6315). 
\item[$\bullet$] On the other hand, Mrk273 (a LINER nucleus with an off-nucleus Seyfert\,2  nebula, cf.\ 
                 Colina et al. (1999)) has a flat NIR flux plateau with a steep rise at about 10\,$\mu$m. 
\item[$\bullet$] 23389-6139, showing a damped power-law flux increase, appears to be in between these 
                 extremes.
\end{itemize}

The power-law shape might be attributed to the central, relatively unobscured AGN. The red J-K colours 
might not be caused by extinction only, rather the hottest dust clouds at a temperature just below the
evaporation temperature of the dust of about 1500\,K, correponding to a peak wavelength of about 2\,$\mu$m, 
might be seen.

In a torus geometry the flat sources like 23389-6139 could be seen more edge-on, so that the hot AGN heated 
dust is more hidden. Then the flat NIR flux plateau -- with blue colours -- may be understood in terms of 
light scattered at bipolar cones. Alternatively, young hot stars in the host galaxies might contribute 
significantly to the NIR flux, shifting the colours towards the blue. Also, the AGN could be weak compared 
with the strength of the starbursts; perhaps Mrk273 belongs to this group.

A somewhat indefinite case is the double source 06206-6315, which is classified by Duc et al. (1997) as a 
Seyfert\,2, but  as starburst via PAH diagnostics (Lutz, Veilleux \& Genzel\ 1999, Rigopoulou et al. 1999), 
although both spectra look quite noisy. The steep optical to NIR 2.2\,$\mu$m flux increase argues in favor 
of a relatively unobscured AGN. 

In the FIR and submm range the Seyfert SEDs also exhibit some diversity, though only a relatively moderate
one. Examples are Mrk463 and 19254-7245, both with double nuclei (and similar power-law NIR flux increase): 
While Mrk463 only shows little FIR flux compared to the MIR flux, 19254-7245 has a strong FIR bump and even 
a considerable amount of mm flux due to cold dust. A similar difference is found for Mrk231 and 05189-2524, 
both with a ``single'' nucleus, i.e.\ probably mergers in an advanced state: Both show a strong FIR bump, 
but Mrk231 has a steeper Rayleigh-Jeans tail and less cold dust than 05189-2524. Thus, the host galaxies of 
AGNs may or may not have cold dust.

\subsubsection{LINERs and HII/SBs}

Our sample contains 11 LINERs (Low Ionisation Nuclear Emission Regions) and 9 HII/SBs. In general, the SEDs 
look very similar for both optical spectral types. Therefore, we discuss them together here.

The SEDs appear quite homogeneous from the NIR to the FIR, with some diversity in the submm range: 
All SEDs show a relatively flat NIR flux plateau with a faint 2\,$\mu$m bump, PAH emission and/or 
Si 9.7\,$\mu$m absorption features, then a steep rise in flux at about 10\,$\mu$m, i.e.\ a 10\,$\mu$m knee. 
The flux peaks in the FIR at about 60-100\,$\mu$m. The only variety appears in the submm range:
Some sources show a steep Rayleigh-Jeans tail, others exhibit a flatter tail providing evidence for 
additional cold dust. Typical sources with and without cold dust are 23230-6926 and 23365+3604, respectively, 
among the LINERs, and 17208-0014 and ESO286-19, respectively, among the HII/SBs. 

The homogeneity and extreme similarity of the NIR to FIR SEDs suggests that the dust in LINERs and HII/SBs 
has similar properties, in particular concerning the spatial distribution and heating mechanism. In fact, 
this similarity places the LINERs closer to HII/SB than AGN dominated ULIRGs. A similar conclusion (that 
LINERs are not AGN dominated) was drawn from the PAH diagnostics (Lutz, Veilleux \& Genzel 1999). 

\subsubsection{The SED shapes and J-K colours as diagnostic tool}

\begin{figure}
\resizebox{9cm}{!} {\includegraphics {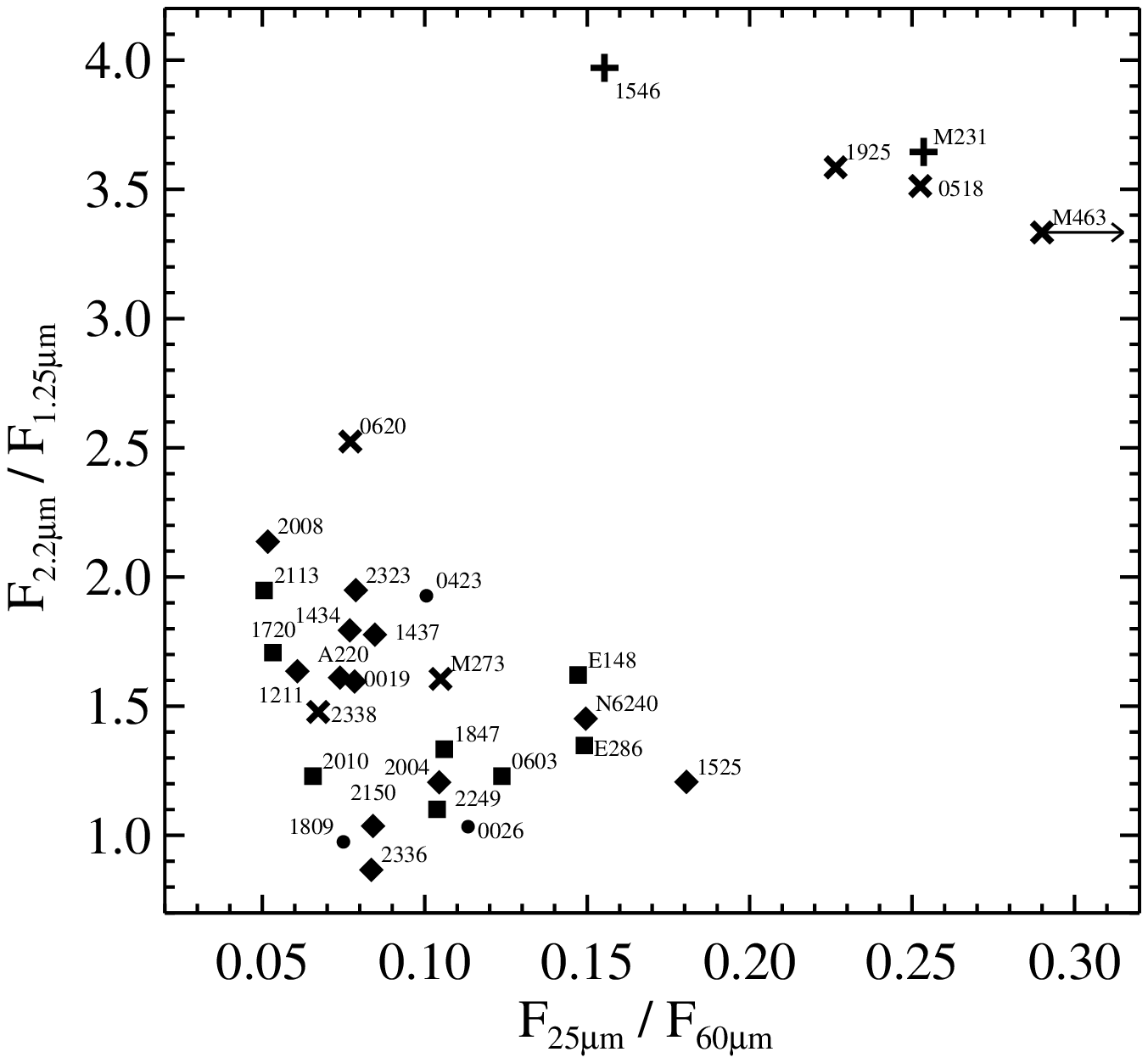} } 
\resizebox{9cm}{!} {\includegraphics {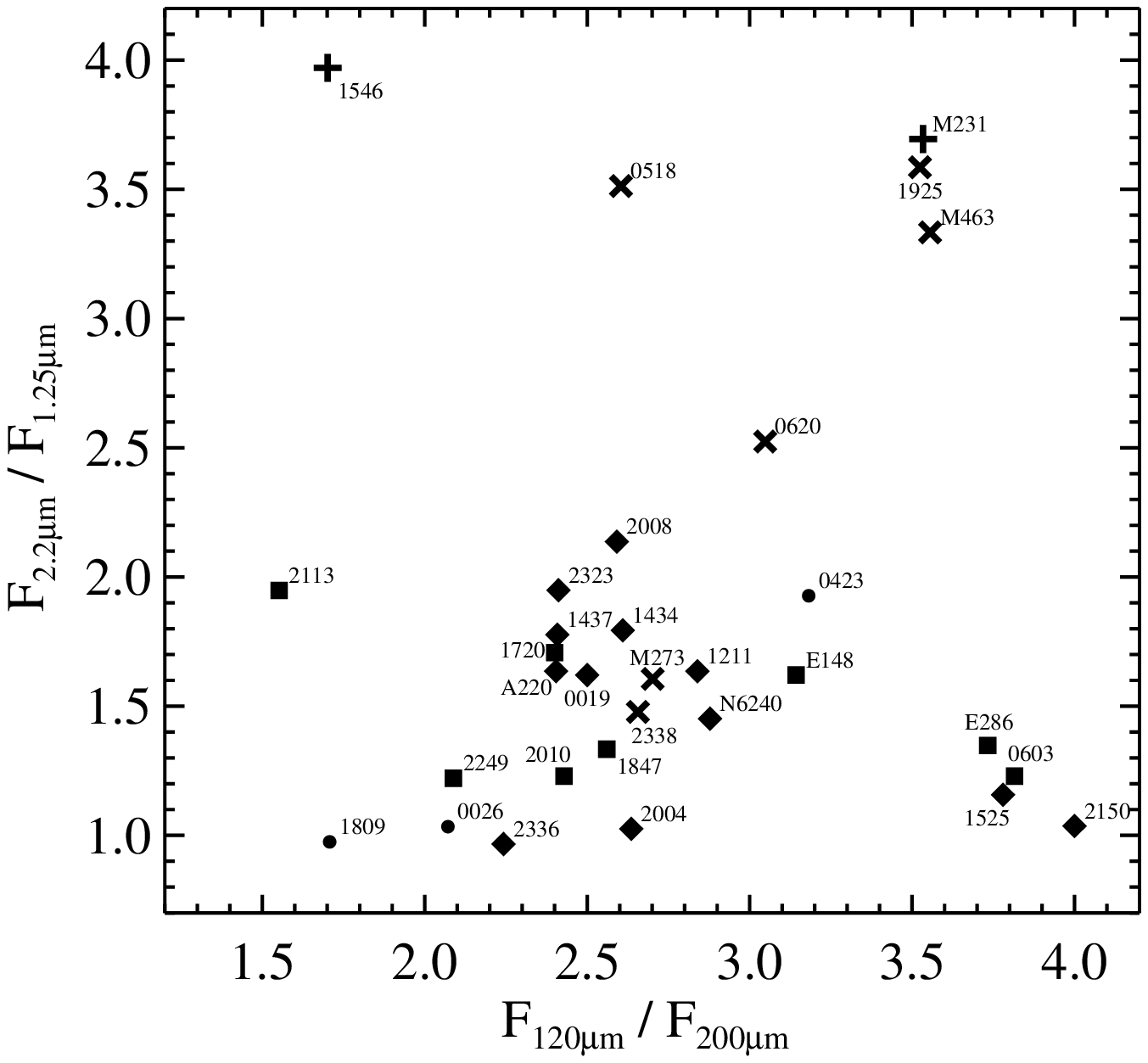} } 
\caption{\label{fig_two_colour_diagrams} 
        Two-colour diagrams 2.2$\mu$m\,/\,1.25$\mu$m versus 25$\mu$m\,/\,60$\mu$m (top) and versus 
        120$\mu$m\,/\,200$\mu$m (bottom). Different symbols for each spectral type as in 
        Fig.~\ref{fig_tau_beta_comparison}. The filled circles label objects with no spectroscopic 
        classification.
}
\end{figure}

\begin{figure} 
\resizebox{9cm}{!}{\includegraphics{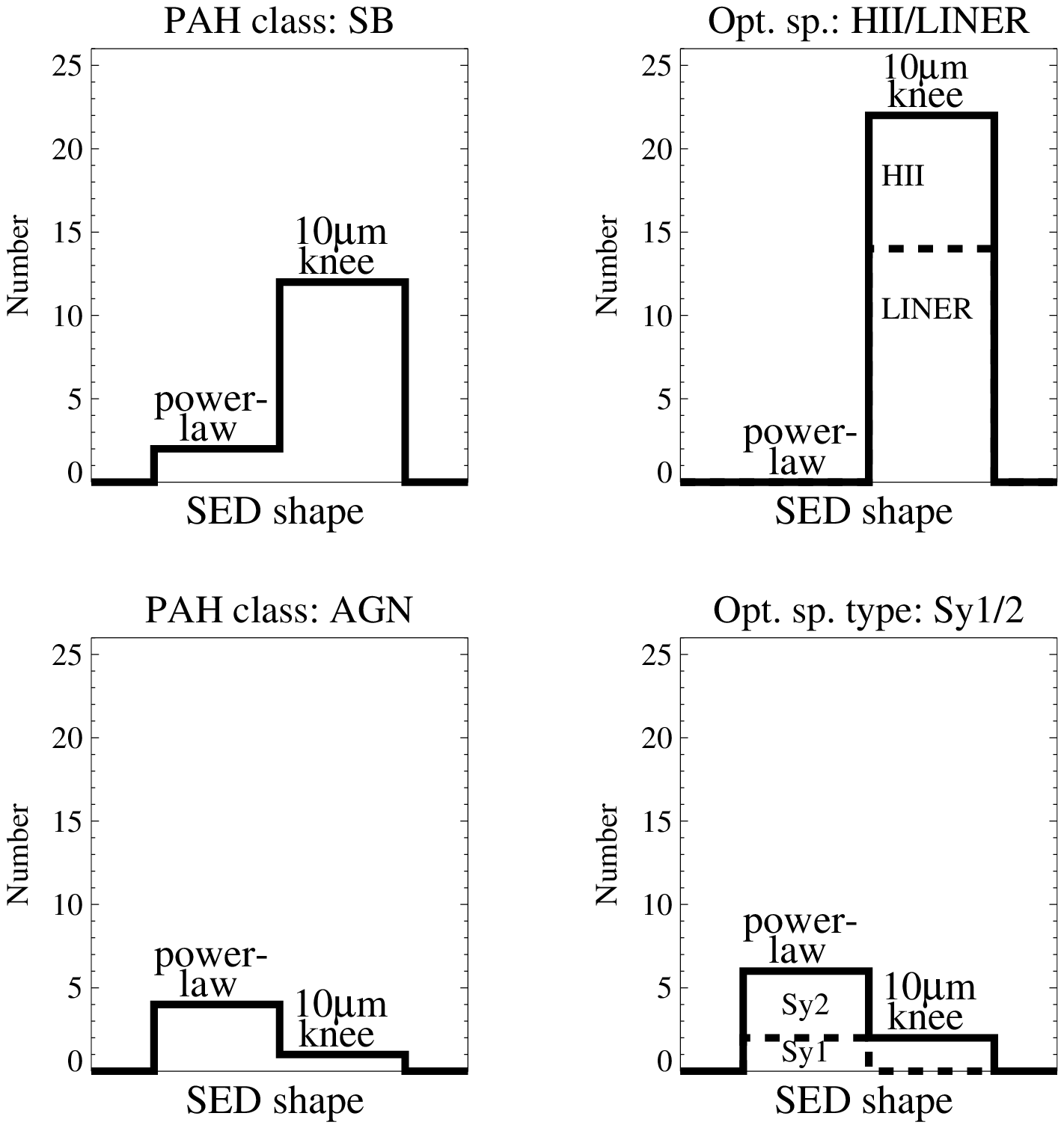}}
\caption{\label{fig_hist_classifications} 
        Histogram showing the correspondence of the  NIR SED shape classification (via 
        the J\,-\,K colour) with the optical emission line diagnostic and MIR (PAH class) spectroscopic 
        classifications.}
\end{figure}

The NIR-MIR power-law flux increase for the AGN-type ULIRGs and the flat NIR flux plateau with a 
10\,$\mu$m knee for the SB-type ULIRGs suggests the utilization of these two different SED 
shapes for a classification scheme. Earlier investigations (e.g.\ Hill et al.\ 1988) showed that red NIR 
colours provide a high probability of finding an AGN among moderately luminous IRAS galaxies. 
As a quantitative characterisation measure for the SED shapes we use the J\,-\,K colours: 
J\,-\,K\,$\ge$\,2\,mag identifies power-law shape SEDs; this value corresponds to a 
2.2$\mu$m\,/\,1.25$\mu$m flux ratio of about 2.34.

Although longer MIR wavelengths are less sensitive to extinction, this range suffers from possible 
confusion by strong PAH emission and 9.7\,$\mu$m silicate absorption which complicates the analysis; 
detailed work has been done by several authors, e.g. Laurent et al. (2000), Imanishi \& Dudley 
(2000) and Tran et al. (2001).

As shown in Fig.\,\ref{fig_two_colour_diagrams}, the J\,-\,K colours (2.2$\mu$m\,/\,1.25$\mu$m flux ratios) 
provide a much clearer separation of AGN- and SB-ULIRGs than the MIR colours (25$\mu$m\,/\,60$\mu$m) 
used to identify warm AGN-type objects (de Grijp et al. 1985), to say nothing of the FIR 
colours (120$\mu$m\,/\,200$\mu$m).

In Fig.~\ref{fig_hist_classifications} we show in histograms the correspondence of this NIR photometry
classification scheme with the spectroscopic optical emission line and the MIR PAH diagnostic classification
as compiled in Table~\ref{tab_spectral_types}. With respect to the optical classification, we find that the 
NIR SED shape confirms for 22 out of 22 (100\,\%) the optical classification as a HII/LINER object and for
6 out of 8 (75\,\%) the classification as Seyfert\,1\,/\,2. With respect to the MIR PAH diagnostics, we find
that the NIR SED shape confirms for 12 out of 14 (86\,\%) the PAH classification as SB and for 4 out of 5 
(80\,\%) the PAH classification as AGN.

These classifications are further confirmed by comparing them with the hard X-ray classifications 
(Table\,\ref{tab_spectral_types}). Thus, the J-K colours represent an attractive tool to explore the 
nature of ULIRGs and other dust rich IR galaxies. This is of particular advantage, if the galaxies are too 
distant or too faint for spectroscopy. Then the intrinsic J-K colours can provide constraints on the nature 
of cosmologically interesting ULIRGs. 

All these methods have some limitations, of course, and in some cases provide indefinite 
classifications. For example, the J-K colour does not reveal any AGN in Mrk273 which is optically classified 
as Seyfert\,2 and has a PAH/continuum flux ratio of 1.9 favouring the starburst dominance (Rigopoulou et al.\ 
1999). Also X-ray observations with Beppo-Sax unveiled an AGN in NGC6240 (Vignati et al. 1999), which is 
optically classified as a LINER, SB dominated according to PAH diagnostics, and shows no AGN-typical 
power-law NIR-MIR flux increase (except for the 1$\arcsec$ nuclear region).

\subsubsection{The nature of the 10\,$\mu$m knee}

Pure starburst objects have a flat NIR flux plateau followed by a steep rise longwards of 10\,$\mu$m.
Part of this pronounced 10\,$\mu$m knee could be due to the 9.7\,$\mu$m silicate absorption.
In contrast, galactic compact HII regions housing O5-O9 stars show SEDs with a  steep power-law-like rise. 
As for AGNs this rise already starts at about 1\,$\mu$m (Chini et al. 1987). Also, the prototype starburst 
galaxy M82 has such a SED shape (e.g. Thuma et al. 2000). Thus, a power-law SED could, in principle, also 
occur in ULIRGs with starburst classification. But this is not observed for our sample.

HST NICMOS images (Scoville et al.\ 2000) reveal that the central 1$\arcsec$ region is redder than the 
outer regions. Fig.\,\ref{fig_seds} shows that for those 9 sources in common with our sample (05189-2524, 
12112+0305, Mrk\,273, 14348-1447, 15250+3609, Arp\,220, NGC\,6240, 17208-0014 and 22491-1808) most of the 
nuclear SEDs (marked with ''$\times$'' symbols) exhibit -- even for LINERs and HIIs -- a power-law-like flux 
increase in the NIR. J-K colours of these nuclear regions are similar to the colours in larger apertures 
found for the AGN-type ULIRGs. For 05189-2524 the NICMOS fluxes (hardly to recognize in Fig.~\ref{fig_seds})
are practically identical with the ground based fluxes showing that this source is completely dominated by 
the central point-like source. For the others the much higher and bluer ground-based fluxes suggest that the 
sources are extended on the 5 - 10$\arcsec$ scale, in particular for 17208-0014 and 22491-1808 for which 
already the NICMOS images indicate that the SEDs are composed of many different components.

With regard to the Balmer decrement LINERs and HII-starburst galaxies have low E$_{\rm B-V}$ 
(Tab.\,\ref{tab_spectral_types}), as found for Seyfert\,1s. The highest E$_{\rm B-V}$ are those for 
Seyfert\,2s, indicating that their emission regions are the most deeply embedded ones. A deeply embedded HII 
region should suffer from extinction in the same way. However, deeply embedded young stars will probably be 
outshone by those located closer to the outer surface of the galaxy and will not dominate the NIR colours.
In conclusion, the relatively flat blue NIR flux plateau comes from the practically unobscured outer regions. 
The strong flux rise longwards of 10\,$\mu$m is caused by more deeply embedded luminous star forming regions. 
This naturally explains the appearance of the 10\,$\mu$m knee.

\subsubsection{Does the AGN power the FIR emission of ULIRGs?}

It is now widely accepted that a powerful (and not obscured) AGN creates a strong NIR-MIR emission 
(e.g. Sanders et al. 1988a, 1988b, Pier \& Krolik 1992, 1993, Rowan-Robinson 1995, Haas et al. 2000a). 
However, it is still a matter of debate whether the FIR/submm emission in ULIRGs (as well as quasars) is 
also mainly powered by the AGN (e.g. Sanders 1999) or by circumnuclear SBs (e.g. Rowan-Robinson 1995, 
Genzel et al. 1998). 

The power-law flux increase of Mrk231 appears similar to that of the quasars PG\,0050+124 and PG\,1613+658 
(Haas et al. 2000a). For these quasars it is not clear whether and in which wavelength range a starburst 
contributes to their power-law SEDs: Energetically, the AGN radiation would be sufficient to heat the dust, 
and since pure starbursts have different SED shapes (with knees, as shown above), the superposition of a 
significant starburst component would probably destroy the smooth power-law shape. For Mrk231, however, 
the FIR luminosity exceeds the MIR- and the (dereddened) UV-optical-luminosity (Downes \& Solomon 1998). 
Downes \& Solomon (1998) found two rotating gas disks with radii of 460 and 1150 pc, respectively, 
and concluded that the outer one is starburst dominated, since it is too extended and FIR-luminous to 
intercept sufficient power from the central region. Although the AGN is clearly visible, it dominates 
neither the FIR luminosity nor the total energy output. However, the PAH/continuum diagnostics indicates 
a dominating AGN in Mrk231 (Genzel et al.\ 1998, Lutz, Veilleux \& Genzel 1999). The reason for this 
might be that the unobscured AGN continuum diminishes the relative PAH strength, mimicking a weak starburst 
and overemphasising the role of the AGN. Thus, it is necessary to consider not only the ratio PAH/continuum, 
but also the absolute PAH and continuum levels, both of which may be affected by different amounts of 
extinction due to different sites of origin in the galaxies. For the diagnostics of moderately luminous 
Seyfert\,1s and Seyfert\,2s this has already been stressed by Clavel et al. (2000).

As shown in Fig.\,\ref{fig_two_colour_diagrams}, the ULIRGs with and without AGN signatures cannot be 
distinguished via their FIR colours (F$_{\rm 120\,\mu m}$\,/\,F$_{\rm 200\,\mu m}$). This suggests that 
either the FIR/submm emission is largely independent of the presence of an AGN, or that every ULIRG 
not classified as AGN contains a hidden AGN. The relative high number of ULIRGs with signs
of strong starbursts versus ULIRGs with AGN signatures in our sample (22:10 from optical emission line 
diagnostics, 13:5 from PAH diagnostics) would argue in favour of the first alternative. Since in 
Sect.~\ref{sect_opacity_for_multiple_bb} we concluded that the bulk FIR component might be moderately opaque
($\tau^{\rm bulk}_{\rm 100\mu m}$\,$\approx$\,1) even IR diagnostics, in particular the PAH\,/\,850$\mu$m 
flux ratio, have limited meaningfullness in this context. An exception seems to be Arp\,220 where the dust
geometry leads to a strong deficiency of PAH strength versus 850\,$\mu$m flux and dereddening would yield
a quasar-like IR continuum (Haas et al. 2001). Again, this is only 1 out of 22 sources of the submm ULIRG 
subsample showing this behaviour. Therefore, we see some evidence from our sample that the role of 
the AGN for powering the FIR/submm emission in nearby ULIRGs is negligible.

\subsection{Cold dust and search for evolutionary trends}

\begin{figure}
\resizebox{9cm}{!}{\includegraphics{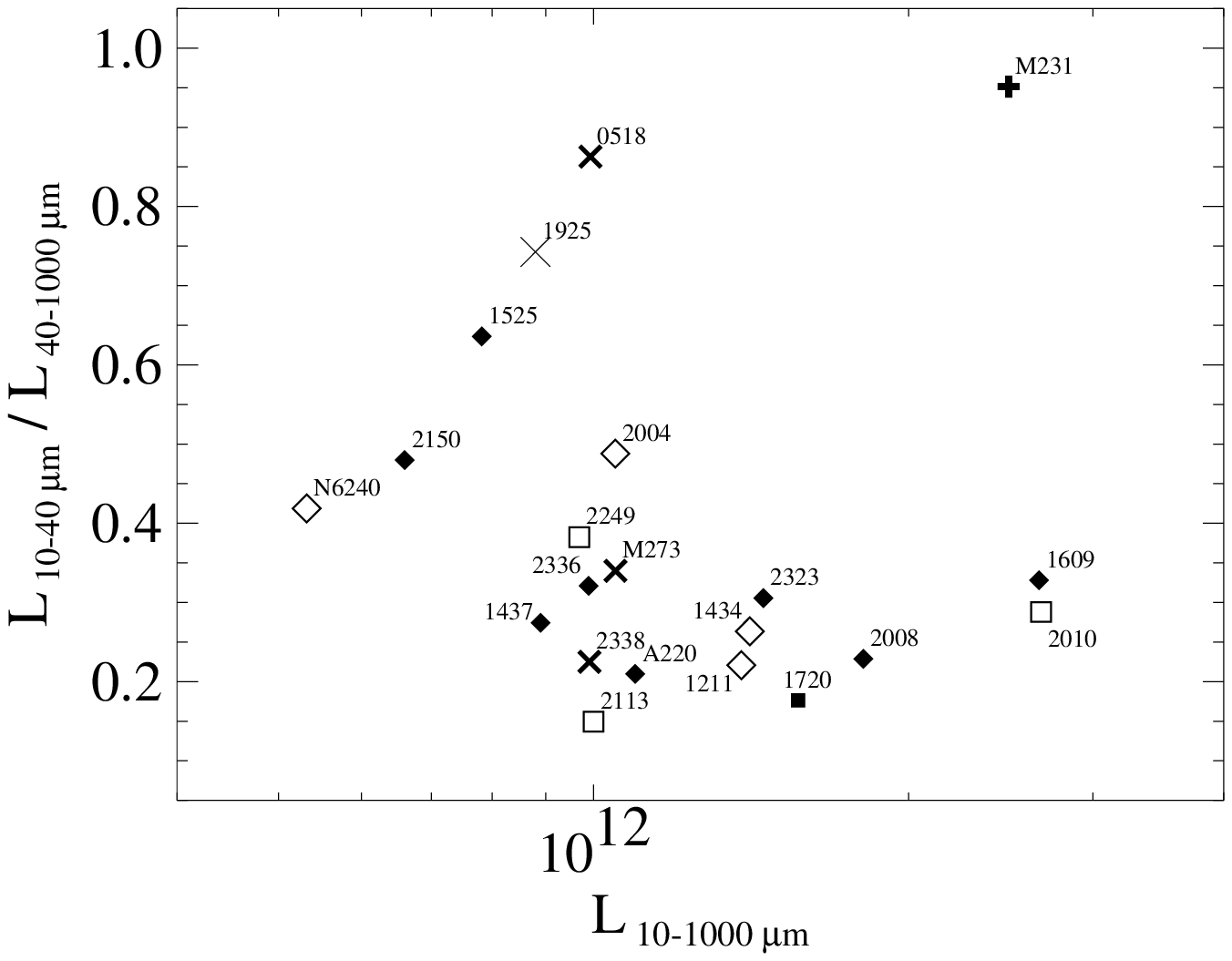}}
\resizebox{9cm}{!}{\includegraphics{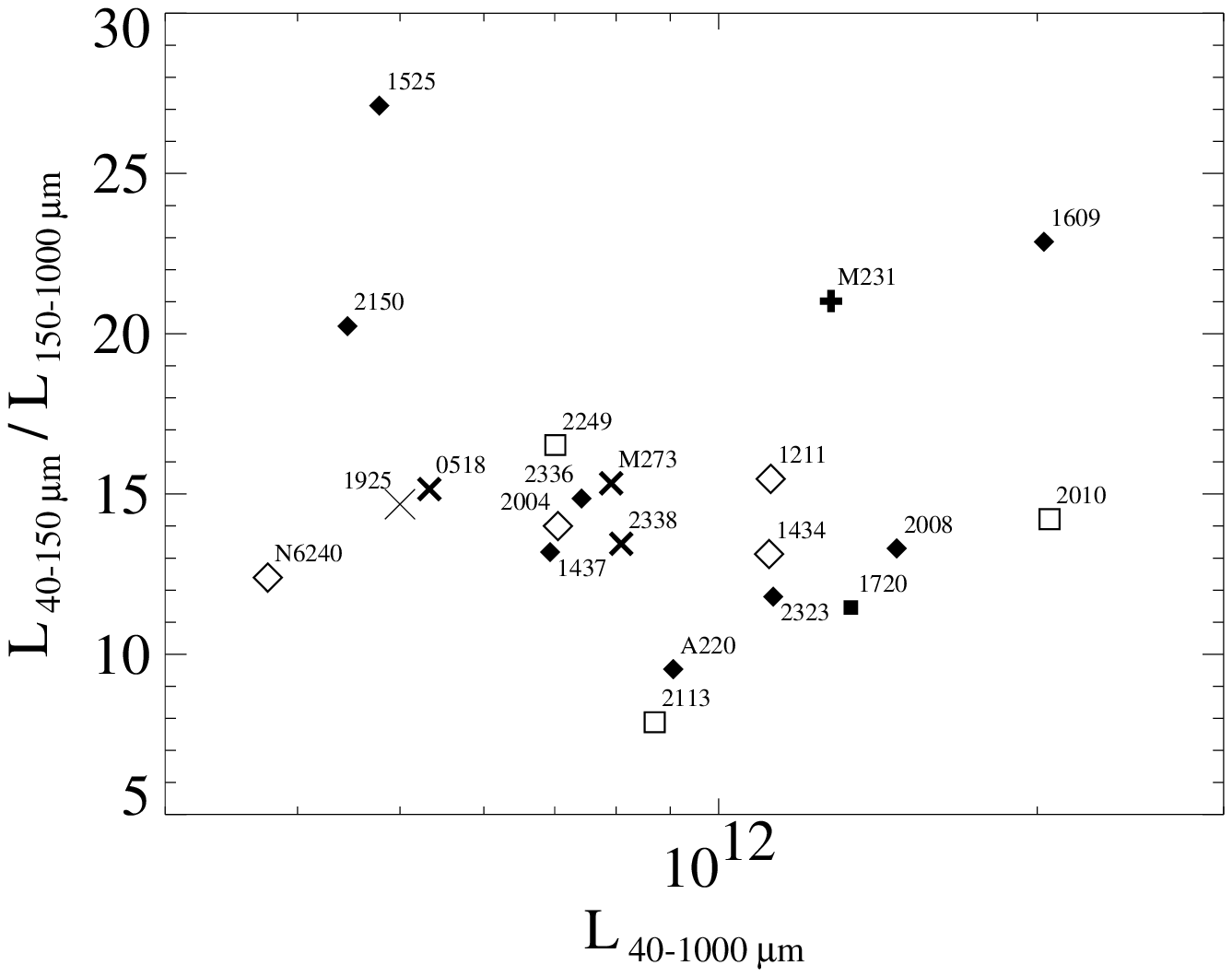}}
\caption[]{
\label{fig_mir_to_submm_vs_morphology} 
Distribution of the MIR/(FIR+submm) luminosity ratio versus total IR luminosity ({\it top}) and 
FIR/submm luminosity ratio versus FIR+submm luminosity ({\it bottom}). Different symbols for each 
spectral type as in Fig.\,\ref{fig_tau_beta_comparison}. The symbol size indicates sources with either a
single nucleus ({\it small and filled or thick}) or a double nucleus ({\it large and open or thin}).}
\end{figure}

ULIRGs may well have considerable amounts of cold dust at 10-30\,K, a temperature typical for dust in less 
active spiral galaxies. If the cold dust (from the parent galaxies) is continuously heated during the 
merging process, then we would expect more advanced mergers to show a relatively larger amount of
warm dust than mergers in a beginning phase. To check this hypothesis, we use the morphological appearance 
and the separation of the galaxy nuclei as a measure for the merger state. This is a simplification, since 
the encounter might not lead to a monotonous approach of the two galaxies. According to simulations by 
Dubinski, Mihos \& Hernquist (1999) the merging process contains repeated approaches with semi-elastical 
collisions followed by a drift apart. Nevertheless, we divided the sources into (1) single and (2) double 
(see Tab.\,\ref{tab_spectral_types}). We use the MIR, FIR and submm luminosities as a measure for the amount 
of warm, cool and cold dust, respectively. Fig.\,\ref{fig_mir_to_submm_vs_morphology} does not show any 
trend of MIR-, FIR- and  submm-luminosity ratios with the simple morphological classifications. Within our 
sample, which covers only a small range in bolometric luminosity, the relative amounts of warm, cool and 
cold dust are not correlated with the merger state inferred from the morphology.

A similar independence was found between the strength of the PAH-feature/MIR-continuum and the (projected) 
separation of the galaxy nuclei (Lutz et al.\ 1998, Rigopoulou et al.\ 1999). Either the (projected) 
separation of the nuclei is not well suited to measure the progress of the merging process, or the dust 
heating does not steadily increase during the merging. It seems likely that while the merging nuclei 
approach each other relatively slowly and/or repeatedly, the dust clouds undergo several phases of 
compression, turning into cold proto-starburst clouds which subsequently undergo heating by starbursts 
(and by an AGN, if present). These distinct phases could occur simultaneously in a merger, but at different 
locations. Photometry of the entire galaxy then shows a mixture of dust complexes at different phases.
This is actually revealed by spatially resolved submm observations of the colliding IR luminous (not yet 
ultra-luminous) galaxy pair NGC\,4038/39 (Haas et al.\ 2000b).

\begin{figure}
\resizebox{9cm}{!}{\includegraphics{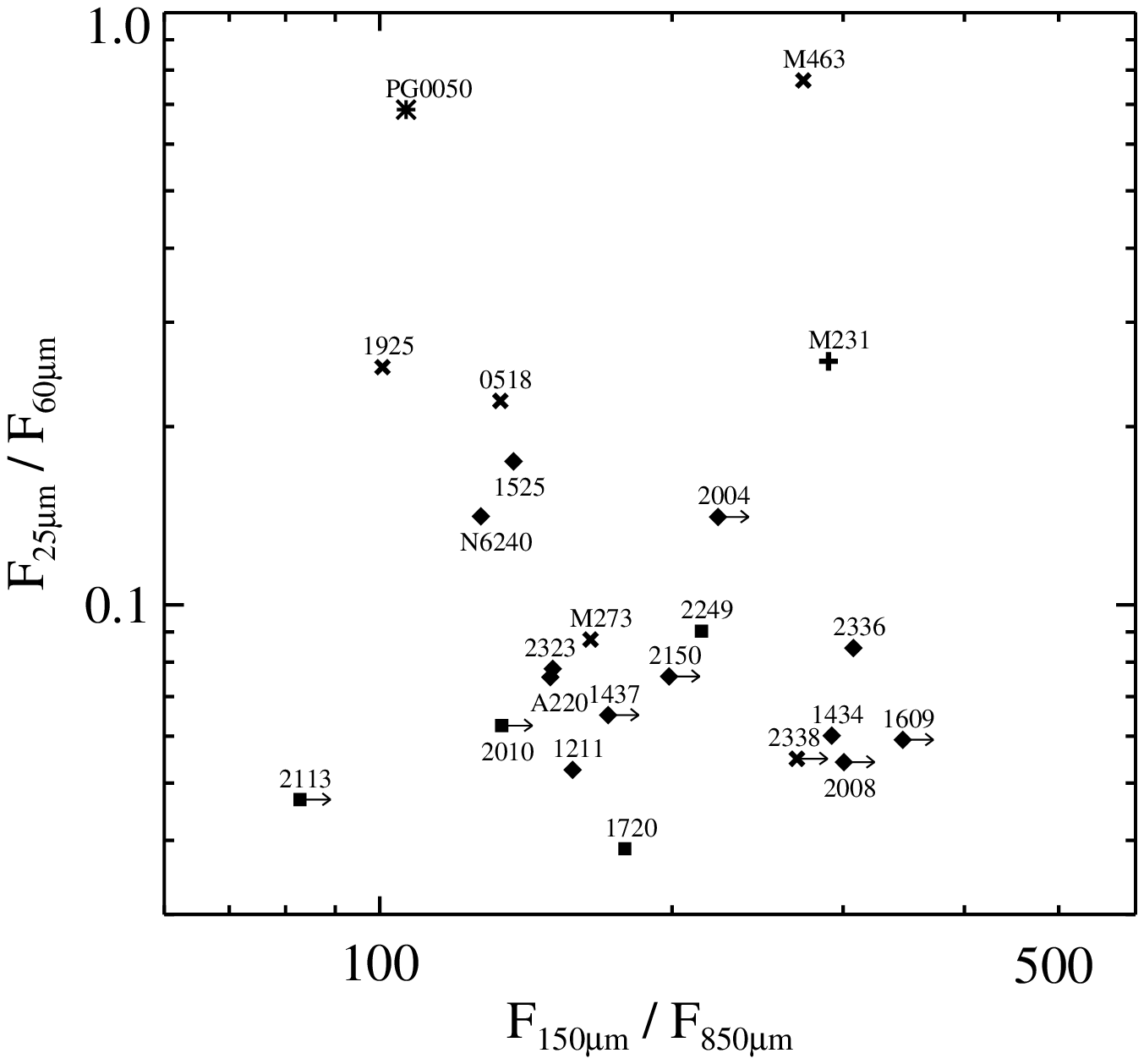}}
\caption{\label{fig_25_to_60_vs_150_to_850}
Two colour diagram (25$\mu$m\,/\,60$\mu$m versus 150$\mu$m\,/\,850$\mu$m) of ULIRGs. Different symbols for 
each spectral type as in Fig.\,\ref{fig_tau_beta_comparison}. Note the relation between this diagram and 
that in Fig.\,\ref{fig_tau_beta_comparison} ($\tau$ versus $\beta$): 25$\mu$m\,/\,60$\mu$m is related to 
1\,/\,$\tau$ and 150$\mu$m\,/\,850$\mu$m to $\beta$.}
\end{figure}

Irrespective of the morphology, the F$_{\rm 25}$/F$_{\rm 60}$ versus F$_{\rm 150}$/F$_{\rm 850}$ two-colour 
diagram (Fig.\,\ref{fig_25_to_60_vs_150_to_850}) illustrates the distribution of our ULIRG mm-subsample.
Schematically, the cold cirrus-like dust is represented by the 850\,$\mu$m emission, the cool SB dust by 
the 150 and 60\,$\mu$m emission, and  the warm AGN dust by the 25\,$\mu$m emission. In this diagram trends 
are recognizable: strong AGNs are located in the upper half (F$_{\rm 25}$/F$_{\rm 60}$\,$\ga$\,0.2), SBs 
in the lower half. The galaxies with and without cold cirrus-like dust lie towards the left and right side, 
respectively. This diagram illustrates that despite similar luminosities the nearby ULIRG sample exhibits 
quite some diversity. Whether or not this is due to evolution is still a puzzle.

Finally, it is interesting to note that the four ULIRGs at medium redshift (z\,$\approx$\,0.3) have similar 
dust temperatures as the low redshift ULIRGs, but exhibit the highest FIR luminosities among the sample.
While their MIR SEDs and their spectral types can be determined soon, the question of whether they also 
contain cold dust will have to be left to future observations with the Herschel Space Observatory or the 
Atacama Large Millimetre Array.

\section{Summary}

Infrared to millimetre spectral energy distributions (SEDs) have been obtained for 41 bright ultra-luminous 
infrared galaxies (ULIRGs). These are the most up-to-date and detailed photometric templates of the nearby 
ULIRGs, which are the fundamental ingredient for cosmological studies. Arp220, which has been considered so 
far as a the archetypal ULIRG, is one of the most peculiar objects in our sample. 

The SED maxima lie between 60 and 100\,$\mu$m, and for those 22 sources with submm detections or upper 
limits the slopes of the Rayleigh-Jeans tails can be well constrained. The FIR and submm parts of the SEDs 
between 60 and 1300\,$\mu$m can be fitted in two ways: 

Firstly, with a single modified blackbody yielding large ranges for both the opacity (0.1 $<$ 
$\tau$$_{\rm 100\mu m}$ $<$ 5) and the emissivity (1.2 $<$ $\beta$ $<$ 2.2). The resulting temperatures 
range from 50 to 70\,K. However, there are several reasons against the physical relevance of one single 
dust component: 
\begin{itemize}
\item[1)] The optical-NIR morphology is generally disturbed, suggesting the presence of
          diffuse outer dust layers in addition to a dense central dust body. 
\item[2)] It is difficult to understand why $\beta$ is significantly $<$\,2 in {\it some} ULIRGs, 
          but $\beta$ $\approx$ 2 for spiral and Markarian galaxies.
\item[3)] The normalised strength of the 7.7\,$\mu$m PAH features shows no correlation 
          with the extinction derived formally from $\tau$$_{\rm 100\mu m}$. 
\item[4)] The derived dust masses lead to unusually high gas-to-dust mass ratios from 500 to 2300.
\end{itemize}

Secondly, the FIR-submm SEDs can be decomposed into two or more $\lambda$$^{\rm -2}$ modified blackbody 
spectra representing cool (30-50\,K) and cold (10-30\,K) dust components. While the cool component 
should be moderately opaque (0.3\,$\la$\,$\tau^{\rm bulk}_{\rm 100\mu m}$\,$\approx$\,1) to match size 
constraints inferred from interferometric CO observations, the cold component should be transparent 
in order to conform to the observed PAH/100\,$\mu$m and PAH/850\,$\mu$m ratios. Remarkably, this suggests 
that the bulk FIR dust component cannot contain the majority of the PAH carriers, rather they are more 
closely coupled with the submm emitting dust.

This second way to fit the SEDs provides evidence favouring two dust stages: 
\begin{itemize}
\item[1)] the cold dust in more quiet regions, which is mainly responsible for the submm emission, and 
\item[2)] the cool dust powered by starbursts (SB) which dominates the FIR and may be partly optically 
          thick even in the FIR. 
\end{itemize}

The dust masses derived from such a multiple component model yield gas-to-dust mass ratios in the 
range between 30 and 300, hence close to the galactic value of 150. ULIRGs classified as either AGNs 
or SBs appear indistinguishable at FIR and submm wavelengths. We suppose that the cool FIR emitting dust 
is not related to the AGN, in particular since SB-type ULIRGs seem to be more than twice as frequent
in our sample as AGN-type ULIRGs.

In the NIR-MIR the SEDs reveal two basic shapes:
\begin{itemize}
\item[1)] A flat NIR-MIR flux plateau followed by a steep rise at about 10\,$\mu$m. 
          All objects optically classified as LINERs and HII/SBs exhibit this SED shape, 
          but so do some of the Seyferts.
\item[2)] A power--law like flux increase starting in the NIR and peaking in the FIR or even MIR.
          This SED shape is only seen in Seyferts.
\end{itemize} 
The NIR-MIR power-law-like emission is due to warm and hot dust powered by the AGN, and this is a 
third dust stage restricted to AGN-type ULIRGs. From this finding, a NIR photometry diagnostic tool is 
established to reveal the presence of an AGN on the basis of the J-K colours (which has red colours, 
while SBs have blue colours). This tool is successful in the sense that if an AGN is seen by other 
diagnostics (optical or MIR spectroscopy), in most cases (6 out of 8) it is also seen via the NIR colours.

The detection or non-detection of cold dust could reflect different evolutionary states of the ULIRGs 
associated with the merger state. However, the current means to determine the merger state, like the 
projected distance of the nuclei, appears to be insufficient to verify evolutionary trends.

\appendix
\section {Near-Infrared images}

\begin{figure*}
\centerline{\hbox{
\epsfig{file=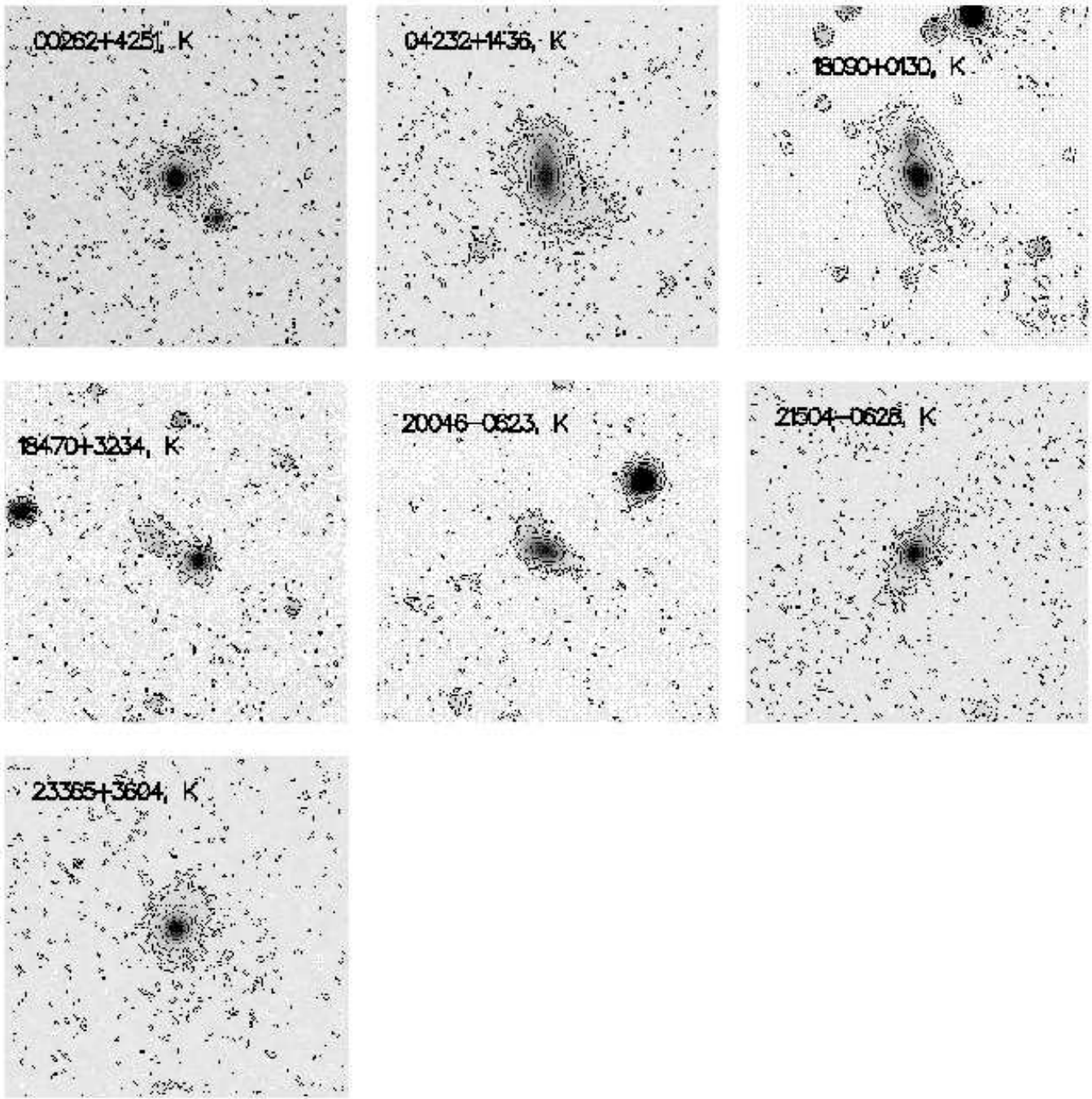,width=18cm,clip=true}
}}
\caption{\label{fig_k_images} 
        K-Band images showing a field of 60$\arcsec$$\times$60$\arcsec$ with the sources located in the 
        image centres, greyscales with logarithmic contours.}
\end{figure*}

The J- and K-band images look very similar, therefore only the K-band images are shown 
(Fig.\,\ref{fig_k_images}). Most of the emission is concentrated, but also irregular/extended tails 
are present. The pointlike blue source 10$\arcsec$ southwest of 00262+4251, and the two ones located 
about 6$\arcsec$ north and south of 18090+0130 are probably stars (and are excluded from the photometry).
The photometry derived from the images is listed in Tab.\,\ref{tab_fluxes}.

\section {Optical spectra}

\begin{figure*}
\centerline{\hbox{
\resizebox{9cm}{6cm}{\includegraphics[clip=true]{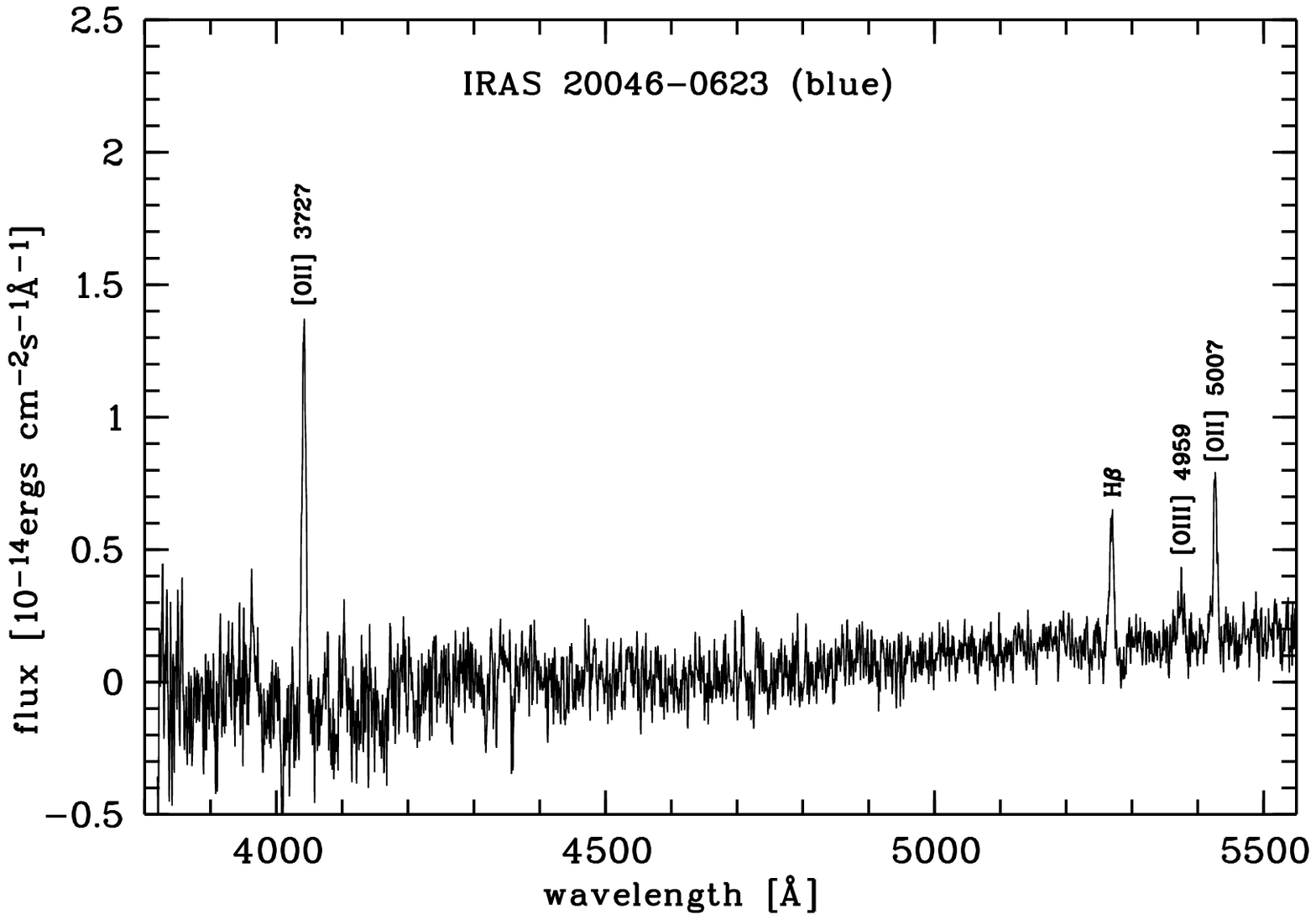}}
\resizebox{9cm}{6cm}{\includegraphics[clip=true]{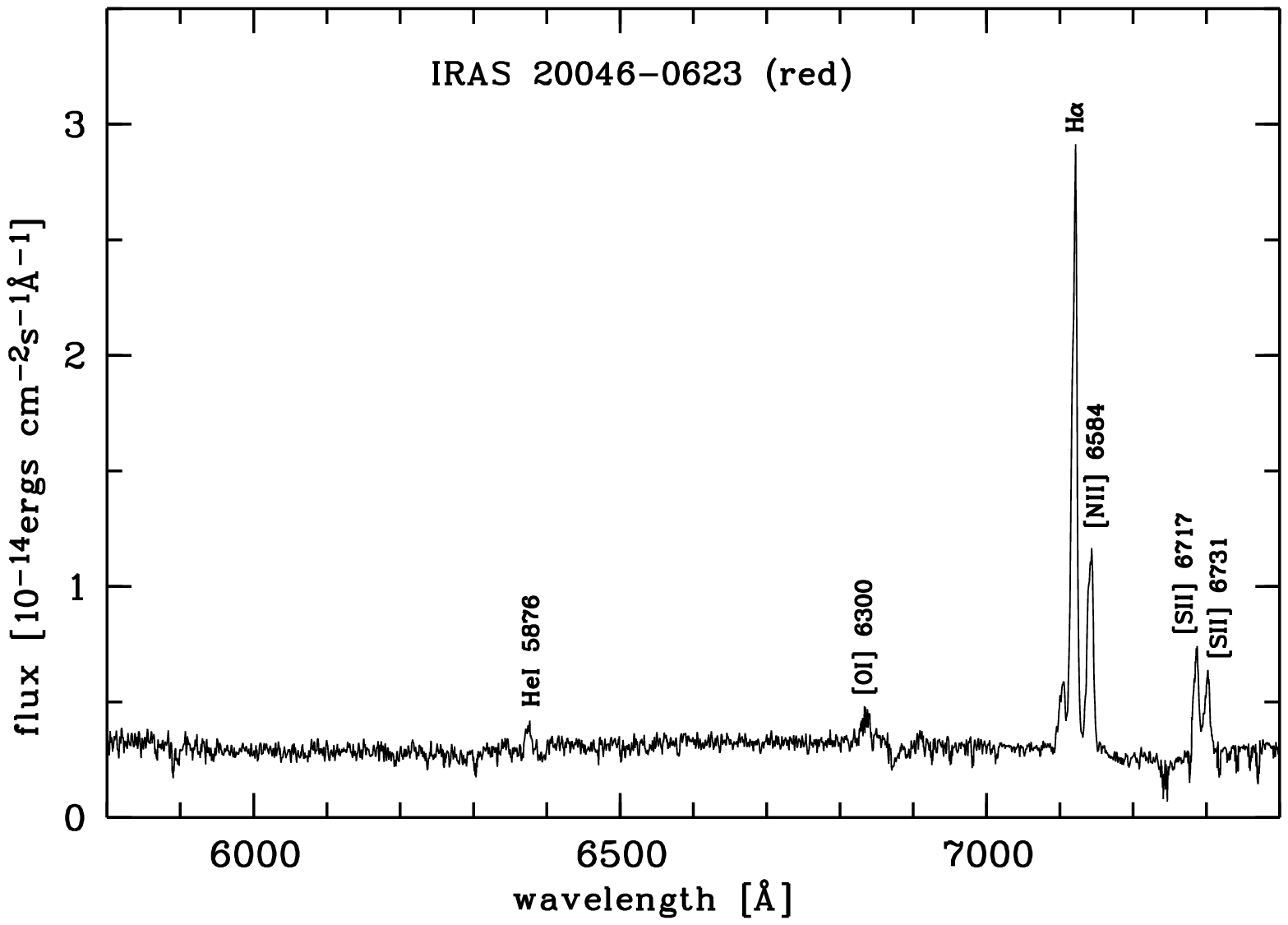}}
}}
\centerline{\hbox{
\resizebox{9cm}{6cm}{\includegraphics[clip=true]{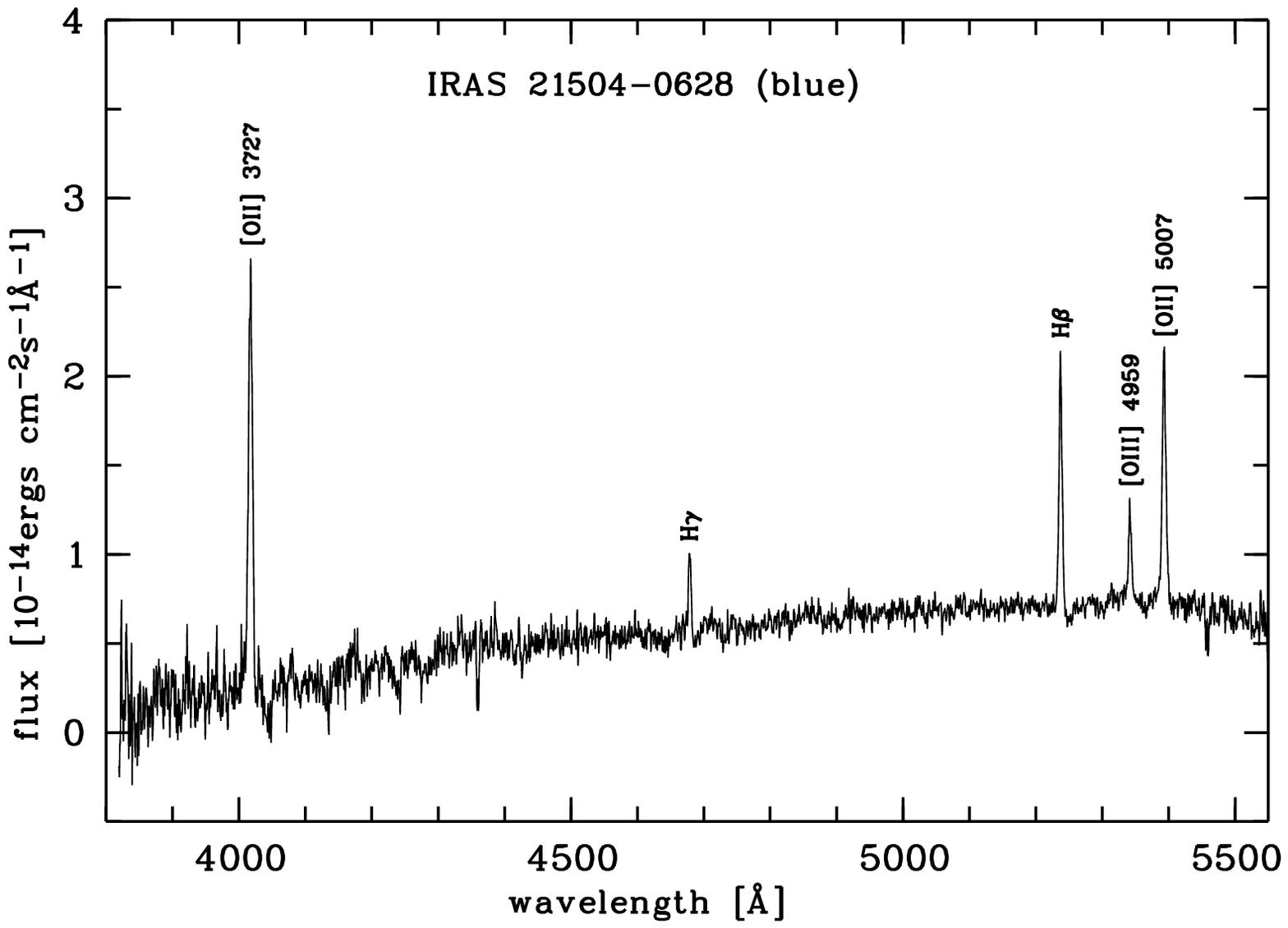}} 
\resizebox{9cm}{6cm}{\includegraphics[clip=true]{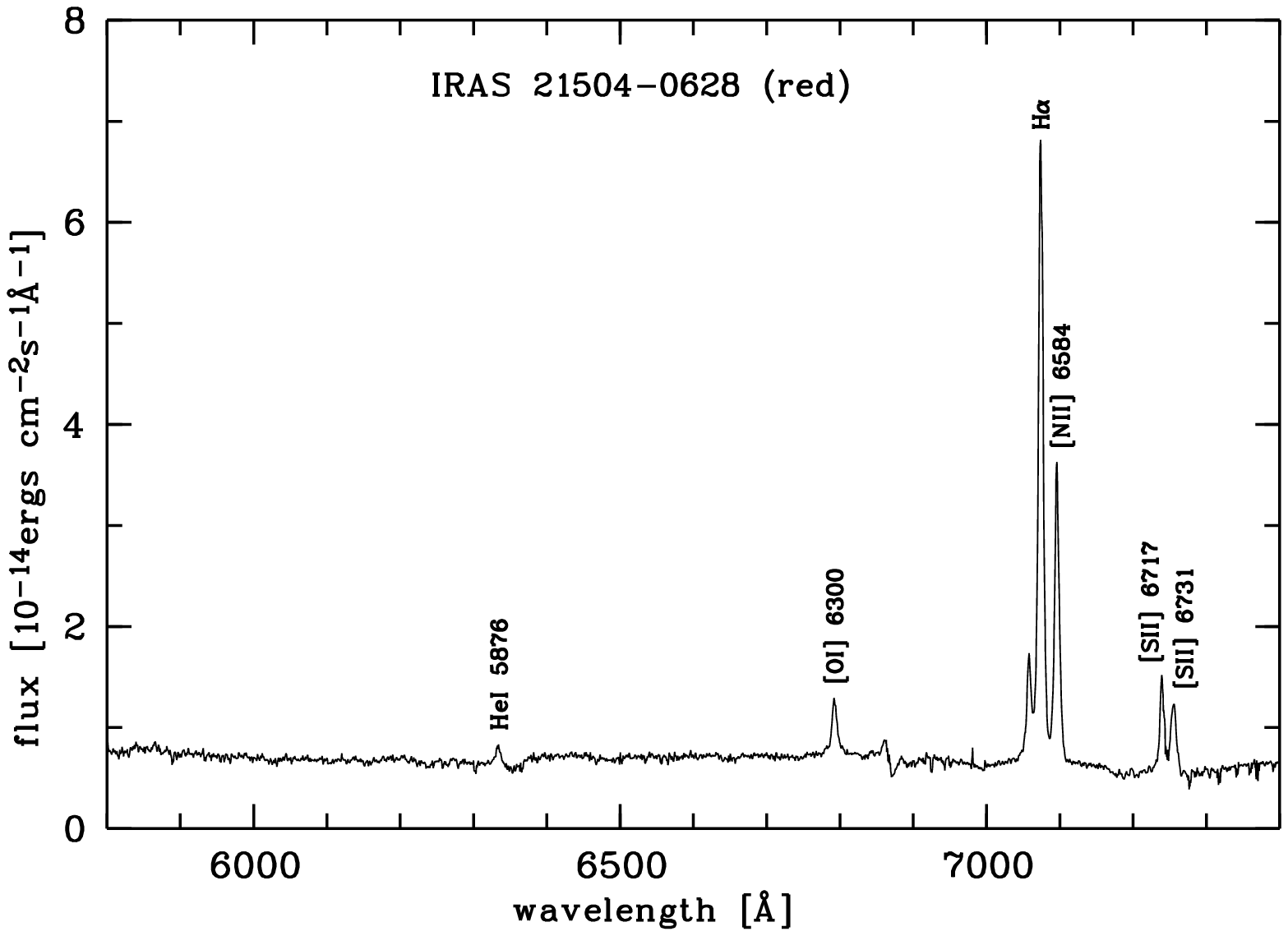}}
}}
\caption{\label{fig_twin_spectra} 
Spectra of 20046-0623 (top) and 21504-0628 (bottom), 
blue (left) and red (right) channel of the TWIN spectrograph.
}
\end{figure*}

Fig.\,\ref{fig_twin_spectra} shows optical spectra for two sources, 
obtained with the TWIN spectrograph at the 3.5\,m\,telescope on 
Calar Alto, Spain. Both sources have a LINER spectrum according 
to the line diagnostics by Veilleux \& Osterbrock (1987). Now all 
sources of our mm-subsample are spectroscopically classified (see Tab.\,\ref{tab_spectral_types}).

\acknowledgements

The development and operation of ISOPHOT were supported by MPIA and funds from Deutsches Zentrum 
f\"ur Luft- und Raumfahrt (DLR, formerly DARA). The ISOPHOT Data Centre at MPIA is supported by 
Deutsches Zentrum f\"ur Luft- und Raumfahrt e.V. (DLR) with funds of Bundesministerium f\"ur Bildung 
und Forschung, grant no. 50\,QI\,9801\,3. The authors are responsible for the contents of this 
publication. JCMT is operated by the Joint Astronomy Centre on behalf of the Particle Physics and 
Astronomy Research Council of the United Kingdom, the Netherlands Organisation for Scientific Research, 
and the National Research Council of Canada.

It is a pleasure for us to thank Dr. Uwe Graser for kindly obtaining the optical spectra during 
operational tests, and Drs. Robert D. Joseph and Jos\'e-M\'{\i}guel Rodr\'{\i}guez-Espinoza for 
stimulating discussions. We thank the referee, Dr. Suzanne Madden, for a comprehensive set of 
valuable comments. 

For literature search and photometry we used the NASA/IPAC Extragalactic Data Base (NED) and the 
NASA Astrophysics Data System (ADS).

\end{document}